\documentclass[acmsmall,screen]{acmart}
\AtBeginDocument{%
  }

\setcopyright{acmlicensed}
\copyrightyear{2026}
\acmYear{2026}
\acmDOI{XXXXXXX.XXXXXXX}

\setcopyright{none}
\renewcommand\footnotetextcopyrightpermission[1]{}
\usepackage{booktabs}
\usepackage{multirow}

\usepackage[dvipsnames]{xcolor}
\usepackage{tcolorbox}

\usepackage{graphicx}
\usepackage[linesnumbered,ruled,vlined]{algorithm2e}

\usepackage[table]{xcolor} 

\usepackage{array}
\usepackage{makecell}
\usepackage{pifont}
\usepackage{ragged2e}

\usepackage{tabularx}
\usepackage{array}
\usepackage{enumitem}

\definecolor{PromptBorder}{RGB}{150,160,175}
\definecolor{PromptTitleBg}{RGB}{236,240,246}
\definecolor{PromptSection}{RGB}{78,96,128}

\newcolumntype{C}[1]{>{\centering\arraybackslash}m{#1}}
\newcolumntype{L}[1]{>{\raggedright\arraybackslash}m{#1}}

\newcommand{\cmark}{\ding{51}}
\newcommand{\xmark}{\ding{55}}

\begin{document}

\title{From Language to Action: Enhancing LLM Task Efficiency with Task-Aware MCP Server Recommendation}
%Hybrid View Alignment for Task-Oriented MCP Recommendation
\author{Shiyu He}
\orcid{0009-0001-7782-9373}
\email{syhe@mails.ccnu.edu.cn}

\author{Zhiman Chen}
\orcid{0009-0007-7381-4581}
\email{chenzhiman@mails.ccnu.edu.cn}

\author{Yuqi Zhao}
\orcid{0000-0002-2642-5109}
\email{yuqizhao@ccnu.edu.cn}

\author{Neng Zhang}
\orcid{0000-0001-8662-5690}
\email{nengzhang@ccnu.edu.cn}

\author{Ran Mo}
\orcid{0000-0001-7556-153X}
\email{moran@ccnu.edu.cn}

\author{Yutao Ma}
\authornote{Corresponding author.}
\orcid{0000-0003-4239-2009}
\email{ytma@ccnu.edu.cn}
\affiliation{%
  \institution{School of Computer Science \& Hubei Provincial Key Laboratory of Artificial Intelligence and Smart Learning, Central China Normal University}
  \city{Wuhan}
  \state{Hubei}
  \country{China}}

\renewcommand{\shortauthors}{He et al.}

\begin{abstract}
The rapid expansion of the model context protocol (MCP) ecosystem enables large language model (LLM)-based agents to access a wide range of external tools via a standardized interface. However, identifying appropriate MCP servers for a specific development task remains challenging. This difficulty arises from several factors, including ecosystem fragmentation, mismatching between task requirements and tool constraints at both the semantic and structural levels, and the absence of transparent rationales for recommendation results. Existing studies primarily focus on measuring the MCP ecosystem or optimizing tool invocation mechanisms, while systematic recommendation frameworks and reproducible benchmarks for real-world development tasks remain largely unexplored.
To address this limitation, we formulate task-oriented MCP server recommendation as a structured retrieval-and-ranking problem that jointly considers semantic relevance and engineering constraints. We first construct Task2MCP, a task-centered dataset that systematically associates taxonomy-grounded development tasks with curated MCP servers. This dataset provides structured supervision and a reproducible evaluation environment for research on MCP tool recommendations. Building on this dataset, we propose T2MRec, a task-to-MCP server recommendation model. It models semantic relevance and structural compatibility to construct an initial candidate set. Then it improves coverage and ranking quality through centroid-based candidate expansion and constrained LLM-based re-ranking.
In addition, we design and implement an interactive MCP server recommendation agent prototype that operates in conversational environments to support dynamic decision-making. The agent assists developers in efficiently evaluating and integrating tools by providing recommended MCP servers together with usage guidelines.
Experimental results indicate that T2MRec consistently improves ranking performance over traditional, graph-based, and LLM-based recommendation models across multiple evaluation metrics. These findings provide practical insights into task-aware MCP server recommendation and contribute a reproducible benchmark for future research on tool selection in LLM-based agent systems.

\end{abstract}






\maketitle

\section{Introduction}
\label{sec:Introduction}

Large language models (LLMs) have demonstrated remarkable capabilities in code generation~\citep{LiuFLWZZX25}, multi-step reasoning~\citep{0002WR0W025}, and instruction following~\citep{ZhangY0HL25}, leading to their increasing adoption in software development workflows~\citep{HeTL25}. Despite these advances, a significant challenge remains: LLM-based agents often struggle to select appropriate tools for specific tasks~\citep{MaH0WZL25}. Rather than lacking reasoning power, these agents typically rely on trial-and-error approaches to identify usable tools. This difficulty is further compounded as the software tool ecosystem continues to expand, creating a complex environment in which LLMs must operate effectively.

The model context protocol (MCP) offers a promising solution by providing a unifying standard for accessing and invoking external tools consistently. By decoupling tool implementation from agent logic, MCP makes it easier to integrate diverse tools into LLM-based agent systems. As a result, the number and diversity of available MCP servers have grown rapidly~\citep{0014W025}. Despite this progress, MCP primarily addresses how tools can be accessed rather than which tools should be selected for a given task. This gap in effective tool recommendation presents several challenges:

(1) \textit{MCP ecosystem fragmentation}. 
Although the MCP ecosystem has grown rapidly, existing MCP servers are distributed across heterogeneous platforms with inconsistent descriptions, incomplete metadata, and varying levels of documentation. \textbf{This fragmentation makes it difficult for developers and agents to obtain a unified view of available MCP servers or to discover reusable tools for specific development tasks systematically}. Therefore, MCP discovery often relies on manual search or prior knowledge, which introduces a strong exposure bias toward well-known MCP servers or frequently referenced MCP tools. Many potentially suitable but less visible MCP servers remain underutilized, leading to redundant tool implementations and inefficient reuse of existing capabilities.

(2) \textit{Semantic–structural mismatch}. 
Most existing tool recommendation approaches emphasize semantic similarity between task descriptions and tool documentation, assuming that semantic relevance is sufficient for effective tool selection. \textbf{However, in practical development scenarios, an MCP server that appears semantically relevant may still be unusable due to mismatches in programming language, system environment, or functional interfaces}. This mismatch arises from treating MCP server recommendations as a purely semantic matching problem while overlooking structural constraints that determine whether a tool can be directly integrated into a development workflow. Consequently, LLM-based agents may select MCP tools that conceptually match the task intent but fail during execution, leading to fragile pipelines and higher trial-and-error costs.

(3) \textit{Opaque recommendation rationale}. 
Many existing MCP server recommendation processes operate as black boxes, providing developers with little explanation of why a particular MCP server is selected for a given task. \textbf{Without transparent rationales, developers cannot easily assess the suitability of recommended MCP servers or understand how to use them in their solutions}. The lack of interpretability undermines developer trust and limits the effectiveness of human-in-the-loop development workflows. In practice, developers often need not only a recommendation result but also concrete reasoning and usage guidance to determine whether an MCP server truly meets their task requirements and to integrate it efficiently into their system implementation.

In response to the above challenges, recent studies have begun to explore techniques to improve tool selection and use in LLM‑based agent systems. Within the MCP ecosystem, prior work has systematically collected and analyzed MCP servers and tools from measurement and dataset-construction perspectives~\citep{LinRLZ25, guo2025measurement}. These efforts provide valuable insights into ecosystem scale, capability distribution, and potential security risks, thereby improving our understanding of what resources exist in the MCP landscape. However, existing studies primarily focus on ecosystem characterization rather than task-oriented decision making. While registries and metadata help catalog available MCP servers, they do not address the core question of which server or tool to invoke for a specific user task. In practice, selecting an appropriate MCP server requires aligning task intent with both semantic relevance and structural constraints, such as programming language compatibility, interface requirements, and execution environments. Current MCP ecosystem datasets often exhibit inconsistent granularity, incomplete capability descriptions, and non-standard naming conventions, further hindering reliable task-to-tool mapping. Despite the rapid growth of MCP infrastructure, LLM-based agent systems still lack principled recommendation mechanisms for automatically retrieving suitable MCP servers and generating executable invocation plans grounded in real-world task requirements. Bridging this gap requires moving beyond static ecosystem measurement toward task-aware, structurally constrained, and interpretable recommendation frameworks.

To address this gap, this article introduces a large-scale, real-world dataset, \textbf{Task2MCP}, for the task-oriented MCP ecosystem, comprising 4,800 development tasks and 5,642 MCP servers. This dataset overcomes the limitations of existing MCP resources by offering a diverse and comprehensive collection of realistic development tasks paired with structured MCP metadata. On average, each task is annotated with a category, subcategory, programming language, theme, and a ranked list of compatible MCP servers. Task2MCP is poised to be an invaluable resource for advancing the development of task-aware MCP recommendation models. To construct this dataset, we first collect MCP server entries from three widely used online directories, which cover a diverse array of tool types, programming languages, and service domains. We then enrich these entries with GitHub repository metadata, perform deduplication and normalization, and apply a validation workflow to ensure data quality. This process results in a high-confidence dataset comprising 5,642 MCP servers and 4,800 realistic development tasks. 

To demonstrate the potential of Task2MCP for improving task-oriented MCP server recommendations, we train a recommendation model, \textbf{T2MRec}, designed to rank MCP servers for realistic development tasks and establish a benchmark for fair comparison. The model combines the semantic relevance of textual descriptions with the structural compatibility of engineering constraints. Specifically, T2MRec first ranks MCP server candidates based on their alignment with the given task, then refines possible recommendations through centroid-based expansion and LLM-based re-ranking. This robust, task-aware approach aims to enhance the accuracy and reliability of tool recommendations, reducing the trial-and-error process traditionally associated with tool selection. Furthermore, we also develop a \textbf{recommender artificial intelligence (AI) agent} for MCP servers designed to operationalize this approach in a real-world, interactive environment. This agent provides developers with dynamic, task-specific MCP server or tool recommendations, along with detailed usage guidelines and explanations for each recommendation. It supports multi-turn dialogue, allowing developers to adjust their tool selections as project requirements and constraints evolve. The agent also delivers a comparative analysis of multiple tool options, improving transparency (or interpretability) and helping developers make more informed decisions. For more details on our dataset and model, please refer to the open-source project available on GitHub\footnote{https://github.com/ssea-lab/Task2MCP}.

The main contributions of our work are as follows:
\begin{itemize}
\item \textbf{New Dataset for MCP Research.} We construct the Task2MCP dataset, a large-scale, task-oriented benchmark dataset that systematically links realistic development tasks to curated MCP servers. The dataset covers thousands of MCP servers and tasks organized under a taxonomy-grounded hierarchy, providing high-quality supervision for MCP server recommendation and other MCP-related research fields.
\item \textbf{New Benchmark for MCP Server Recommendation.} We propose T2MRec, a task-to-MCP recommendation model that jointly models semantic relevance and structural compatibility, and further refines candidate rankings through a constrained LLM-based re-ranking module. The model is designed to be scalable, reproducible, and suitable for real-world engineering scenarios. We then conduct extensive experiments and analyses on the Task2MCP dataset, demonstrating that the T2MRec model consistently outperforms selected traditional, graph-based, and LLM-based baselines. Our results provide empirical insights into the roles of semantic modeling, structural constraints, and LLM refinement in task-oriented MCP server recommendation.
\item \textbf{New Recommender AI Agent for System Implementation.} We design and implement an interactive recommender AI agent for MCP server recommendations, which operationalizes the T2MRec model in a real-world conversational environment for developing LLM-based agent systems. The agent recommends MCP servers, explains their suitability for a given task, and provides actionable guidance on tool usage and execution workflows, thereby supporting human-in-the-loop decision-making to facilitate system development.
\end{itemize}

The remainder of this article is organized as follows. Section~\ref{2} reviews the related work. Section~\ref{3} introduces the construction of the Task2MCP dataset. Section~\ref{4} details the proposed T2MRec model. Section~\ref{5} presents the experimental studies we conducted and analyzes the experimental results in depth. Section~\ref{6} shows a case study demonstrating the recommender AI agent's practical application. Section~\ref{7} discusses potential threats to the validity of our work. Finally, Section~\ref{8} concludes the article and outlines our future work.

\section{Related Work}
\label{2}

\subsection{MCP Ecosystem}

MCP has recently emerged as a practical interface layer for connecting LLM-based agents with external tools, services, and data sources~\citep{abs-2503-23278}. This standardization has accelerated the growth of the MCP ecosystem by simplifying the integration of agents and tools~\citep{WangZLFJZLX26}. However, MCP servers are distributed across heterogeneous platforms, and there are significant variations in metadata structures, resource quality, and document consistency. To address these challenges, researchers are currently exploring the MCP ecosystem from three complementary perspectives: ecosystem measurement, tool management, and benchmark evaluation.

Early studies mainly focus on understanding the MCP ecosystem itself. For example, MCPCorpus~\citep{LinRLZ25} constructs a large-scale, evolvable dataset of MCP servers and clients, enabling empirical analysis of protocol adoption, implementation diversity, and potential security risks. In a similar spirit, Guo et al.~\citep{guo2025measurement} presented a market-level measurement view of MCP by crawling and normalizing artifacts from multiple MCP marketplaces, thereby revealing the MCP ecosystem's structural properties and growth patterns. Complementing these measurement-oriented efforts, MCPZoo~\citep{abs-2512-15144} emphasizes the collection of runnable MCP servers and provides a resource that is better aligned with execution-level evaluation and agent-tool interaction studies. These previous studies established an important empirical foundation for understanding what kinds of MCP resources exist, how they are organized, and how they may support downstream LLM-based agent systems.

A second line of work investigates MCP from the perspective of tool discovery and retrieval. Rather than treating MCP servers as static protocol endpoints, these efforts examine how LLM-based agents can identify, rank, and compose relevant tools under natural language intents. For example, MCP-Zero~\citep{fei2025mcpzero} formulates proactive toolchain construction as a retrieval-oriented problem and compiles an MCP tool corpus for evaluating server and tool discovery. Building upon such structured MCP resources, HumanMCP~\citep{laddha2025humanmcp} further introduces human-like queries for evaluating tool retrieval, thereby moving beyond template-style requests toward more realistic user expressions. This line of research highlights that, in practical deployments, the bottleneck of LLM-based agents using MCP often lies not only in tool execution but also in accurately locating the right tool under ambiguity, scale, and heterogeneous descriptions.

The third research direction is the construction of benchmarks for capability and robustness evaluation. For example, MCP-RADAR~\citep{abs-2505-16700} and MCP-Universe~\citep{abs-2508-14704} develop representative task suites over MCP-enabled environments to assess how well LLM-based agents can use tools in realistic scenarios. MCP-Bench~\citep{abs-2508-20453} and MCPMark~\citep{abs-2509-24002} further extend this paradigm by stressing multi-step reasoning, cross-tool coordination, and more comprehensive MCP workflows. MCP-AgentBench~\citep{GuoXZHWM26} further evaluates real-world agent performance with MCP-mediated tools. MCPTox~\citep{WangGWLSCSDL26} benchmarks the risks of tool poisoning on real-world MCP servers. In domain-specific settings, IoT-MCP~\citep{YangLMLLGYZC025} studies MCP-enabled interaction in Internet of Things environments and evaluates agent performance over both basic and complex tasks. In parallel with capability evaluation, MCPSecBench~\citep{abs-2508-13220} shifts attention to the attack surface introduced by MCP and provides a systematic security benchmark that covers multiple threat categories, clients, and platforms. These efforts collectively suggest that MCP should be evaluated not merely as a protocol abstraction, but as an end-to-end agentic infrastructure involving discovery, planning, invocation, and defense.

In brief, the above studies have primarily focused on ecosystem-wide collection, measurement, and characterization of MCP resources~\citep{abs-2504-03767}. Although these studies improve our understanding of the resources available in the MCP ecosystem, they offer limited guidance on selecting task-time servers in practical development settings. In real-world use, developers and LLM-based agents must still identify appropriate servers from a large and noisy candidate pool, verify whether their capabilities align with the target task, evaluate interface and dependency compatibility, and endure substantial trial-and-error due to invalid, low-quality, or incompatible resources. Consequently, there remains a clear gap between ecosystem-level resource awareness and dependable, on-demand selection and invocation of MCP servers.

\subsection{Task-Oriented Recommendation}

Task-oriented recommendation in software engineering concerns the automatic matching of task demands to candidate objects capable of fulfilling them, such as Web services, tools, application programming interfaces (APIs), or development resources~\citep{LiuZPYWS23}. In contrast to traditional recommender systems that primarily focus on estimating user-item preferences, task-oriented recommendations must account not only for task relevance but also for feasibility constraints~\citep{MaWKBP024, YuYXCLH24}. The central challenge lies in accurately aligning tasks with users' needs to achieve precise, personalized task assignment~\citep{HeZLM26SPE}. Accordingly, the literature has evolved from early heuristic and collaborative filtering methods to graph-based representation learning and, more recently, to LLM-driven retrieval and reasoning.

\subsubsection{Traditional Methods}

Traditional task-oriented recommendation methods can be broadly categorized into content-based, collaborative filtering, and matrix factorization approaches. Content-based methods rank candidate resources by estimating their relevance to a given task description, typically through textual or code-level matching~\citep{LiaoSW22, ZengGC23}. This paradigm has remained competitive due to its simplicity and interpretability, and recent studies have substantially strengthened its semantic modeling capabilities~\citep{WangR0YLCXYZRC024}. For example, Wei et al.~\citep{wei2022clear} proposed CLEAR, which combines BERT-based sentence representations with contrastive learning to improve semantic matching for API recommendation. Li et al.~\citep{LiLTHGLNWHZ24} further developed PTM-APIRec, which leverages pretrained source-code models to encode program context and improve context-aware recommendation. Although these methods can improve the modeling of task semantics, they still depend heavily on the quality and explicitness of task descriptions and resource documentation. Their effectiveness, therefore, tends to degrade when the task goal (or intent) is under-specified or when resource capability boundaries are not clearly documented.

To reduce this dependence on explicitly expressed task semantics, collaborative filtering and matrix factorization models recommend items based on historical interaction data~\citep{SilvaMH24, AhmadianBRFMJ25}. Collaborative filtering infers task relevance from users' past behaviors and shared preference patterns. For instance, Pu et al.~\citep{PuC00LC24} proposed the RG2 loss for collaborative filtering, improving recommendation performance and convergence speed. Matrix factorization builds on this idea by decomposing sparse interaction matrices into low-dimensional latent factors, thereby uncovering implicit associations among users, tasks, and candidate resources~\citep{ZhengNZL24}. For example, Yuen et al.~\citep{YuenKL21} proposed a time-aware task recommendation framework that integrates probabilistic matrix factorization with temporal preference modeling and online updating. Such matrix factorization methods are particularly effective when explicit features are limited and sufficient interaction data are available~\citep{YuYWYH24}. However, both collaborative filtering and matrix factorization remain vulnerable to cold-start settings and sparse supervision, which restricts their applicability in open and fast-evolving task recommendation scenarios~\citep{WuHWZW23}.

\subsubsection{Graph Neural Network-Based Methods}

Graph-based methods have become an important paradigm for task-oriented recommendation because they capture structured dependencies among tasks, resources, and historical interactions~\citep{AnandM25}. Their advantages are particularly evident in sparse, heterogeneous, and cold-start settings. Prior studies have shown that graph-based modeling can improve crowdsourcing task matching~\citep{TianHZCLY25}, enhance few-shot recommendation~\citep{YangY25}, and strengthen strict cold-start recommendation~\citep{CaoYWLPYY23}, suggesting that relational context provides useful signals for inferring task-resource suitability.

In software engineering, this line of research has evolved from semantic matching to explicit relational modeling~\citep{KimK23}. Until now, neural graph collaborative filtering (NGCF)~\citep{Wang0WFC19} and LightGCN~\citep{he2020lightgcn} have established strong graph-based recommendation baselines through neighborhood propagation. In task-oriented settings, Req2Lib~\citep{SunLCYC20} performs requirement-to-library matching, but it does not model higher-order relations among tasks, libraries, and projects. MLTaskKG~\citep{LiuZPYWS23} advances this direction by organizing tasks, models, implementations, and repositories into a knowledge graph for task-conditioned recommendation. GRec~\citep{LiHC0LGY21} and HGNRec~\citep{LiGWQLXZ24} further strengthen this line by learning graph representations over library interaction data, enabling more effective modeling of higher-order dependency signals.

Compared with traditional feature-driven methods, graph-based approaches better exploit relational evidence underlying task-resource matching. However, their effectiveness still depends on informative and sufficiently complete graph structures~\citep{JuYWXMLGQYWLYZ26}. In open and fast-changing development environments, graph construction may be noisy or incomplete, and graph propagation alone is often insufficient to resolve semantic ambiguity in task descriptions.

\subsubsection{LLM-Based Methods}

Recently, the rise of LLMs has significantly transformed the paradigm of recommending and selecting tools or developers for specific tasks in software development~\citep{ZhaoWWTWR24, JinLHJJH24}. LLMs can directly comprehend natural-language task objectives and parse contextual constraints, enabling their widespread application in tool use~\citep{LuoGHKLM25}, function calling~\citep{ParamanayakamKA25}, development task allocation~\citep{HeZLM26SPE}, and AI agent systems~\citep{HeTL25}. Unlike traditional methods that primarily rely on explicit task features or user behavior, LLMs can analyze the semantic information in task descriptions to generate more personalized and precise task recommendations~\citep{RajputMSKVHHT0S23, XuHZ24}. Moreover, LLMs can not only directly generate recommendations based on task-level linguistic descriptions but also optimize recommendation strategies using user feedback, thereby granting LLM-based approaches a greater advantage over traditional methods in dynamic development environments~\citep{ZhengHLCZCW24, Dong0L25, YangWR25}. By integrating LLMs and fine-tuning strategies, task recommendation models or systems can process complex task descriptions and multi-modal information, delivering more precise and personalized recommendations to developers or LLM-based agents~\citep{ZhangCWLWKJWY26}. Generally speaking, this approach shows promising results by addressing the limitations of traditional methods in handling intricate task scenarios.

Compared with earlier methods, LLM-based approaches are better suited to coarse-grained, compositional, and under-specified tasks because they can directly interpret task intent and perform semantic decomposition during the recommendation process~\citep{PengLHYYS25, WuZQWGSQZZLXC24}. However, their recommendations may remain unstable when candidate resources have ambiguous capability boundaries or when structural compatibility and execution constraints are not explicitly modeled. Therefore, although LLMs substantially strengthen task understanding, effective task-oriented recommendation still requires explicit grounding in resource structure and operational feasibility~\citep{abs-2601-04690}.

\subsection{MCP Server Recommendation}

Research on task-oriented MCP server recommendations aims to address bottlenecks arising from the rapid expansion of the MCP ecosystem. The same task often corresponds to a large number of candidate MCP servers or tools, and their quality varies widely. Directly injecting a large-scale tool set into the prompt will incur significant token costs and introduce noise, thereby reducing the task success rate and increasing interaction latency.

In this context, existing work generally models the selection process as a retrieval problem and proposes mechanisms from various perspectives, including dynamic retrieval, large-scale filtering, and system routing. For instance, ScaleMCP~\citep{abs-2505-06416} transforms tool selection from a static configuration to runtime autonomy, equips the agent with an MCP tool retriever that dynamically retrieves and synchronizes tools across multiple task rounds, and emphasizes key fields of tool documentation through weighted embedding strategies to improve retrieval quality and call effectiveness. HGMF (short for hierarchical Gaussian mixture framework)~\citep{xing2025hgmf} focuses on the scalability of retrieval and selection by using a two-stage probabilistic pruning strategy to reduce the candidate space, thereby decreasing context noise and invocation overhead while improving selection accuracy. Complementing this, NetMCP~\citep{li2025netmcp} further argues that semantic relevance alone is insufficient, as real-world deployments are also affected by network latency. Therefore, it performs routing optimization by jointly considering semantic relevance and quality of service to improve task completion rates.

Some recent studies also focus on context inflation caused by the expansion of MCP-related tool ecosystems. For instance, JSPLIT~\citep{antonioni2025jsplit}, a taxonomy-based solution for prompt bloating in MCP, includes only the subset of MCP tools highly relevant to the current task, thereby reducing token overhead and mitigating interference from irrelevant tools or servers. In contrast, the engineering-oriented MCP Bridge~\citep{ahmadi2025mcp} places greater emphasis on providing general-purpose infrastructure for task execution. It unifies access to multiple MCP servers through a proxy layer and introduces confirmation mechanisms and isolated execution to improve usability and facilitate secure deployment. In addition, several empirical studies provide task sets and evaluation methodologies~\citep{abs-2508-07575, luo2025mcp, song2025help}. Through experiments on real-world tasks, they show that MCP server-recommendation scenarios pose specific challenges, including multi-step reasoning, large tool spaces, and performance volatility. These findings indicate that task-driven MCP server recommendation is a promising approach and also provide a reproducible benchmark for comparing subsequent methods.

Overall, existing studies have explored several strategies to improve MCP tool selection, such as retrieval optimization and routing mechanisms. However, research on task-oriented MCP server recommendation remains limited. Most previous approaches primarily rely on semantic matching between tasks and tool descriptions, while overlooking structural constraints in practical development environments. Consequently, effectively recommending suitable MCP servers for complex development tasks is still a challenging problem.

\begin{table}[t]
\centering
\caption{Overall comparison of related work on task-aware MCP server recommendation.}
\label{tab:comparison}
\scriptsize
\setlength{\tabcolsep}{2pt}
\renewcommand{\arraystretch}{1.0}

\resizebox{\textwidth}{!}{%
\begin{tabular}{L{1.8cm}
                L{1.9cm}
                L{1.8cm}
                L{4cm}
                C{0.6cm}
                C{0.6cm}
                C{1.3cm}
                C{1cm}
                C{1cm}
                C{1cm}}
\toprule
\multicolumn{1}{C{1.8cm}}{\textbf{Category}} &
\multicolumn{1}{C{1.9cm}}{\textbf{Representative Study}} &
\multicolumn{1}{C{1.8cm}}{\textbf{Key Approach}} &
\multicolumn{1}{C{4cm}}{\textbf{Description}}&
\multicolumn{1}{C{0.6cm}}{\textbf{MCP Data} }&
\multicolumn{1}{C{0.6cm}}{\textbf{Task Data}} &
\multicolumn{1}{C{1.3cm}}{\textbf{Task–MCP Interaction}} &
\multicolumn{1}{C{1cm}}{\textbf{S \& S}} &
\multicolumn{1}{C{1cm}}{\textbf{MCP-Specific}}&
\multicolumn{1}{C{1cm}}{\textbf{Guidance} }
% \textbf{Description} 
\\
\midrule

\multirow{10}{1.8cm}{MCP Ecosystem Analysis}
& MCP Ecosystem Measurement
& MCPCorpus~\citep{LinRLZ25}; MCPCrawler~\citep{GuoXZHWM26}; MCPZoo~\citep{abs-2512-15144}.
& Provides large-scale ecosystem visibility, but remains descriptive rather than decision-oriented for MCP selection.
& \cmark & \xmark & \xmark & \xmark & \cmark & \xmark
\\

& Tool Retrieval Resources
& MCP-Zero~\citep{fei2025mcpzero}; HumanMCP~\citep{laddha2025humanmcp}.
& Improves discovery on realistic queries, but lacks task-grounded supervision to determine which retrieved servers are truly suitable.
& \cmark & \cmark & \xmark & \xmark & \cmark & \xmark
\\

& Capability and Tool-Use Benchmarks
& MCP-RADAR~\citep{abs-2505-16700}; MCP-Bench~\citep{abs-2508-20453}; MCPMark~\citep{abs-2509-24002}.
& Evaluates MCP-enabled tool use and workflows, but benchmarks execution rather than recommendation quality.
& \cmark & \xmark & \xmark & \xmark & \xmark & \xmark
\\
&\textit{Task-Specific MCP Recommendation Dataset}& \textit{Task2MCP}
& \multicolumn{1}{L{4cm}}{\textit{A task-oriented MCP dataset that systematically aligns realistic development tasks with MCP servers.}}
& \textbf{\cmark} & \textbf{\cmark} & \textbf{\cmark}& \textbf{\cmark} & \textbf{\cmark} &\textbf{\cmark}
\\

\midrule
%recommendation
\multirow{9}{1.8cm}{Task-Oriented Recommendation}
& Traditional Recommendation
& CLEAR~\citep{wei2022clear}; PTM-APIRec~\citep{LiLTHGLNWHZ24}; NeuMF~\citep{HeLZNHC17}.
& Models relevance effectively, but does not design for the heterogeneous constraints of MCP ecosystems.
& \xmark & \xmark & \xmark & \xmark & \xmark & \xmark
\\

& Graph-Based Recommendation
& LightGCN~\citep{he2020lightgcn}; NGCF~\citep{Wang0WFC19}; Req2Lib~\citep{SunLCYC20}; MLTaskKG~\citep{LiuZPYWS23}.
& Uses relational signals, but depends on graph completeness, which is hard to guarantee in fragmented MCP settings.
& \xmark & \xmark & \xmark & $\triangle$ & \xmark & \xmark
\\

& LLM-Based Recommendation
& OpenP5~\citep{XuHZ24}; DSKIPP~\citep{YangWR25}; DiMA~\citep{ZhangCWLWKJWY26}; LiS~\citep{ParamanayakamKA25}.
& Understands task intent well, but often produces semantically plausible yet operationally incompatible recommendations.
& \xmark & \xmark & \xmark & $\triangle$ & \xmark & \xmark
\\

\midrule
\multirow{5}{1.8cm}{MCP Server Recommendation}
& MCP-Oriented Retrieval and Selection
& ScaleMCP~\citep{abs-2505-06416}; HGMF~\citep{xing2025hgmf}; NetMCP~\citep{li2025netmcp}; JSPLIT~\citep{antonioni2025jsplit}.
& Improves retrieval and routing, but does not fully integrate semantic relevance, structural feasibility, and supervised task–MCP alignment.
& \xmark & \xmark & \xmark & $\triangle$ & \cmark & \xmark
\\
& \textit{Task-Aware Recommendation}
& \textit{T2MRec}
& \multicolumn{1}{L{4cm}}{\textit{A task-to-MCP recommendation model that jointly models semantic relevance and structural feasibility.}}
& \textbf{\cmark} & \textbf{\cmark} &\textbf{\cmark} & \textbf{\cmark} & \textbf{\cmark} &\textbf{\cmark}
\\
\bottomrule

\end{tabular}%
}

\vspace{1mm}
\begin{minipage}{0.98\textwidth}
\scriptsize
Note: S \& S: Semantic \& Structural; \cmark: explicitly covered; $\triangle$: partially covered; \xmark: not explicitly covered.
\end{minipage}
\end{table}

\subsection{Summary}

As shown in Table~\ref{tab:comparison}, prior work has advanced MCP ecosystem analysis, task-oriented recommendation, and MCP-specific tool retrieval, but these directions remain only partially connected. Ecosystem studies improve resource visibility but provide limited guidance on selecting an MCP server for a specific development task. General task-oriented recommendation methods improve task-resource matching, but they are not tailored to the fragmented, noisy, and constraint-sensitive nature of MCP ecosystem environments. Existing MCP-specific recommendation methods improve retrieval efficiency and system usability, but most still treat MCP server recommendation as a semantic retrieval problem.

Consequently, a clear gap remains in task-aware MCP server recommendation for realistic development workflows. What is still missing is a benchmark that systematically aligns real development tasks with curated MCP servers, together with a recommendation framework that jointly models semantic relevance and structural feasibility. This gap motivates the construction of Task2MCP and the design of T2MRec in this article.

\section{Task2MCP Dataset}
\label{3}

\subsection{Data Source and Motivation}

LLMs demonstrate strong capabilities in instruction-following, problem-solving, and multi-step reasoning. However, in real-world software development, many tasks fail at the practical stage of selecting appropriate MCP tools. As the MCP ecosystem continues to expand, developers often lack clear guidance on which tools are most suitable for a given task, making tool selection increasingly difficult and error-prone. This challenge motivates the construction of a structured dataset for task-aware MCP tool recommendation and evaluation.

In this work, the focus is placed on tool selection within the MCP ecosystem. MCP is a standard that enables LLM-based agents to access external tools via MCP servers in a consistent, interoperable manner. The number and diversity of MCP servers are growing rapidly; however, to the best of our knowledge, few public benchmarks currently provide a systematic mapping between real software development tasks and suitable MCP servers. The absence of such data makes it difficult to train task-aware MCP server recommendation models and hinders fair, reproducible evaluation. To address this gap, we construct a large-scale, high-quality task–MCP interaction dataset that maps taxonomy-grounded development tasks to curated MCP server candidates. The following sections describe the dataset construction pipeline and the dataset content in detail.

\subsection{Dataset Construction}

\begin{figure*}[t]
  \centering
  \includegraphics[width=\linewidth]{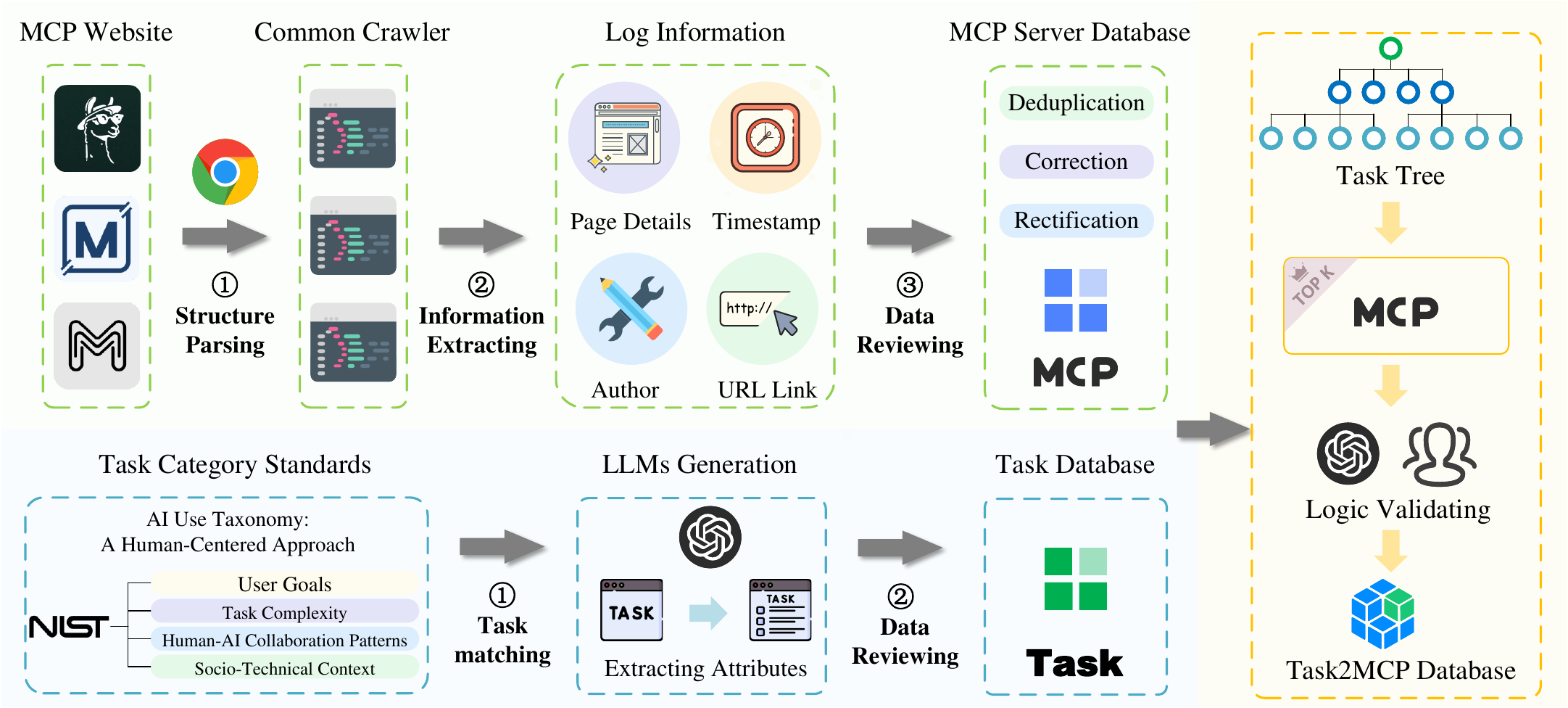}
  \caption{Overview of the construction pipeline of the Task2MCP dataset, including MCP set construction, task set construction, and task-MCP interaction set construction.}
  \label{fig:Dataset}
  \Description{Overview of the dataset construction pipeline of Task2MCP.}
\end{figure*}

The Task2MCP dataset is designed to capture (i) the landscape of actively used MCP servers and their tool capabilities, (ii) a taxonomy-grounded collection of realistic development tasks, and (iii) supervised task–MCP relevance signals represented by the top-ranked MCP server candidates for each task. Specifically, this dataset comprises three components: raw MCP data, raw task data, and interaction data between tasks and MCP servers. Figure~\ref{fig:Dataset} illustrates the overall construction pipeline of this dataset.

\subsubsection{MCP Set}

The MCP server data collection includes the following three steps.

(1) \textit{Crawling MCP directories}. We crawled three widely used MCP directories, including Glama\footnote{https://glama.ai}, mcp.so\footnote{https://mcp.so}, and mcpservers.org\footnote{https://mcpservers.org}, to build an initial candidate set of MCP servers. For each directory, we developed automated crawling scripts that download the relevant pages and parse their web structures into structured records. We then normalized the extracted fields across directories to ensure consistent formatting. For each directory entry, we extracted the MCP server name and, when provided, the corresponding GitHub repository's uniform resource locator (URL). We also removed duplicate entries across directories and kept a single canonical record for each server.

(2) \textit{Enriching metadata from GitHub}. The MCP directories we crawled typically provide only short summaries and lack sufficient information to support reliable task-to-tool matching. Therefore, after extracting the GitHub repository URL for each MCP entry, we treated GitHub as the primary source of ground-truth metadata and capability descriptions. For each repository, we automatically retrieved and normalized repository-level signals that reflect both usability and maintenance status, including the repository owner/author, project description, license information, primary programming language, and popularity indicators such as stars and watch counts. In addition to these high-level attributes, we extracted tool-relevant textual content crucial for recommendations, such as the stated functionality of an MCP server, the list of supported tools or endpoints (when available), and other usage-oriented descriptions in the repository documentation. When the same MCP server appeared across multiple directories, we consolidated records by using the GitHub URL as the canonical identifier and kept a unified representation of its metadata. This GitHub-based enrichment step substantially improved the completeness and consistency of MCP server profiles, providing richer signals for both semantic matching and structure-awareness in downstream recommendation models.

(3) \textit{Cleaning and normalizing MCP entries}. MCP entries collected from different directories are often noisy. The same server may appear multiple times under slightly different names, some records contain incomplete fields, and others point to outdated or invalid repositories. To ensure this corpus is reliable for model training and evaluation, we applied a unified cleaning and normalization workflow following GitHub enrichment. We first validated each record by resolving its GitHub URL and removing entries that could not be accessed or did not correspond to an MCP server implementation. We then deduplicated MCP servers across sources using their GitHub URLs as the canonical key, merging duplicate records while preserving the most informative text description and the most complete metadata. During this process, we corrected inconsistent fields by cross-checking directory entries against GitHub metadata and filled in missing values whenever they could be inferred from the corresponding repository (e.g., license, primary programming language, and owner). Finally, we normalized categorical attributes, such as category and subcategory, to a consistent schema and standardized tool/capability text to a unified format, reducing variance caused by different naming conventions across directories. After these steps, we obtained a cleaned MCP corpus with consistent identifiers and complete metadata.

\subsubsection{Task Set}
\label{sec:taskset}

The task data collection includes the following three steps.

\begin{figure*}[t]
  \centering
  \includegraphics[width=\textwidth]{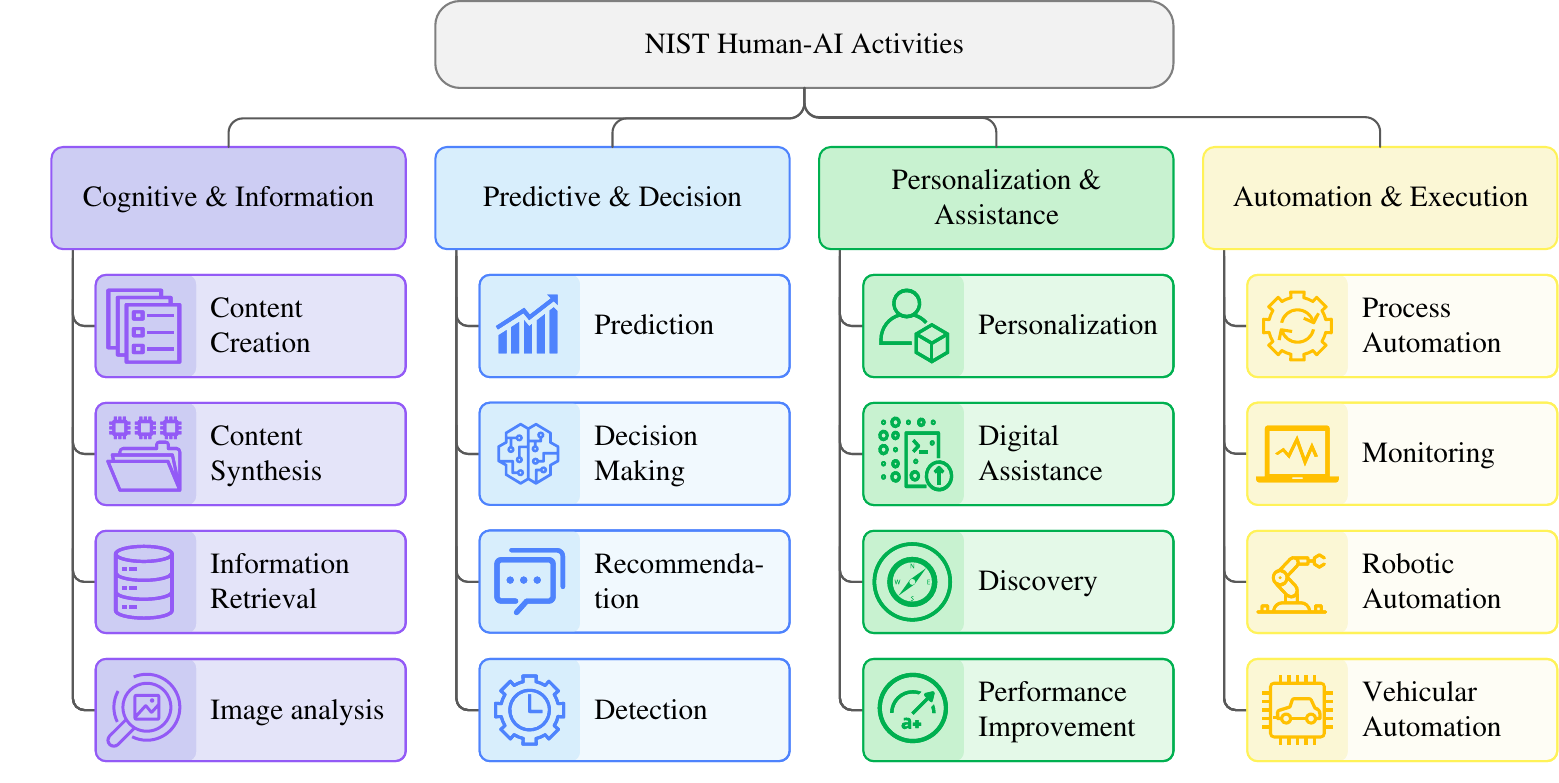}
  \caption{Two-Level task taxonomy based on the NIST human–AI activity framework.}
  \label{fig:NIST}
  \Description{NIST Human–AI Activities.}
\end{figure*}

(1) \textit{Task taxonomy definition}. To avoid an ad hoc task set and to ensure that our dataset covers a broad range of realistic development scenarios, we grounded task set construction in an external, well-defined taxonomy. Specifically, we adopted the National Institute of Standards and Technology (NIST) specification\footnote{https://www.nist.gov/publications/ai-use-taxonomy-human-centered-approach} of 16 human–AI activity types as our task subcategories, and we further organized them into four higher-level groups to form a two-level task hierarchy. As shown in Figure~\ref{fig:NIST}, this taxonomy serves two purposes. First, it provides a consistent way to define ``task diversity'' in our dataset, ensuring that tasks are not concentrated in a few popular areas. Second, it enables structured analysis and evaluation, as we can report results not only overall but also by category and subcategory. Throughout the dataset construction, each task is assigned a unique position in this concept hierarchy, which later supports task-tree–guided annotation and helps align tasks with MCP servers whose capabilities are most relevant to that part of the task space.

(2) \textit{Task generation}. After defining the above taxonomy, we populated each of the 16 activity types with concrete software development tasks that are specific enough to guide tool selection. We used an LLM as a drafting assistant to propose candidate tasks under each activity type, conditioned on the category definition and the requirement that tasks reflect realistic developer needs. For each activity type, the model generated task descriptions that included a clear objective, practical constraints, and an expected outcome; we also ensured coverage across different application themes and programming contexts. Each generated task was stored in a structured format with a short title and a detailed requirement specification, along with optional implementation constraints such as the intended programming language, a higher-level category label, a domain theme, and a unique identifier. This step produces a diverse pool of candidate tasks that are aligned with the NIST taxonomy while remaining close to real-world development workflows.

(3) \textit{Quality control}. Since the task set is intended to serve as a basis for supervising and evaluating MCP tool recommendations, we applied strict quality control to ensure that each task is clear, feasible, and correctly categorized. We first conducted automated checks to detect obvious issues, such as near-duplicate tasks across categories, overly generic descriptions that lack specific inputs or expected outputs, and mismatches between task descriptions and their assigned NIST activity types. Then, we performed manual reviews to verify that each task represents a realistic development request, contains sufficient detail to support tool selection, and does not rely on hidden assumptions or unclear requirements. The review process was conducted by multiple authors using shared criteria, achieving high inter-annotator agreement (Cohen’s $\kappa > 0.8$). We resolved disagreements in ambiguous cases through discussion. During this process, we revised or removed tasks that were ambiguous, redundant, or difficult to implement as stated. After filtering out unqualified tasks and further refinement, we retained only those that passed both automated screening and human validation, resulting in a high-quality task corpus with consistent labels and reliable task descriptions for downstream recommendation and analysis.

\subsubsection{Task–MCP Interaction Set}

To train and evaluate task-oriented MCP server recommendation models, we require supervised signals that indicate which MCP servers are most likely to be useful for implementing a given development task. To achieve this, we linked each task in the task set to a curated set of MCP server candidates, taking into account both the task's semantics and its position within the structured task taxonomy. First, we categorized each task within a two-level hierarchy, consisting of four top-level groups and sixteen NIST human–AI activity subcategories. This classification helps narrow the search for relevant MCP servers that typically support the identified activity type. Next, we analyzed each task description, including its objective, constraints, and expected outcomes. We matched it with the descriptions of MCP server capabilities and the list of tools available in the corresponding repositories. This process produced a ranked list of MCP servers for each task. These task-MCP links were established primarily according to whether an MCP server could realistically support task completion, while clearly inactive or outdated MCP servers had already been removed during preprocessing. They were selected not only for their direct functional relevance but also for their complementary nature, as they are often used together in a development workflow. 

To ensure the accuracy of these labels, we employed a human-in-the-loop verification system. This system used an LLM-based checker to identify mismatches, redundancies, and missing essential capabilities. Any ambiguous cases were reviewed manually, and the final curated, relevant set of MCP servers was confirmed. The task–MCP labels generated through this process can provide high-quality, task-grounded supervision for training recommendation models and offer a reproducible benchmark for evaluating retrieval and ranking performance.

\begin{table*}[t]
    \centering 
    \caption{Primary attributes of the MCP and task sets.}
    \label{tab:dataset}
    \begin{tabular}{llll}
        \toprule
        \textbf{Data Set} & \textbf{Attribute} & \textbf{Field Type} & \textbf{Count} \\
        \midrule
        \multirow{7}{*}{MCP}
            & mcp\_num     & Integer (numeric)      & 5,642 \\
            & license      & Categorical (string)   & 21   \\
            & language     & Categorical (string)   & 95   \\
            & subcategory  & Categorical (string)   & 52   \\
            & category     & Categorical (string)   & 22   \\
            & system       & Categorical / Boolean  & 3    \\
            & official     & Boolean                & 2    \\
        \midrule
        \multirow{5}{*}{Task}
            & task\_num    & Integer (numeric)      & 4,800 \\
            & language     & Categorical (string)   & 21   \\
            & category     & Categorical (string)   & 4    \\
            & subcategory  & Categorical (string)   & 16   \\
            & theme        & Text / Categorical     & 606  \\
        \bottomrule
    \end{tabular}
\end{table*}

\subsection{Ethical Compliance}

The Task2MCP dataset is constructed from publicly available MCP directories and GitHub repositories, as well as LLM-assisted task descriptions and task–MCP links. Only public metadata is utilized, and no additional personally identifiable information is intentionally collected beyond what is already accessible on GitHub. During the construction and annotation process, we ensure that the LLM used did not generate harmful, illegal, or discriminatory content. Additionally, tasks and labels that could potentially lead to misuse or contain explicit or offensive language were manually filtered out. The dataset is made available on GitHub for research on MCP tool recommendation and LLM-based agent systems, and we expect all users to adhere to relevant ethical and legal guidelines.

\subsection{Dataset Characteristics}

\begin{figure*}[t]
  \centering
  \includegraphics[width=\textwidth]{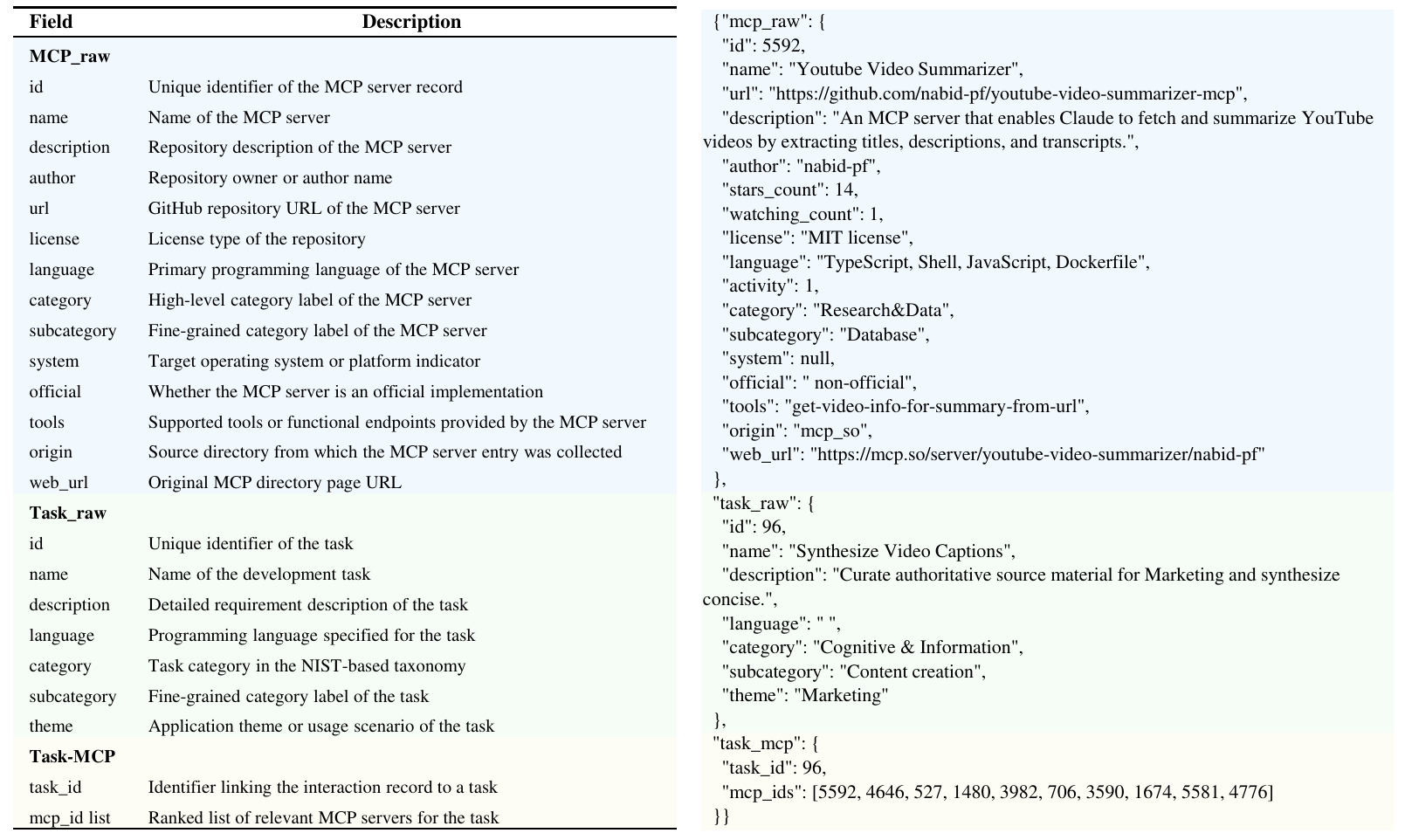}
  \caption{Schema and illustrative example of the Task2MCP dataset.}
  \label{fig:schema}
  \Description{Schema and illustrative example of the Task2MCP dataset.}
\end{figure*}

Our dataset comprises three key components: the MCP server corpus (denoted as MCP\_raw), the task corpus (denoted as Task\_raw), and the task–MCP interaction labels. In total, we have curated a collection of 5,642 MCP servers and 4,800 validated tasks. Each task is associated with a curated set of relevant MCP servers. Table~\ref{tab:dataset} presents a summary of the primary attributes and their respective cardinalities. To facilitate understanding, Figure~\ref{fig:schema} also shows the dataset schema together with an illustrative example.

On the MCP server side, MCP\_raw encompasses a diverse and rapidly evolving ecosystem. The 5,642 MCP servers are distributed across 22 high-level categories and 52 subcategories, representing a wide array of application domains. These servers are implemented using 95 different programming languages and are licensed under 21 distinct types. Additionally, we categorized each MCP server by whether it is an official implementation or community-maintained (with two boolean values: ``official'' and ``non-official''), and by the system type it targets (with three specific values for operating systems: ``Windows,'' ``iOS,'' and ``Linux''). These varied metadata fields offer multiple dimensions for modeling task–tool compatibility, extending beyond simple textual matching.

On the task side, Task\_raw is organized according to a NIST-based taxonomy, as outlined in Section~\ref{sec:taskset}. The 4,800 tasks are distributed across four top-level categories and 16 subcategories, covering 21 programming languages. Furthermore, we annotated 606 distinct theme values to capture finer-grained domains and use cases, such as data analysis, automation, and productivity tools. This well-defined structure enables detailed analysis of recommendation performance across various task types, languages, and development themes.

Overall, the MCP server metadata, taxonomy-grounded tasks, and dense task–MCP links form a comprehensive, structured dataset and benchmark for training and evaluating task-oriented MCP server recommendation models.

\section{T2MRec: Task-to-MCP Server Recommendation Model}
\label{4}

This section introduces T2MRec, a task-aware recommendation model that assists developers and LLM-based agents in selecting appropriate MCP servers for software development tasks. As the MCP ecosystem rapidly expands, identifying suitable tools becomes increasingly challenging due to the large number of available servers and their heterogeneous capabilities. The T2MRec model addresses this challenge by combining semantic retrieval, structural compatibility scoring, and constrained LLM-based refinement within a unified recommendation pipeline.

Given a task description and its associated development constraints, T2MRec generates a ranked list of MCP server candidates likely to support task implementation. The model adopts a multi-stage retrieval-and-refinement paradigm, progressively narrowing the search space from coarse semantic matching to fine-grained capability reasoning. Specifically, T2MRec consists of three main components: (i) \textbf{Semantic Relevance Scoring}, which learns semantic correspondences between tasks and MCP server capabilities using a dual-tower encoder; (ii) \textbf{Structural Compatibility Scoring}, which evaluates engineering feasibility by aligning structured attributes such as category, programming language, and theme; (iii) \textbf{Candidate Refinement}, which improves coverage and ranking quality through centroid-based expansion and constrained LLM re-ranking. This multi-stage design ensures both retrieval efficiency and recommendation quality, while restricting computationally expensive reasoning to a small-scale candidate set. Figure~\ref{fig:T2MRec} illustrates the overall architecture of the T2MRec model.

\begin{figure*}[t]
  \centering
  \includegraphics[width=\textwidth]{Figures/4-1T2MRec.pdf}
  \caption{Architecture of the T2MRec model. The model progressively narrows the search space of MCP servers through a multi-stage pipeline, including three core components: Semantic Relevance Scoring, Structural Compatibility Scoring, and Candidate Refinement.}
  \label{fig:T2MRec}
  \Description{Overview of the T2MRec model.}
\end{figure*}

\subsection{Problem Formulation}
\label{4.1}

We formulate task-to-MCP server recommendations as a task-oriented top-$K$ ranking problem. Given a task specification and associated engineering constraints, the objective is to generate a fixed-length candidate list from a rapidly evolving MCP ecosystem to assist developers or LLM-based agents in selecting and composing appropriate MCP servers. Let the MCP server set be $\mathcal{M}=\{m_j\}_{j=1}^{M}$ and the task set be $\mathcal{T}=\{t_i\}_{i=1}^{N}$.

Each MCP server $m \in \mathcal{M}$ is represented as a two-tuple $m = (x_m, a_m)$, where $x_m$ contains the textual description (e.g., name, description, and tools) and $a_m$ contains the structured metadata (e.g., category, subcategory, language, license, system, and official). Similarly, each task $t \in \mathcal{T}$ is represented as $t = (x_t, a_t)$, where $x_t$ denotes the concatenated textual description (e.g., name, description, and theme) and $a_t$ denotes a representation of structured attributes (e.g., category, subcategory, and language).

For each task $t$, our dataset provides a curated list of MCP servers that are deemed relevant for implementing $t$. We denote the corresponding positive set as:
\begin{equation}
\label{eq-1}
\mathcal{M}^{+}(t) = \left\{\,m \in \mathcal{M} \mid m \ \text{is annotated as relevant to}\ t\,\right\},
\end{equation}
where $|\mathcal{M}^{+}(t)|$ denotes the number of annotated relevant MCP servers for the task. For convenience, we define a binary relevance label as follows:
\begin{equation}
\label{eq-2}
y(m, t) =
\begin{cases}
1, & \text{if } m \in \mathcal{M}^{+}(t), \\
0, & \text{otherwise}.
\end{cases}
\end{equation}
Here, $y(m, t) = 0$ should be interpreted as an unlabeled or non-relevant instance in a weak-supervision setting, rather than a definitive claim that all remaining MCP servers are strictly irrelevant to the target task.

Our research goal is to learn a scoring function:
\begin{equation}
\label{eq-3}
s: \mathcal{M} \times \mathcal{T} \to \mathbb{R}, \quad (m, t) \mapsto s(m, t),
\end{equation}
which estimates the suitability of MCP server $m$ for task $t$ based on both textual attributes $(x_m, x_t)$ and structured attributes $(a_m, a_t)$. During inference, given a new task $t'$, our model ranks all MCP server candidates by $s(m, t')$ and outputs a top-$K$ recommendation list.

During training, we optimize this scoring function using the task–MCP server interaction set by contrasting positive pairs $(m, t)$ with negative MCP servers sampled from $\left\{\,m' \in \mathcal{M} \mid y(m', t) = 0\,\right\}$, encouraging the model to assign higher scores to annotated MCP servers than to the sampled negatives. In T2MRec, the scoring function integrates semantic relevance and structural compatibility, and is further refined through candidate expansion and LLM-based re-ranking.

\subsection{Semantic Relevance Scoring}

Semantic retrieval aims to identify MCP servers whose capabilities are semantically aligned with the task intent. Because task descriptions primarily express requirements, whereas MCP tool descriptions emphasize capabilities, a learned representation is needed to bridge the semantic gap between them. Recall from Section~\ref{4.1} that each MCP server $m$ and each task $t$ are represented as $(x_m, a_m)$ and $(x_t, a_t)$, respectively. Let $x_m$ denote the concatenated MCP text and $x_t$ denote the concatenated task text. In this section, we focus on the text pair $(x_m, x_t)$ and learn a semantic scoring function such that MCP servers that are better aligned with the task intent will receive higher scores.

\subsubsection{Text Representation}

We first convert the concatenated task text $x_t$ and the concatenated MCP text $x_m$ into sparse lexical vectors:
\begin{equation}
\label{eq-4}
\mathbf{v}_m = \mathbf{v}(x_m) \in \mathbb{R}^{d}, \qquad
\mathbf{v}_t = \mathbf{v}(x_t) \in \mathbb{R}^{d},
\end{equation}
where $\mathbf{v}(\cdot)$ denotes a vectorized operation, and $d$ is the shared vocabulary size. Tasks and MCP servers share the same vocabulary to ensure that they lie in a common feature space. We then apply an $L_2$ normalization for text representations:
\begin{equation}
\label{eq-5}
\tilde{\mathbf{v}}_m = \frac{\mathbf{v}_m}{\|\mathbf{v}_m\|_2}, \qquad
\tilde{\mathbf{v}}_t = \frac{\mathbf{v}_t}{\|\mathbf{v}_t\|_2}.
\end{equation}

This step reduces sensitivity to document length and provides well-scaled inputs for the subsequent encoder. It also preserves informative lexical signals, such as tool names, APIs, and library identifiers, which frequently appear in both task requirements and MCP tool descriptions.

\subsubsection{Semantic Encoder}

Directly comparing the normalized sparse vectors ($\tilde{\mathbf{v}}_m$ and $\tilde{\mathbf{v}}_t$) is often insufficient for task-to-tool matching. Task texts typically encode intent, constraints, and desired outcomes, whereas MCP server descriptions enumerate capabilities, interfaces, and supported tools. Sparse lexical overlap is brittle under synonymy and paraphrasing, and it cannot adequately model the asymmetric correspondence between requirement expressions on the task side and capability expressions on the MCP tool side. Therefore, we introduce a learnable projection that maps both sides into a shared dense space, where the model can measure semantic compatibility more robustly.

To capture semantic correspondences between tasks and MCP server capabilities, we employ a dual-tower encoder architecture with two side-specific encoders, $g_m(\cdot)$ and $g_t(\cdot)$. The task tower emphasizes requirement-bearing signals, and the MCP tower focuses on capability and tool descriptors. This design results in independent embeddings for tasks and MCP servers, enabling large-scale retrieval by pre-computing and indexing all MCP server embeddings offline and computing only the task embedding online at query time. This decoupling is essential when $|\mathcal{M}|$ is large, and the model must return top-$K$ candidates under strict latency constraints. Given the normalized sparse inputs, we encode a task and an MCP server as:
\begin{equation}
\label{eq-6}
\mathbf{z}_m = g_m(\tilde{\mathbf{v}}_m) \in \mathbb{R}^{d'}, \qquad
\mathbf{z}_t = g_t(\tilde{\mathbf{v}}_t) \in \mathbb{R}^{d'}, 
\end{equation}
where $d'$ is the output embedding dimension. 

In this study, each tower is implemented as an $L$-layer multilayer perceptron (MLP). The $L-1$ layers use affine transformations to compress the high-dimensional sparse input and learn weighted combinations of lexical features. Rectified linear unit (ReLU) introduces nonlinearity, allowing the encoder to capture cross-feature interactions beyond linear matching. Finally, dropout regularizes the hidden units, preventing co-adaptation and overfitting and improving robustness while keeping the architecture lightweight. The final layer outputs a $d'$-dimensional embedding.

We further normalize the tower outputs to unit length:
\begin{equation}
\label{eq-7}
\hat{\mathbf{z}}_m = \frac{\mathbf{z}_m}{\|\mathbf{z}_m\|_2}, \qquad 
\hat{\mathbf{z}}_t = \frac{\mathbf{z}_t}{\|\mathbf{z}_t\|_2}.
\end{equation}

Normalization removes scale ambiguity in embeddings and ensures that scores are comparable across tasks. The semantic relevance score is then computed by:
\begin{equation}
\label{eq-sim8}
s_{\mathrm{sem}}(m,t) = \hat{\mathbf{z}}_m^\top \hat{\mathbf{z}}_t,
\end{equation}
which equals cosine similarity due to normalization.

During inference, we pre-compute $\hat{\mathbf{z}}_m$ for all MCP servers and store them in an embedding index. Given a new task $t'$, we compute $\hat{\mathbf{z}}_{t'}$ once and obtain possible candidates by ranking MCP servers according to $s_{\mathrm{sem}}(m,t')$. This design enables efficient large-scale retrieval because MCP server embeddings can be precomputed offline and indexed.

Overall, the two-tower semantic encoder reduces the representation gap between requirement-centric task texts and capability-centric MCP tool texts. By learning semantic correspondences between tasks and MCP servers, the encoder improves robustness beyond raw lexical overlap. It also enables scalable retrieval by decoupling offline MCP server encoding from online task encoding.

\subsubsection{Contrastive Alignment Training}
\label{4.2.3}

The dual-tower encoder is trained using task–MCP interaction labels under a contrastive learning objective. Recall that for each task $t$, $\mathcal{M}^{+}(t) \subseteq \mathcal{M}$ denotes the curated set of positive MCP servers. During model training, we construct a mini-batch of tasks $\mathcal{B} = \{t_i\}_{i=1}^{B}$, where $B = |\mathcal{B}|$ is the batch size. For each task $t_i \in \mathcal{B}$, we sample one positive MCP server $m_i^{+} \in \mathcal{M}^{+}(t_i)$, yielding $B$ matched pairs $\{(m_i^{+}, t_i)\}_{i=1}^{B}$. All MCP servers paired with other tasks in the same batch are treated as implicit negatives for $t_i$, enabling efficient contrastive learning without explicitly enumerating the full candidate set $\mathcal{M}$.

Since $y(m, t) = 0$ denotes unlabeled or non-relevant pairs rather than strictly irrelevant ones, some in-batch negatives may actually be useful MCP servers that are unlabeled. Nevertheless, the contrastive objective provides a scalable and effective alignment signal in practice.
Let $\hat{\mathbf{z}}_{t_i} \in \mathbb{R}^{d'}$ and $\hat{\mathbf{z}}_{m_j} \in \mathbb{R}^{d'}$ denote the $L_2$-normalized embeddings produced by the task tower and the MCP tower for task $t_i$ and MCP server $m_j$, respectively. We define the semantic relevance score used for training as the dot product of normalized embeddings, as shown in Equation~\eqref{eq-sim8}. For each task $t_i$, let $\mathcal{M}_{\mathcal{B}} = \{m_j^{+} \mid t_j \in \mathcal{B}\}$ denote the set of MCP servers appearing in the current mini-batch (including $m_i^{+}$). The temperature-scaled contrastive loss for task $t_i$ is then defined as:
\begin{equation}
\label{eq-9}
\mathcal{L}_{t_i} = -\log
\frac{\exp\!\left(s_{\mathrm{sem}}(t_i, m_i^{+})/\tau\right)}
{\sum_{m \in \mathcal{M}_{\mathcal{B}}} \exp\!\left(s_{\mathrm{sem}}(t_i, m)/\tau\right)},
\end{equation}
where $\tau > 0$ is a temperature hyperparameter controlling the sharpness of the softmax distribution. This objective increases the semantic relevance score assigned to the matched pair $(t_i, m_i^{+})$ relative to all other MCP servers in $\mathcal{M}_{\mathcal{B}}$, thereby separating positives from in-batch negatives. A smaller $\tau$ yields a sharper distribution and places greater emphasis on hard negatives.

The overall mini-batch loss is the average over tasks:
\begin{equation}
\label{eq-10}
\mathcal{L} = \frac{1}{B} \sum_{i=1}^{B} \mathcal{L}_{t_i}.
\end{equation}

We optimize the parameters of the two towers by minimizing $\mathcal{L}$ using stochastic gradient-based optimization. Once the training process ends, the dual-tower encoder is frozen, and for any task–MCP server pair $(m, t)$, the cosine similarity $s_{\mathrm{sem}}(m, t)$ in Equation~\eqref{eq-sim8} is used as the final semantic relevance score throughout inference. This objective encourages semantically relevant MCP servers to be closer to the task embedding than unrelated ones.

\subsection{Structural Compatibility Scoring}

While semantic similarity captures textual relevance between $x_m$ and $x_t$, it may overlook engineering feasibility and ecosystem compatibility. For example, an MCP server may be semantically relevant to a task but still be unusable due to a mismatch in programming language, platform requirements, or taxonomy. To account for such factors, we define a structural compatibility score $s_{\mathrm{str}}(m, t)$ based on the structured attributes ($a_m$ and $a_t$). This score complements semantic retrieval by introducing lightweight constraint-aware signals beyond pure textual similarity.

\subsubsection{Compatibility Feature Design}

Recall that each MCP server and task are represented as $m=(x_m,a_m)$ and $t=(x_t,a_t)$, respectively. For notational convenience, we denote the structured attributes as:
\begin{equation}
\label{eq-11}
a_m = (c_m, \tilde{c}_m, \ell_m, p_m), \qquad
a_t = (c_t, \tilde{c}_t, \ell_t, h_t),
\end{equation}
where $c$ denotes category, $\tilde{c}$ denotes subcategory, $\ell$ denotes programming language, $p_m$ denotes the operating system in which an MCP server runs, and $h_t$ denotes the task theme. 

Our attribute design is inspired by prior multi-dimensional taxonomies of recommendation settings, which characterize tasks and methods from complementary perspectives~\cite{LinDXLCZLWLZGYTZ25}. The attributes $p_m$ and $h_t$ characterize compatibility from the system and task sides, respectively. The former specifies the operating environment supported by an MCP server, whereas the latter describes the application setting of the given task. In practice, tasks within the same theme often share similar deployment assumptions and software interfaces. Hence, the compatibility between $p_m$ and $h_t$ serves as a useful signal of practical deployability. By leveraging these attributes, we define a small set of interpretable compatibility features:
\begin{equation}
\label{eq-12}
\boldsymbol{\phi}_{\mathrm{str}}(m,t) =
\bigl[
\phi_{\mathrm{cat}}(m,t),\ 
\phi_{\mathrm{lan}}(m,t),\ 
\phi_{\mathrm{the}}(m,t)
\bigr].
\end{equation}

\begin{itemize}
\item \textbf{Category feature $\phi_{\mathrm{cat}}(m,t)$.}
This feature measures taxonomy proximity between an MCP server and the given task based on $(c_m,\tilde{c}_m)$ and $(c_t,\tilde{c}_t)$, reflecting whether they reside in a similar functional ecosystem.

\item \textbf{Language feature $\phi_{\mathrm{lan}}(m,t)$.}
This feature captures whether the task-side language constraint $\ell_t$ is compatible with the MCP-side declared programming language $\ell_m$.

\item \textbf{Theme feature $\phi_{\mathrm{the}}(m,t)$.}
This feature measures whether the task theme $h_t$ is compatible with the MCP-side system environment $p_m$, serving as a lightweight signal of ecosystem-level feasibility beyond pure lexical similarity.
\end{itemize}

Overall, these structured features provide a lightweight and interpretable complement to semantic matching. They inject signals of taxonomy consistency, language compatibility, and deployability into the ranking process of possible MCP servers, thereby improving constraint awareness while preserving inference efficiency.

\subsubsection{Taxonomy-Based Distance Modeling}

To quantify taxonomy proximity, we represent the category taxonomy as a rooted tree $\mathcal{T}_{\mathrm{cat}}$, whose root is the taxonomy root. In this tree, first-level nodes correspond to categories, and second-level nodes correspond to subcategories. For an MCP server $m$ and a task $t$, let $v_m$ and $v_t$ denote the subcategory nodes uniquely determined by $(c_m,\tilde{c}_m)$ and $(c_t,\tilde{c}_t)$, respectively. We define the taxonomy distance as the tree distance:
\begin{equation}
\label{eq-13}
d_{\mathrm{cat}}(m,t) = \operatorname{depth}(v_m) + \operatorname{depth}(v_t) - 2\times\operatorname{depth}\bigl(\operatorname{LCA}(v_m,v_t)\bigr),
\end{equation}
where $d_{\mathrm{cat}}(m,t)$ denotes the tree distance between nodes $v_m$ and $v_t$ in $\mathcal{T}_{\mathrm{cat}}$, and $\operatorname{LCA}(v_m,v_t)$ denotes the lowest common ancestor of $v_m$ and $v_t$. Since both $v_m$ and $v_t$ are depth-2 nodes in our two-level taxonomy, $d_{\mathrm{cat}}(m,t) \in \{0,2,4\}$. Specifically, $d_{\mathrm{cat}}(m,t)=0$ when $(c_m,\tilde{c}_m)=(c_t,\tilde{c}_t)$, $d_{\mathrm{cat}}(m,t)=2$ when $c_m=c_t$ but $\tilde{c}_m \neq \tilde{c}_t$, and $d_{\mathrm{cat}}(m,t)=4$ otherwise.

We then normalize this distance to obtain a bounded proximity score:
\begin{equation}
\label{eq-14}
\phi_{\mathrm{cat}}(m,t) = 1 - \frac{d_{\mathrm{cat}}(m,t)}{4}.
\end{equation}

This formulation preserves $d_{\mathrm{cat}}(m,t)$ as a proper tree metric while separating distance measurement from proximity-score normalization. We further define two binary compatibility checks:
\begin{align}
\label{eq-15}
\phi_{\mathrm{lan}}(m,t) &= \mathbb{I}(\ell_t, \ell_m), \\
\phi_{\mathrm{the}}(m,t) &= \mathbb{I}(h_t , p_m),
\end{align}
where $\mathbb{I}(\cdot,\cdot)$ denotes a binary indicator function defined on an attribute pair. It takes the value ``1'' if the two attributes are regarded as compatible under the predefined matching rules, and ``0'' otherwise. In particular, $\mathbb{I}(\ell_t,\ell_m)=1$ indicates that the task language is compatible with the language specification of an MCP server. Similarly, $\mathbb{I}(h_t,p_m)=1$ indicates that the task theme is compatible with the platform or system environment.

Finally, we calculate the structural compatibility score as a weighted combination of the above signals:
\begin{equation}
\label{eq-17}
s_{\mathrm{str}}(m,t)
=
w_{\mathrm{cat}}\times\phi_{\mathrm{cat}}(m,t)
+
w_{\mathrm{lan}}\times\phi_{\mathrm{lan}}(m,t)
+
w_{\mathrm{the}}\times\phi_{\mathrm{the}}(m,t),
\end{equation}
where $w_{\mathrm{cat}}, w_{\mathrm{lan}}, w_{\mathrm{the}} > 0$ are feature weights. In this work, we use a simple heuristic setting and assign equal weights to all structural signals. This choice keeps the method simple and reproducible, while avoiding extra hyperparameter tuning.

\subsubsection{Final Matching Score}

We combine semantic relevance and structural compatibility to obtain the final score for initial ranking:
\begin{equation}
\label{eq-18}
s(m,t) =
\alpha_{\mathrm{sem}}\, s_{\mathrm{sem}}(m,t)
+
\alpha_{\mathrm{str}}\, s_{\mathrm{str}}(m,t)
\qquad
\alpha_{\mathrm{sem}}+\alpha_{\mathrm{str}}=1,
\end{equation}
where $\alpha_{\mathrm{sem}}$ and $\alpha_{\mathrm{str}}$ denote the fusion weights assigned to semantic relevance and structural compatibility, respectively.
By default, we set $\alpha_{\mathrm{sem}} > \alpha_{\mathrm{str}}$, reflecting that semantic relevance is the primary signal, while structural compatibility provides complementary guidance on feasibility. The resulting $s(m,t)$ is used to generate the initial candidate list, which is further refined by candidate expansion and LLM-based re-ranking.

\subsection{Candidate Refinement}

After semantic retrieval and structural compatibility scoring, T2MRec obtains an initial ranking of MCP server candidates for a given task. However, this ranking may still be subject to two limitations. First, high-precision retrieval may miss complementary MCP servers that are not explicitly mentioned in the task description. Second, lightweight matching scores may not fully capture fine-grained alignment between task intent and MCP tool capabilities. To address these issues, we introduce a candidate refinement stage that transforms the learned semantic and structural signals into the final top-$K$ recommendation list.

Given a task $t$, we first compute the semantic score $s_{\mathrm{sem}}(m,t)$ and the structural score $s_{\mathrm{str}}(m,t)$ for each MCP server candidate $m \in \mathcal{M}$. We then apply a multi-stage pipeline comprising anchor retrieval, neighborhood expansion, and constrained LLM re-ranking. The initial ranking stage fuses semantic and structural evidence to produce a high-confidence anchor set $\mathcal{C}_{K_1}(t)$. The centroid expansion stage enlarges this set to form an expanded candidate pool $\mathcal{C}_{K_2}(t)$, thereby improving coverage of MCP servers near the task in the semantic space. Finally, the LLM re-ranking stage, which refines the candidate pool and selects the final recommendation set $\mathcal{C}_{K}(t)$ under practical constraints, aims to capture better fine-grained alignment between task intent and MCP tool capabilities.

\subsubsection{Anchor Candidate Retrieval}

Given the fused matching score $s(m,t)$ defined in Section~\ref{4.1}, we first construct a high-precision anchor set to support downstream expansion. Concretely, we sort all MCP server candidates $m \in \mathcal{M}$ by $s(m,t)$ and retain the top-$K_1$ MCP servers:
\begin{equation}
\label{eq-19}
\mathcal{C}_{K_1}(t) = \mathrm{TopK}_{m \in \mathcal{M}}\big(s(m,t),\, K_1\big).
\end{equation}

This anchor set prioritizes precision. It retains MCP servers that are both semantically relevant and structurally feasible, serving as the reference set for the subsequent centroid expansion step. Using a small $K_1$ helps concentrate the anchor set on high-confidence candidates, preventing later expansion from drifting toward noisy regions of the MCP server space.

\subsubsection{Capability Neighborhood Expansion}

Although the anchor set $\mathcal{C}_{K_1}(t)$ typically contains highly relevant MCP servers, it may not cover complementary tools required in realistic development workflows. To improve candidate coverage without excessively increasing $K_1$, we design a centroid-based expansion mechanism.

Let $\hat{\mathbf{z}}_m$ denote the MCP server embedding and $\hat{\mathbf{z}}_t$ denote the task embedding produced by the semantic encoder. We construct an expanded query representation by aggregating the task embedding with the embeddings of the anchor MCP servers:
\begin{equation}
\label{eq-20}
\mathbf{c}(t) = \frac{1}{K_1+1}\left(\hat{\mathbf{z}}_t + \sum_{m \in \mathcal{C}_{K_1}(t)} \hat{\mathbf{z}}_m\right).
\end{equation}

Conceptually, $\mathbf{c}(t)$ moves the query representation from the task text alone to a composite embedding that reflects both the task and its high-confidence capability neighborhood. This representation helps retrieve MCP servers that are not only semantically related to the task description but also coherent with the anchor set.

We then retrieve additional candidates from the remaining MCP server pool by ranking them according to centroid similarity:
\begin{equation}
\label{eq-21}
\mathcal{C}_{K_2}(t)
=
\mathrm{TopK}_{m \in \mathcal{M} \setminus \mathcal{C}_{K_1}(t)}
\big(\hat{\mathbf{z}}_m^{\top}\mathbf{c}(t),\, K_2-K_1\big),
\end{equation}
and form the expanded candidate set:
\begin{equation}
\label{eq-22}
\mathcal{C}_{K'}(t) = \mathcal{C}_{K_1}(t) \cup \mathcal{C}_{K_2}(t).
\end{equation}

The centroid expansion step is computationally lightweight. It only requires a simple aggregation to form $\mathbf{c}(t)$ and an additional similarity search over the existing embedding index, without any expensive cross-encoding. Despite this low cost, it broadens retrieval coverage by prioritizing MCP servers that are close to the anchor capability set in the embedding space. This bias tends to introduce MCP servers that are semantically coherent with the anchors and potentially complementary, thereby increasing candidate diversity without sacrificing consistency with the task intent.

Take a simple task ``\emph{read a PDF file and extract all tables}'' as an example. The anchor set retrieves MCP servers for PDF parsing and table extraction. Centroid expansion then introduces a complementary MCP server for optical character recognition (OCR), which is often needed when PDF files contain scanned pages. Although OCR may not be explicitly mentioned in the task text, it lies within the anchor neighborhood in the embedding space by our approach, thereby improving the completeness of the candidate set.

\begin{figure}[t]
\centering
\footnotesize

\begin{tcolorbox}[
    width=0.98\linewidth,
    colback=white,
    colframe=PromptBorder,
    coltitle=black,
    colbacktitle=PromptTitleBg,
    title=\textbf{Prompt Template for Constrained LLM Re-ranking},
    boxrule=0.5pt,
    arc=1mm,
    left=6pt,
    right=6pt,
    top=6pt,
    bottom=6pt
]

{\centering \normalsize\color{PromptSection}\textbf{Role}\par}
You are a constrained list-wise re-ranking module for task-oriented MCP server recommendation. Given one development task and a fixed candidate pool, rank the candidates by overall suitability. Select identifiers only from the provided candidate pool.

{\centering \normalsize\color{PromptSection}\textbf{Input}\par}

\textbf{(a) Task Data}. \\
\begin{tabularx}{\linewidth}{@{}>{\raggedright\arraybackslash\bfseries}p{0.30\linewidth}X@{}}
ID & $\{i\}$ \\
Name & $\{t_i\}$ \\
Description & $\{x_t\}$ \\
Attributes & $\{a_t\}$ \\
\end{tabularx}

\vspace{4pt}
\textbf{(b) Ranking Criteria}. \\
Rank candidates by overall suitability for the task. Prefer candidates that:
\begin{itemize}
    \item Closely match the task intent and required capabilities.
    \item Satisfy the specified engineering constraints.
    \item Fit naturally into the intended execution workflow.
    \item Compose effectively with other candidates when coordination or dependencies are required.
    \item Avoid redundancy, weak relevance, and overly generic functionality.
\end{itemize}

\vspace{2pt}
\textbf{(c) Candidate Card.}\\
For each candidate MCP server $m_j \in \mathcal{C}_{K'}(t)$, provide:

\vspace{2pt}
\begin{tabularx}{\linewidth}{@{}>{\raggedright\arraybackslash\bfseries}p{0.30\linewidth}X@{}}
ID & $\{j\}$ \\
Name & $\{m_j\}$ \\
Description & $\{x_m\}$ \\
Metadata & $\{a_m\}$ \\
\end{tabularx}

\vspace{4pt}
\textbf{(d) Rules.}
\begin{itemize}
    \item Return exactly $K$ MCP server identifiers in ranked order.
    \item Every identifier must appear in the provided candidate pool.
    \item Preserve each identifier exactly as given.
    \item Do not output any content outside the required JSON object.
    \item Output must follow the JSON list format shown below.
\end{itemize}
\vspace{2pt}
{\centering \normalsize\color{PromptSection}\textbf{Output}\par}
\textbf{JSON Object.}\\
Return a single JSON object of the form:\\
\{\\
"Task": $t_i$,\\
"MCP\_servers": $[m_1, m_2, ..., m_k]$,\\
"Explanation": "Briefly explain why these MCP servers are ranked in this order."\\
\}

\end{tcolorbox}

\caption{Prompt template used by the constrained list-wise LLM re-ranker.}
\label{fig:rerank-prompt}
\Description{Prompt template for constrained LLM re-ranking}
\end{figure}

\subsubsection{Constrained LLM Re-ranking}

The final stage employs an LLM-based re-ranker to refine the ordering of the candidate pool $\mathcal{C}_{K'}(t)$. Crucially, we use an LLM as a constrained ranking component rather than an open-ended retriever. It operates on a small, pre-filtered candidate set and returns only a reordered list, thereby improving controllability and reducing bounds inference latency.

We formalize the re-ranking operator as:
\begin{equation}
\label{eq-23}
\pi_{K}(t) = \mathrm{LLM\text{-}Rerank}\big(t, \mathcal{C}_{K'}(t)\big),
\end{equation}
where $\pi_{K}(t)$ denotes the final top-$K$ ranked tool list for task $t$. The LLM model receives the task text $x_t$ together with a compact summary of structural constraints derived from $a_t$. For each candidate MCP server $m \in \mathcal{C}_{K'}(t)$, it also receives a standardized \emph{candidate card} containing the MCP server identifier, name, description, tool keywords, and key metadata. This LLM model then reorders MCP servers by assessing fine-grained intent-capability fit that is difficult to capture using lightweight similarity metrics and rule-based compatibility checks alone.

We constrain the LLM model to re-rank only the retrieved set $\mathcal{C}_{K'}(t)$, ensuring deterministic and machine-parseable outputs. It must return exactly $K$ MCP server identifiers in ranked order; if replacement is allowed, at most two items may be substituted, and only with alternatives from $\mathcal{C}_{K'}(t)$. Decoding is deterministic, and any output violating length, membership, or formatting constraints is discarded in favor of the pre-LLM ordering. This constrained design allows T2MRec to leverage LLMs' reasoning capabilities while maintaining stable inference latency and predictable behavior. To improve reproducibility and make the re-ranking stage auditable, we instantiate LLMs with a fixed prompt template rather than a free-form instruction. The template serializes the task context, structural constraints, candidate cards, ranking objective, and output schema into a deterministic input format. Figure~\ref{fig:rerank-prompt} shows the prompt used in our design.

This prompt operationalizes Equation~\eqref{eq-23} as a constrained list-wise ranking problem over a bounded candidate pool. The LLM model is therefore responsible only for reordering the retrieved MCP servers, not for expanding the search space or generating new candidates. This design allows the re-ranker to leverage the LLM model's reasoning capabilities while keeping the recommendation process grounded in the upstream retrieval and compatibility-scoring stages.

\subsection{Model Training}

The T2MRec model is trained by optimizing the dual-tower semantic encoder on the task–MCP interaction data. In contrast, the structural compatibility scoring module and the constrained LLM-based re-ranking module are inference-stage components, which do not introduce additional trainable parameters. This design keeps model learning focused on the semantic alignment between task requirements and MCP server capabilities, while leaving structural compatibility filtering and fine-grained ranking refinement to later stages.

For model training, we construct positive instances from annotated task–MCP pairs in the interaction set. For each task $t$, the MCP servers in $\mathcal{M}^{+}(t)$ are treated as positive instances.  Since the dataset provides curated positive associations rather than exhaustive binary relevance labels, the remaining unlabeled MCP servers are not treated as guaranteed true negatives. Instead, negative supervision is introduced implicitly via mini-batch contrast, in which MCP servers paired with other tasks in the same batch serve as negatives for the current task, and vice versa. This strategy is well-suited to the weakly supervised nature of the recommendation setting and avoids imposing overly strong assumptions on unlabeled task–MCP pairs.

The semantic encoder is optimized with the contrastive objective introduced in Equation~\eqref{eq-10} in Section~\ref{4.2.3}, which encourages matched task–MCP pairs to be close in the shared embedding space while separating mismatched pairs. After model training, the encoder is frozen and used to produce semantic relevance scores for downstream retrieval and ranking.

\subsection{Model Inference}

Algorithm~\ref{alg:T2MRec-infer} summarizes the complete inference pipeline of T2MRec. The algorithm first performs semantic retrieval to identify MCP servers relevant to the task description, then incorporates structural compatibility scoring to filter out engineering-incompatible candidates. The candidate pool is subsequently expanded using centroid-based capability neighborhood exploration, and finally refined through constrained LLM re-ranking.

The inference cost of T2MRec is dominated by corpus-level retrieval and candidate selection. Let $M$ denote the number of MCP servers and $d'$ the embedding dimension. Since MCP server embeddings are precomputed offline, online inference only requires encoding the input task once, followed by semantic relevance scoring over the MCP corpus, which costs $O(Md')$ under exact retrieval. Structural compatibility scoring introduces only linear overhead, i.e., $O(M)$, because each compatibility check is constant-time. The centroid-based refinement stage adds one additional retrieval pass of the same order, while candidate selection incurs $O(M\log M)$ under exact sorting. Finally, the constrained LLM re-ranker is applied only to a bounded candidate pool and thus contributes an additional cost of $C_{\mathrm{LLM}}(K)$ without changing the asymptotic dependence on $M$. Therefore, the overall inference complexity of T2MRec is $O(Md' + M\log M + C_{\mathrm{LLM}}(K))$ under exact retrieval, indicating that the model remains scalable with respect to corpus size while restricting expensive LLM reasoning to a small late-stage re-ranking step.

\begin{algorithm}[t]
\caption{Inference Algorithm of T2MRec}
\label{alg:T2MRec-infer}
\KwIn{Task $t=(x_t,a_t)$, MCP server set $\mathcal{M}=\{m_j\}_{j=1}^{M}$ with precomputed embeddings $\{\hat{\mathbf z}_m\}_{m\in\mathcal M}$}
\KwOut{Ranked top-$K$ MCP server list $\mathcal{C}_{K}(t)$}

Initialize hyperparameters $\alpha_{\mathrm{sem}}, \alpha_{\mathrm{str}}, w_{\mathrm{cat}}, w_{\mathrm{lan}}, w_{\mathrm{the}}, K_1, K_2$, and $K$

Compute sparse lexical vector $\mathbf{v}_t \gets \mathbf{v}(x_t)$ using Equation~\eqref{eq-4}

Normalize task input $\tilde{\mathbf{v}}_t \gets \mathbf{v}_t / \|\mathbf{v}_t\|_2$ using Equation~\eqref{eq-5}

Compute task embedding $\mathbf{z}_t \gets g_t(\tilde{\mathbf{v}}_t)$ using Equation~\eqref{eq-6} 

Normalized task embedding $\hat{\mathbf{z}}_t \gets \mathbf{z}_t / \|\mathbf{z}_t\|_2$  using Equation~\eqref{eq-7}

\ForEach{$m \in \mathcal{M}$}{
    Compute semantic relevance score $s_{\mathrm{sem}}(m,t) \gets \hat{\mathbf{z}}_m^\top \hat{\mathbf{z}}_t$ using Equation~\eqref{eq-sim8}

    Compute structural compatibility score $s_{\mathrm{str}}(m,t)$ using Equation~\eqref{eq-17}

    Compute fused score $s(m,t)$ using Equation~\eqref{eq-18}
}

Select anchor set $\mathcal{C}_{K_1}(t)$ using Equation~\eqref{eq-19}

Compute centroid $\mathbf{c}(t)$ using Equation~\eqref{eq-20}

Retrieve expanded candidates $\mathcal{C}_{K_2}(t)$ using Equation~\eqref{eq-21}

Construct final candidate pool $\mathcal{C}_{K'}(t)\gets \mathcal{C}_{K_1}(t)\cup\mathcal{C}_{K_2}(t)$ using Equation~\eqref{eq-22}

Compute $\mathcal{C}_{K}(t)\gets \mathrm{LLM\mbox{-}Rerank}(t,\mathcal{C}_{K'}(t))$ using Equation~\eqref{eq-23}

\Return $\mathcal{C}_{K}(t)$
\end{algorithm}

\section{Experiments and Result Analysis}
\label{5}

\subsection{Research Questions}
To evaluate the proposed T2MRec model, we design experiments on the Task2MCP dataset to answer the following research questions (RQs). 
\begin{itemize}
\item \textbf{RQ1:} How does T2MRec perform compared with competitive baselines?

\item \textbf{RQ2:} How do the specific component designs of T2MRec affect model performance?

\item \textbf{RQ3:} How do the setting choices of different parameters affect model performance?
\end{itemize}

\subsection{Baseline Methods}

To assess the effectiveness of T2MRec, we compare it against selected competitive baselines from three categories: traditional, graph-based, and LLM-based recommendation methods. All baselines are adapted to the same Task2MCP formulation and evaluated under identical data splits, candidate spaces, and evaluation protocols. In this setting, each task is treated as an input query, and each MCP server is regarded as a candidate item. For methods originally designed for user-item recommendation, we reformulate each task–MCP association as an interaction and use the curated relevant MCP servers of each task as positive instances. For text-aware and LLM-based methods, we use the same task and MCP textual representations as model inputs and require all methods to produce a ranked list of MCP servers for each task under the same candidate pool.

\subsubsection{Traditional Recommendation Models}

Our experiment first includes traditional recommendation models (e.g., collaborative filtering and matrix factorization) as non-graph baselines. These methods provide strong reference points for assessing whether T2MRec's performance gains in recommendation stem from improved modeling of task–MCP associations rather than from increased architectural complexity alone.
\begin{itemize}
\item \textbf{NeuMF}\footnote{https://github.com/hexiangnan/neural\_collaborative\_filtering} (2017)~\citep{HeLZNHC17} is a neural collaborative filtering model that combines generalized matrix factorization with a multi-layer perceptron to capture both linear and non-linear interaction patterns.

\item \textbf{MSBPR}\footnote{https://github.com/Delta-zl/Code-MSBPR} (2023)~\citep{ZengGC23} extends Bayesian personalized ranking for one-class collaborative filtering by partitioning non-interacted items with auxiliary feedback and item/user-based similarity and by learning multiple pairwise preferences over these partitions.

\item \textbf{ReCAFR}\footnote{https://github.com/smufang/ReCAFR.git} (2025)~\citep{Dong0L25} is a review-centric contrastive alignment framework that aligns user, item, and review representations in a unified space to alleviate sparsity and improve robustness of personalized recommendations.
\end{itemize}

\subsubsection{GNN-based Recommendation Models}

We further consider graph-based recommendation methods, which are widely used for modeling relational dependencies in user-item interaction data. In our setting, these methods operate on the task–MCP interaction graph and serve as strong baselines for evaluating whether explicit graph modeling improves recommendation quality.
\begin{itemize}
\item \textbf{NGCF}\footnote{https://github.com/huangtinglin/NGCF-PyTorch} (2019)~\citep{Wang0WFC19} incorporates the bipartite interaction graph into representation learning through neural message passing, enabling the modeling of collaborative signals and higher-order connectivity.

\item \textbf{LightGCN}\footnote{https://github.com/kuandeng/LightGCN} (2020)~\citep{he2020lightgcn} is a simplified graph collaborative filtering model that retains only neighborhood aggregation, learning node representations by linearly propagating embeddings over the interaction graph.

\item 
\textbf{GDSRec}\footnote{https://github.com/MEICRS/GDSRec}(2023)~\citep{ChenXLHL23} is a graph-based distributed collaborative filtering recommendation method that models bias information in user and item representations.

\item 
\textbf{CVH-REC}\footnote{https://zenodo.org/records/18346231} (2026)~\citep{11400603} models mashup-API interactions and tag usage records as a multi-view knowledge graph, and it employs cross-view hypergraph neural networks with contrastive learning and multi-task learning to learn entity representations.
\end{itemize}

\subsubsection{LLM-based Recommendation Models}

Our experiment also includes recent LLM-based recommendation methods, enabling a comparison of T2MRec against new approaches that exploit LLMs for recommendation and ranking. These baselines are particularly relevant to our problem setting because task and MCP server descriptions are both naturally expressed in text.
\begin{itemize}
\item \textbf{TiGER}\footnote{https://github.com/XiaoLongtaoo/TIGER.git} (2023)~\citep{RajputMSKVHHT0S23} formulates recommendation as a generative retrieval problem by assigning semantic identifiers to items and autoregressively generating the identifier of the target item.

\item \textbf{OpenP5}\footnote{https://github.com/agiresearch/OpenP5} (2024)~\citep{XuHZ24} is an open-source platform for developing and evaluating LLM-based generative recommender systems, which we use as a representative generative recommendation baseline in this study.

\item \textbf{LC-Rec}\footnote{https://github.com/RUCAIBox/LC-Rec/} (2024)~\citep{ZhengHLCZCW24} adapts LLMs to recommendation by integrating language semantics with collaborative semantics through learned item indices and alignment-oriented tuning.

\item \textbf{MPT-Rec}\footnote{https://github.com/BAI-LAB/MPT-Rec} (2025)~\citep{BaiHYYHZS25} improves task recommendation for emerging tasks through a two-stage prompt-tuning framework, which disentangles shared and task-specific knowledge and transfers it to new tasks via task-aware prompts with reduced training cost.
\end{itemize}

\subsection{Evaluation Metrics}

We evaluate all methods using top-$k$ ranking metrics with $k \in \{5, 10\}$. Since the task-to-MCP recommendation aims to return a short list that both covers useful MCP servers and prioritizes the most suitable ones, we report Recall@$k$, Precision@$k$, F1@$k$, and NDCG@$k$. Among these metrics, Recall@$k$ and NDCG@$k$ are treated as the primary indicators for model performance comparison, and Precision@$k$ and F1@$k$ are reported as supplementary statistics.

For a test task $t \in \mathcal{T}_{\mathrm{test}}$, let $\mathcal{M}^{+}(t)$ denote the annotated relevant MCP server set and $\mathcal{C}_{K}(t) = [m_1, \ldots, m_k]$ the top-$k$ ranked list returned by a method. We first define the number of relevant MCP servers retrieved within the top-$k$ positions as
\begin{equation}
H_k(t) = \left| \mathcal{C}_{K}(t) \cap \mathcal{M}^{+}(t) \right|.
\end{equation}

Then, the per-task Recall@$k$ and Precision@$k$ are defined as
\begin{equation}
R_k(t) = \frac{H_k(t)}{\left|\mathcal{M}^{+}(t)\right|},
\qquad
P_k(t) = \frac{H_k(t)}{k}.
\end{equation}

Their macro-averaged forms are computed as
\begin{equation}
\mathrm{Recall}@k = \frac{1}{|\mathcal{T}_{\mathrm{test}}|} \sum_{t \in \mathcal{T}_{\mathrm{test}}} R_k(t),
\qquad
\mathrm{Precision}@k = \frac{1}{|\mathcal{T}_{\mathrm{test}}|} \sum_{t \in \mathcal{T}_{\mathrm{test}}} P_k(t).
\end{equation}

Recall@$k$ measures the fraction of annotated relevant MCP servers covered by the top-$k$ list, whereas Precision@$k$ measures the proportion of returned MCP servers that are relevant. To jointly reflect coverage and exactness, we further report F1@$k$, which is defined from the per-task precision and recall values:
\begin{equation}
F1_k(t) =
\begin{cases}
\frac{2 P_k(t) R_k(t)}{P_k(t) + R_k(t)}, & \text{if } P_k(t) + R_k(t) > 0,\\
0, & \text{otherwise},
\end{cases}
\end{equation}
and macro-averaged as
\begin{equation}
\mathrm{F1}@k = \frac{1}{|\mathcal{T}_{\mathrm{test}}|} \sum_{t \in \mathcal{T}_{\mathrm{test}}} F1_k(t).
\end{equation}

Since Recall@$k$ does not distinguish whether relevant MCP servers appear at the top or near the bottom of the ranked list, we additionally report NDCG@$k$ to evaluate position-sensitive ranking quality. Under binary relevance, the discounted cumulative gain for task $t$ is
\begin{equation}
\mathrm{DCG}@k(t) = \sum_{i=1}^{k} \frac{\mathbb{I}\!\left[m_i \in \mathcal{M}^{+}(t)\right]}{\log_2(i+1)},
\end{equation}
and the corresponding ideal DCG is
\begin{equation}
\mathrm{IDCG}@k(t) = \sum_{i=1}^{\min\left(k, \left|\mathcal{M}^{+}(t)\right|\right)} \frac{1}{\log_2(i+1)}.
\end{equation}

The macro-averaged NDCG@$k$ is then defined as
\begin{equation}
\mathrm{NDCG}@k = \frac{1}{|\mathcal{T}_{\mathrm{test}}|} \sum_{t \in \mathcal{T}_{\mathrm{test}}}
\frac{\mathrm{DCG}@k(t)}{\mathrm{IDCG}@k(t)}.
\end{equation}

A higher NDCG@$k$ value indicates that MCP servers relevant to the target task are ranked the top on average, which is particularly important in practice because developers usually inspect only the first few recommendations.

\subsection{Implementation and Settings}

All the experiments for T2MRec\footnote{https://github.com/ssea-lab/Task2MCP/T2MRec} were conducted on a server equipped with dual Intel Xeon Silver 4210 central processing units (2.20 GHz), two NVIDIA A6000 graphics processing units, and 128 GB of random access memory, using Python 3.8 and PyTorch 2.1.2 with CUDA Toolkit\footnote{https://developer.nvidia.com/cuda/toolkit} 12.1. For baseline methods, we followed the settings reported in their original papers and reproduced them based on publicly available source code. Since different baseline implementations have their own dependency requirements, they were executed in the corresponding software environments specified by their original implementations or released code. The Task2MCP dataset was partitioned into training, validation, and test sets in a 60:20:20 ratio. The validation set was used for hyperparameter tuning, and all final results were reported on the held-out test set.

For text modeling, each task $t$ was represented as a unified text field $x_t$ by concatenating its task name, description, programming language, category, and theme. Similarly, each MCP server $m$ was represented by $x_m$ as a concatenation of its name, description, supported programming language, system type, tool list, category, and subcategory. To model semantic alignment between task requirements and MCP tool capabilities, we trained a dual-tower projection network over the textual representations of tasks and MCP servers. Both the task tower and the MCP tower were implemented as three-layer MLPs, with a hidden dimension of $512$ and an output embedding dimension of $256$. ReLU was used as the activation function between layers, and dropout with a rate of $0.2$ was applied for regularization. The two towers were trained using a symmetric contrastive objective with in-batch negative samples. We optimized the T2MRec model with AdamW~\citep{adamw}, using a learning rate of $1\times10^{-3}$, a batch size of $256$, and a training duration of $200$ epochs. The temperature parameter in the contrastive loss was set to $0.07$. After model training, the task encoder and MCP encoder projected their inputs into a shared semantic embedding space. During model inference, semantic relevance and structural compatibility were optimally weighted at 0.9 and 0.1, respectively, and the fused scores were used to form the initial candidate set.

For the final refinement stage, we adopted a constrained LLM-based re-ranking procedure with GPT-5.2. Given a task description and a structured list of candidate MCP servers, the LLM model was instructed to reorder the candidates by the degree of alignment between the task requirements and the MCP servers' capabilities. To improve output consistency, we enforced a strict output format and used deterministic decoding during model inference. We further applied a validity check to verify the returned ranking before producing the final recommendation list. In our method, LLMs are used only as a constrained re-ranking component over a compact candidate set, which keeps inference overhead manageable. A more substantive role appears in the downstream interactive recommendation system, where it grounds recommended MCP servers in developers' concrete requirements and produces practical guidance for adoption and use.

\begin{table}[t]
\caption{Overall performance comparison of T2MRec and selected baselines on the Task2MCP dataset. The best results are shown in bold, and the best baseline results are underlined. Gains indicate the relative improvement of T2MRec over the strongest baseline for each metric.}
\label{tab:overall}
\centering
\small
\setlength{\tabcolsep}{2pt}
\resizebox{\textwidth}{!}{
\begin{tabular}{llcccccccc}
\toprule
\textbf{Category} & \textbf{Model} & \textbf{Recall@5} & \textbf{Recall@10} & \textbf{Precision@5} & \textbf{Precision@10} & \textbf{F1@5} & \textbf{F1@10} & \textbf{NDCG@5} & \textbf{NDCG@10}
\\
\midrule
\multirow{3}{*}{Traditional}
& NeuMF  & 0.5375 & 0.6423 & 0.1101 & 0.0674 & 0.1902 & 0.1226 & 0.4377 & 0.4568\\
& MSBPR  & 0.5463 & 0.6489 & 0.1093 & 0.0654 & 0.2247 & 0.1190 & 0.3994 & 0.4346\\
& ReCAFR & 0.4803 & 0.5957 & 0.0966 & 0.0600 & 0.1609 & 0.1090 & 0.3539 & 0.3916\\
\midrule
\multirow{4}{*}{GNN-based}
& NGCF     & 0.4541 & 0.5355 & 0.0904 & 0.0559 & 0.1937 & 0.1016 & 0.3352 & 0.3634\\
& LightGCN & 0.4047 & 0.5047 & 0.0809 & 0.0505 & 0.1349 & 0.0918 & 0.3035 & 0.3358\\
& GDSRec   & 0.3841 & 0.6454 & 0.1775 & 0.1209 & 0.1444 & 0.1417 & 0.0834 & 0.0927\\
& CVH-REC  & 0.1550 & 0.2853 & 0.1384 & 0.1165 & 0.1165 & 0.1682 & 0.3359 & 0.1572\\
\midrule
\multirow{5}{*}{LLM-based}
& TiGER    & 0.5631 & 0.5827 & 0.1115 & 0.0585 & 0.1859 & 0.1064 & \underline{0.5203} & 0.5267\\
& OpenP5   & 0.0081 & 0.0097 & 0.0017 & 0.0009 & 0.0029 & 0.0016 & 0.0061 & 0.0063\\
& LC-Rec   & 0.4677 & 0.5076 & 0.0945 & 0.0508 & 0.1575 & 0.0924 & 0.3872 & 0.3985\\
& MPT-Rec  & \underline{0.5838} & \underline{0.6533} & \underline{0.1804} & \underline{0.1463} & \underline{0.2690} & \underline{0.2456} & 0.4440 & \underline{0.5501}\\
& T2MRec   & \textbf{0.5971} & \textbf{0.6702} & \textbf{0.4246} & \textbf{0.3952} & \textbf{0.4963} & \textbf{0.4973} & \textbf{0.5452} & \textbf{0.6633} \\
\cmidrule{1-10}
& Gains & \textbf{2.23\%}$\uparrow$ & \textbf{2.52\%}$\uparrow$ & \textbf{57.51\%}$\uparrow$ & \textbf{62.89\%}$\uparrow$ & \textbf{45.80\%}$\uparrow$ & \textbf{50.61\%}$\uparrow$ & \textbf{4.57\%}$\uparrow$ & \textbf{17.07\%}$\uparrow$
\\
\bottomrule
\end{tabular}
}
\end{table}

\subsection{Experimental Results}

\subsubsection{Performance Comparison (RQ1)}

We analyzed the recommendation performance of T2MRec and selected competitive baselines on the Task2MCP dataset, using multiple evaluation metrics, to assess the overall effectiveness of the proposed method. We compared T2MRec with traditional, graph-based, and LLM-based recommendation methods using the same data split, candidate space, and evaluation protocol. Table~\ref{tab:overall} presents the comparison results, where the underlined values denote the best results achieved by the baseline methods.

Overall, T2MRec achieved the best overall performance over all baselines. Specifically, it obtained the highest Recall@5 and Recall@10, indicating a stronger ability to retrieve relevant MCP servers within a limited top-$K$ recommendation budget. It also achieved the highest NDCG@5 and NDCG@10, suggesting that relevant MCP servers were ranked earlier in the recommendation list. Furthermore, T2MRec outperformed all baselines regarding Precision@5, Precision@10, F1@5, and F1@10, confirming its superiority in both retrieval effectiveness and ranking quality. These results demonstrate that T2MRec provides more accurate and better-ordered recommendations for task-oriented MCP tool selection in realistic development scenarios.

First, traditional recommendation methods showed limited effectiveness in this setting. Methods such as MSBPR, NeuMF, and ReCAFR primarily model task–MCP associations from observed interaction patterns and therefore rely heavily on historical co-occurrence signals to infer relevance. Such modeling is useful for capturing coarse-grained matching regularities. Still, it is less effective for task-oriented MCP server recommendation, where the key challenge lies in aligning task intent with MCP tool capability. In our scenario, task descriptions usually emphasize user goals, functional requirements, constraints, and expected outcomes, whereas MCP server descriptions focus more on supported tools, interfaces, execution environments, and service capabilities. This mismatch makes it difficult for interaction-driven methods to establish accurate correspondences when semantic understanding is required. Among the traditional baselines, NeuMF and MSBPR were the most competitive, with NeuMF showing relatively stronger ranking quality and MSBPR achieving slightly better Recall and F1 values. However, both methods still consistently underperformed T2MRec across the overall evaluation. These results suggest that learning from interaction signals alone is insufficient for accurately recommending MCP servers for complex development tasks.

Second, graph-based recommendation methods improved the modeling of structural relations, but their gains remained limited. NGCF and LightGCN extend traditional recommender systems by exploiting higher-order connectivity on the task–MCP interaction graph, allowing information to propagate through multi-hop neighborhoods. GDSRec and CVH-REC further incorporate graph-based structural information from different perspectives, reflecting the potential of graph modeling for recommendation. For recommendation problems with dense interactions and strong relational patterns, such graph propagation is often highly effective. However, the task–MCP server recommendation scenario considered in this study differs from these settings. The interaction graph is relatively sparse, and the connections between tasks and MCP servers depend not only on structural proximity but also on whether an MCP server semantically satisfies the target task's functional requirements. Under these conditions, higher-order propagation may introduce noisy or weakly relevant signals, rather than providing reliable evidence for recommendations. In the experiment, although graph-based methods captured structural information more effectively than traditional interaction-driven approaches, they still failed to achieve competitive overall performance. This result suggests that structural modeling alone is insufficient for task-to-MCP recommendation. 

Third, LLM-based recommendation methods achieved stronger overall performance, highlighting the importance of semantic understanding in this specific recommendation scenario. Methods such as TiGER, LC-Rec, MPT-Rec, and OpenP5 are better equipped to capture the semantic relationships between task descriptions and MCP server descriptions. They are therefore generally more suitable for task-oriented MCP tool recommendation than traditional or graph-based methods. Among them, MPT-Rec was the strongest baseline on most evaluation metrics, and TiGER achieved the best baseline result on NDCG@5, showing that LLM-based methods can improve both retrieval and ranking quality when semantic matching is critical. However, their overall performance remained below that of T2MRec. Without an effective retrieval and filtering mechanism, LLM-based recommendations may suffer from instability, sensitivity to prompting strategies, and difficulty in maintaining consistent performance.  These findings indicate that while LLMs exhibit strong semantic reasoning capabilities, relying on them alone is insufficient to produce stable, accurate MCP server recommendations in realistic, large-scale development-task settings.

The overall comparison indicates the effectiveness of T2MRec for task-oriented MCP server recommendation. By consistently outperforming competitive baselines on both retrieval and ranking metrics, T2MRec demonstrates that accurate MCP tool recommendations for development tasks benefit from the joint use of semantic modeling, structural information, and ranking refinement.

\begin{tcolorbox}[title=\textbf{Answer to RQ1},colback=blue!3!white,colframe=blue!45!black]
T2MRec consistently outperforms selected competitive baselines in recommendation performance on the Task2MCP dataset. It demonstrates greater effectiveness in both retrieving relevant MCP servers and ranking them ahead of competing candidates, yielding the best overall performance across evaluation metrics. These results provide clear evidence that T2MRec is more effective than existing representative methods and can better capture the complex relationships between task requirements and MCP tool capabilities.
\end{tcolorbox}

\subsubsection{Ablation Study (RQ2)}

To answer RQ2, we conducted an ablation study to examine the contribution of each key component in T2MRec. Specifically, we evaluated four variants by removing or replacing one component at a time, including the lexical representation module, the two-tower encoder, contrastive learning, and the LLM-based re-ranking stage. Table~\ref{tab:ablation} reports the ablation study results on the test set. Overall, removing any component degraded model performance, confirming that all components contribute to T2MRec's overall performance. Among them, the largest performance drops were observed when the two-tower encoder or contrastive learning was removed, indicating that semantic representation learning is the core of our approach and that the re-ranking stage primarily contributes to finer-grained ordering.

\begin{table}[b]
\caption{Performance comparison among T2MRec variants.}
\label{tab:ablation}
\centering
\small
\setlength{\tabcolsep}{2pt}
\resizebox{\linewidth}{!}{%
\begin{tabular}{lcccccccc}
\toprule
\textbf{Metric} & \textbf{Recall@5} & \textbf{Recall@10} & \textbf{Precision@5} & \textbf{Precision@10} & \textbf{F1@5} & \textbf{F1@10} & \textbf{NDCG@5} & \textbf{NDCG@10} \\
\midrule
w/o CL         & 0.3824 & 0.4568 & 0.2761 & 0.2514 & 0.3220 & 0.3218 & 0.3527 & 0.4413 \\
w/o Two-tower  & 0.4726 & 0.5481 & 0.3423 & 0.3148 & 0.3979 & 0.3970 & 0.4385 & 0.5297 \\
w/o Lexical    & 0.5754 & 0.6482 & 0.4129 & 0.3836 & 0.4805 & 0.4805 & 0.5287 & 0.6463 \\
w/o Re-ranking & 0.5932 & 0.6668 & 0.4213 & 0.3925 & 0.4946 & 0.4947 & 0.5389 & 0.6574 \\
T2MRec         & \textbf{0.5971} & \textbf{0.6702} & \textbf{0.4246} & \textbf{0.3952} & \textbf{0.4963} & \textbf{0.4973} & \textbf{0.5452} & \textbf{0.6633} \\
\bottomrule
\end{tabular}
}
\end{table}

(1) \textit{Contrastive Learning}.
To examine the effect of the training objective, we constructed the \emph{w/o CL} variant by replacing the contrastive loss with a binary cross-entropy (BCE) style point-wise objective. As shown in Table~\ref{tab:ablation}, this variant suffered the largest performance drop among all ablated versions. Specifically, Recall@10 decreased from 0.6702 to 0.4568, Precision@10 dropped from 0.3952 to 0.2514, F1@10 declined from 0.4973 to 0.3218, and NDCG@10 decreased from 0.6633 to 0.4413. Similar degradation was consistently observed across the top-5 metrics. These results show that contrastive learning is a critical component of T2MRec and makes an essential contribution to both retrieval and ranking performance.

The effectiveness of contrastive learning is closely related to the characteristics of the task-to-MCP server matching problem. In this task, unobserved task–MCP server pairs cannot be treated as reliable negative samples, since some unlabeled MCP servers may still be potentially relevant to the target task. Under such conditions, a BCE-style point-wise objective may impose inaccurate negative supervision, thereby distorting the learned matching space. In contrast, the contrastive objective emphasizes the relative discrimination between positive and negative candidates within a mini-batch, making it better suited to handle label ambiguity. This component enables our model to learn a more discriminative compatibility space, thereby improving both candidate retrieval and ranking quality.

(2) \textit{Two-tower Encoder}.
To examine the effect of the dual-tower architecture, we constructed the \emph{w/o two-tower} variant, which removed the task and MCP projection networks and instead matched them directly in the original representation space. This variant consistently underperformed the complete model on all evaluation metrics. For example, Recall@10 decreased from 0.6702 to 0.5481, F1@10 decreased from 0.4973 to 0.3970, and NDCG@10 decreased from 0.6633 to 0.5297. These results show that the dual-tower architecture is important for model performance. 

\begin{figure*}[t]
  \centering
  \includegraphics[width=\textwidth]{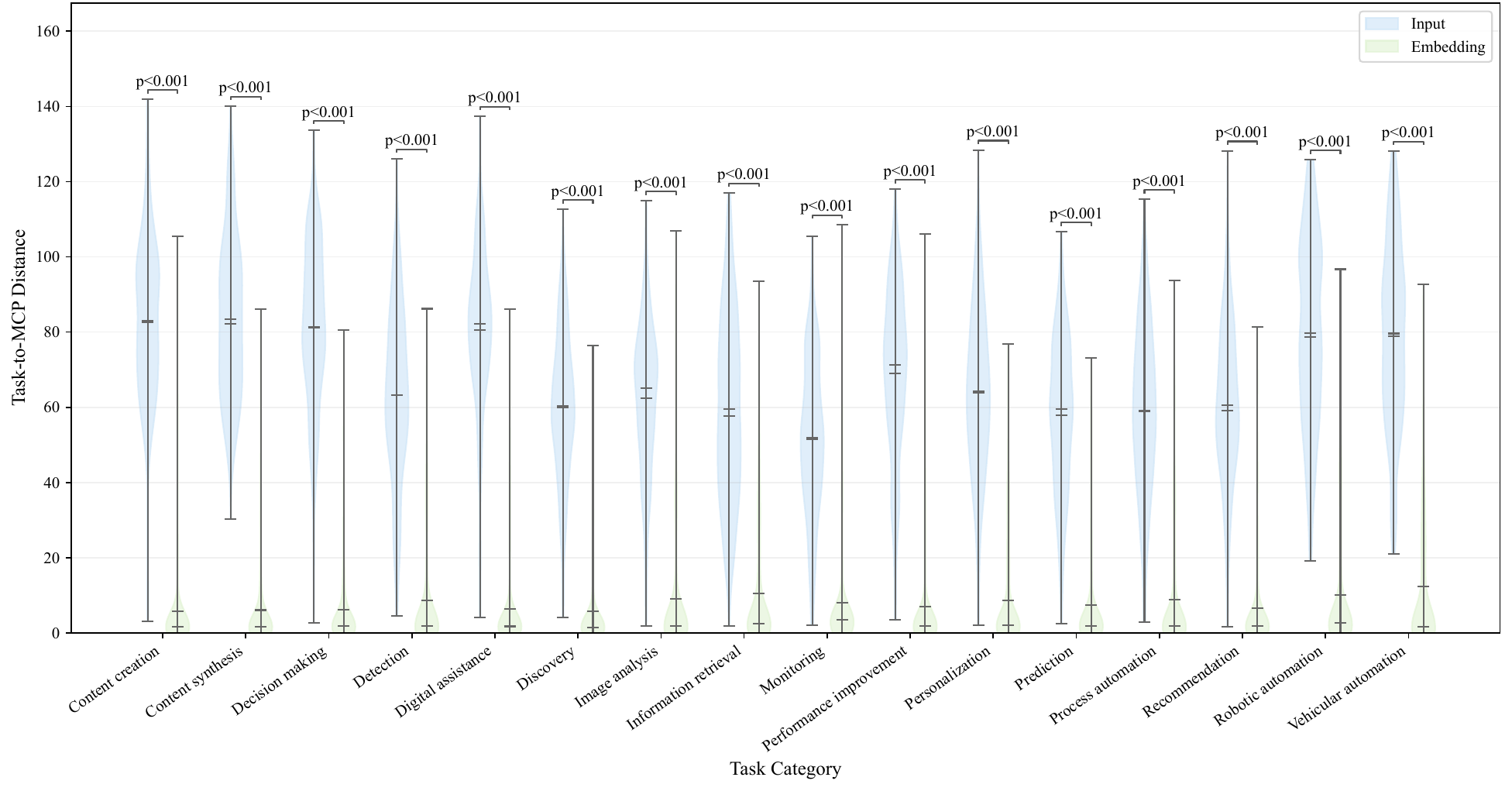}
  \caption{Effect of the two-tower encoder on task-to-MCP alignment in the shared embedding space using violin plots. Statistical significance is evaluated using paired one-sided Wilcoxon signed-rank tests with 95\% confidence intervals, where $p$-value denotes the category-wise difference between the input and embedding distances. $p<0.001$ indicates a very significant statistical difference.}
  \label{fig:alignment}
  \Description{Effect of the two-tower encoder on task-to-MCP alignment.}
\end{figure*}

To further examine whether the two-tower encoder improves alignment at the representation level, we compared the distributions of task-to-MCP distances before and after embedding learning across all task categories, as shown in Figure~\ref{fig:alignment}. For each category, statistical significance was evaluated using a paired one-sided Wilcoxon signed-rank test with a confidence interval of 95\%, where $p$ measured whether the embedding space significantly reduced the distance of matched task-to-MCP pairs relative to the input space. Figure~\ref{fig:alignment} shows that the embedding distance distributions were consistently far lower than the input distance distributions, and all categories satisfied $p<0.001$. This statistical result also confirms that the two-tower encoder learns a more effective shared space for aligning tasks with MCP servers.

This improvement can be attributed to the heterogeneous semantics of tasks and MCP servers. Task descriptions focus on requirements and goals, whereas MCP tool descriptions emphasize capabilities and execution support. The dual-tower architecture helps bridge this gap by learning separate mappings and projecting both sides into a shared space for compatibility modeling. Without this projection process, our model cannot align task requirements with MCP server capabilities as effectively, which results in weaker recommendation performance.

(3) \textit{Lexical Representation}.
To examine the effect of lexical-aware input representation, we designed the \emph{w/o lexical} variant by replacing the lexical-aware sparse vectors defined in Equations~\eqref{eq-4} and~\eqref{eq-5} with SBERT-based dense semantic embeddings for both tasks and MCP servers, while keeping the dual-tower encoder unchanged. As shown in Table~\ref{tab:ablation}, this change caused a consistent performance drop across all evaluation metrics. Specifically, Recall@10 decreased from 0.6702 to 0.6482, Precision@10 decreased from 0.3952 to 0.3836, F1@10 decreased from 0.4973 to 0.4805, and NDCG@10 decreased from 0.6633 to 0.6463. Similar degradation was also observed in the top-5 metrics. These results indicate that lexical representation remains an important component in task-oriented MCP server recommendation.

This finding is consistent with the characteristics of MCP tool descriptions. In many cases, MCP servers are described using capability-related lexical cues, such as tool names, library names, framework terms, API keywords, and other domain-specific identifiers, which often provide direct evidence for matching task requirements with MCP tool capabilities. Although dense semantic representations are effective in capturing overall semantic similarity, they may smooth over exact lexical signals when texts are mapped into a more abstract embedding space. Instead, lexical-aware representations explicitly preserve these distinctive cues, making them better suited to identifying concrete capability indicators in MCP server descriptions. Consequently, removing this component weakened our model's ability to capture fine-grained task–MCP server alignment, leading to lower retrieval effectiveness and poorer ranking quality.

(4) \textit{LLM Re-ranking}.
To examine the contribution of the final refinement stage, we constructed the \emph{w/o re-ranking} variant by removing the constrained LLM-based re-ranking module and directly using the pre-reranking order for MCP server recommendations. As shown in Table~\ref{tab:ablation}, this variant performed slightly worse than the complete model across all evaluation metrics. In particular, Recall@5 decreased from 0.5971 to 0.5932, Recall@10 decreased from 0.6702 to 0.6668, NDCG@5 decreased from 0.5452 to 0.5389, and NDCG@10 decreased from 0.6633 to 0.6574. Similar declines were also observed in Precision and F1. These results indicate that although the re-ranking stage can consistently improve both retrieval and ranking performance, its effect is more moderate than that of the contrastive objective or the dual-tower architecture.

This finding is consistent with the role of re-ranking in T2MRec. After the earlier stages have produced a strong candidate set, the remaining challenge is to more effectively distinguish among highly relevant MCP servers that differ subtly in their capability alignment. The constrained LLM-based re-ranking module provides this finer-grained refinement by leveraging more detailed semantic matching between task requirements and MCP server descriptions. Therefore, while its contribution is relatively small, it still yields more accurate and better-ordered recommendations, especially at the top of the ranked MCP server list.

\begin{tcolorbox}[title=\textbf{Answer to RQ2},colback=blue!3!white,colframe=blue!45!black]
All component designs of T2MRec contribute to its overall performance, but their effects differ in magnitude. The contrastive learning objective and the dual-tower architecture have the largest impact, %showing that learning a discriminative shared matching space is essential for recommendation quality. 
the lexical-aware representation further improves performance by preserving capability-related cues in MCP tool descriptions, and the LLM-based re-ranking stage provides additional gains by refining the order of highly relevant candidates. %Overall, T2MRec's strong performance stems from the effective collaboration among these components rather than from any single design choice.
\end{tcolorbox}

\subsubsection{Parameter Sensitivity Analysis (RQ3)}

\begin{figure*}[t]
  \centering
  \includegraphics[width=\textwidth]{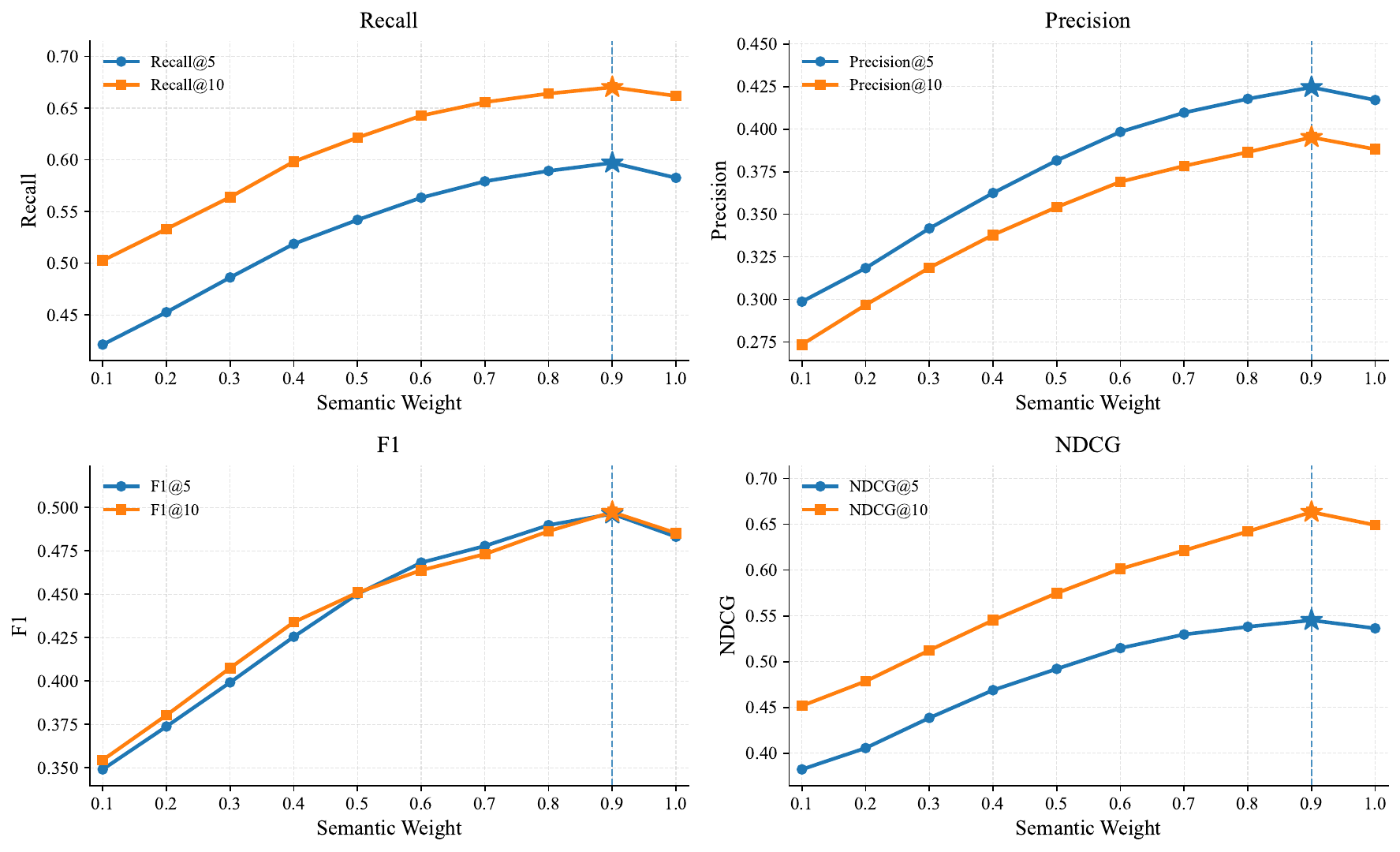}
  \caption{Effect of different semantics–structure weights.}
  \label{fig:semantic}
  \Description{Semantic weight.}
\end{figure*}

\begin{figure*}[t]
  \centering
  \includegraphics[width=\textwidth]{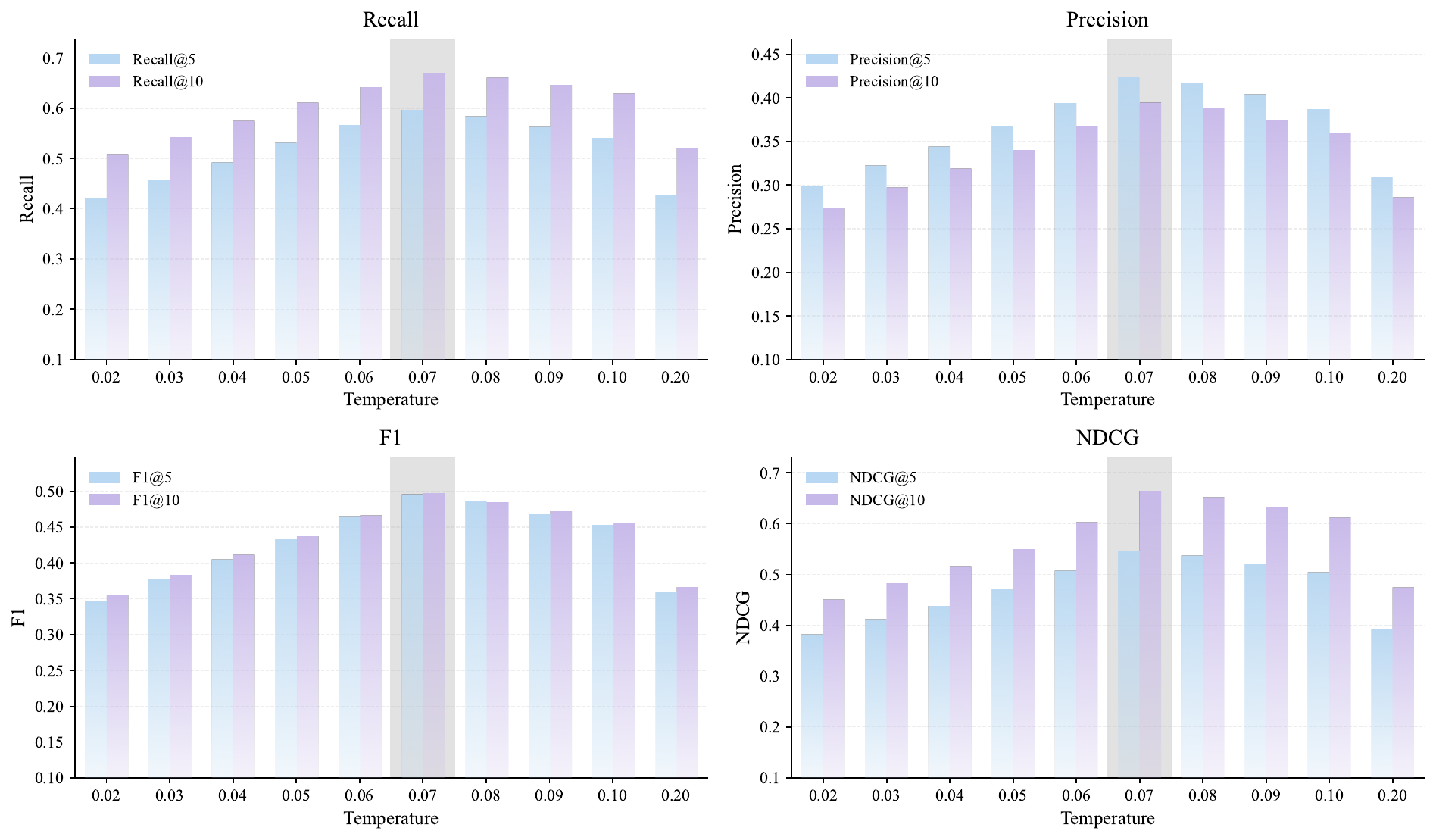}
  \caption{Effect of different temperature values in the contrastive objective.}
  \label{fig:temperature}
  \Description{Temperature.}
\end{figure*}

\begin{figure*}[t]
  \centering
  \includegraphics[width=\textwidth]{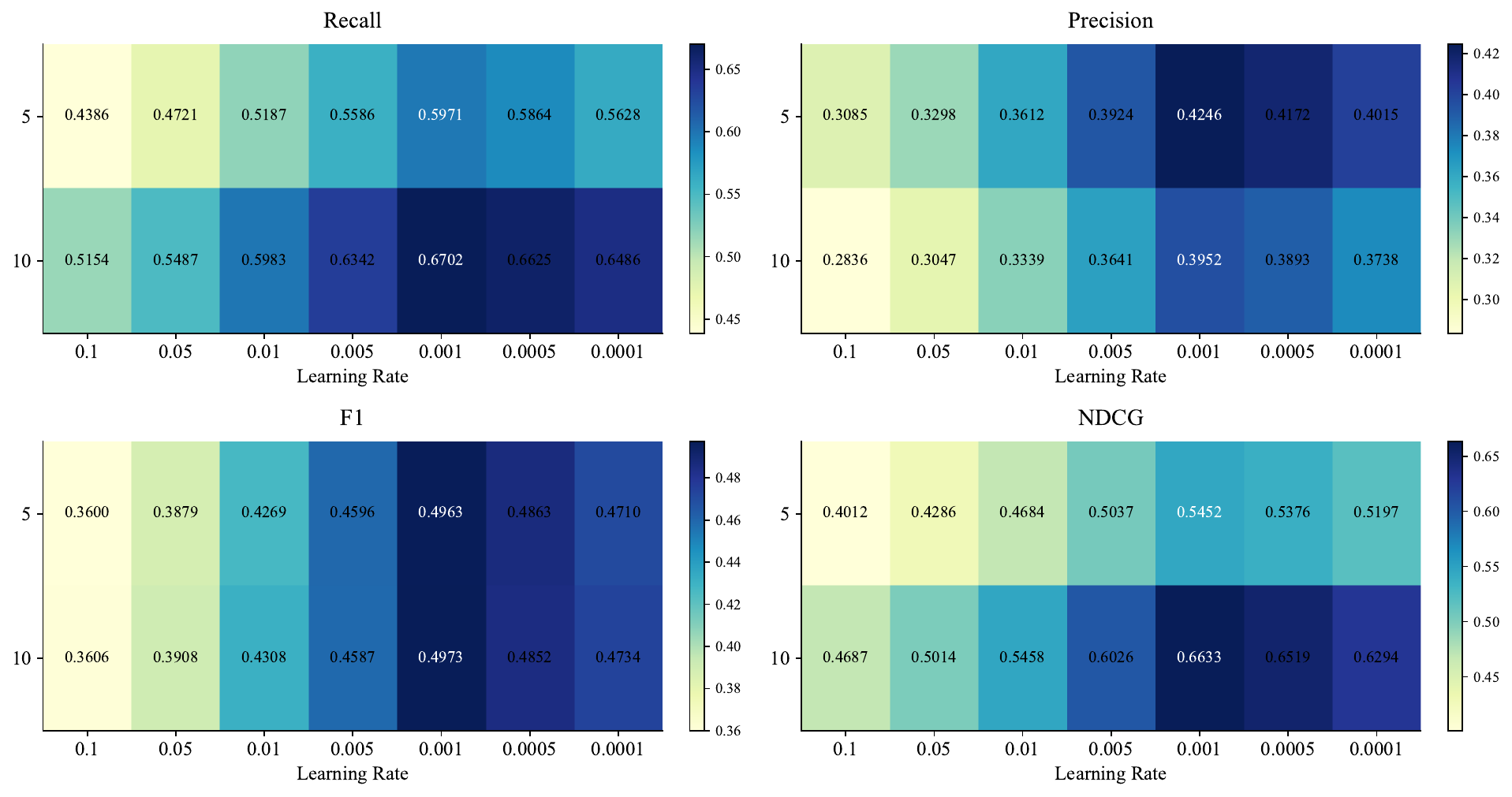}
  \caption{Effect of different learning rates.}
  \label{fig:LR}
  \Description{LR}
\end{figure*}

To answer RQ3, we analyzed the sensitivity of T2MRec to three key hyperparameters: the fusion weights for semantic relevance and structural compatibility, the temperature parameter in contrastive learning, and the learning rate used to optimize the dual-tower encoder. Figures~\ref{fig:semantic}, \ref{fig:temperature}, and \ref{fig:LR} present the analysis results of these hyperparameters. Overall, T2MRec's effectiveness does not depend on fragile parameter choices, but rather on a principled balance among its primary components.

(1) \textit{Semantics–Structure Weighting}.
Figure~\ref{fig:semantic} presents the impact of varying the fusion weights between semantic relevance and structural compatibility, with the constraint that the two weights sum to one. The results show that a higher weight on semantic relevance consistently improves the model's performance across all metrics, indicating that semantic relevance is the leading factor in task-oriented MCP server recommendation. Since both task and MCP descriptions are primarily textual, successful recommendation depends first on whether the model can correctly capture the semantic alignment between task requirements and MCP tool capabilities.

The best performance was achieved with a semantic weight of 0.9 and a structural weight of 0.1. This result is highly consistent with the design intuition of T2MRec. On the one hand, semantic relevance should dominate the matching process because it directly reflects whether an MCP server is functionally suitable for a given task. On the other hand, structural compatibility still provides useful complementary information. Although its contribution is smaller, it helps regularize the matching process by incorporating additional relational signals, thereby reducing the risk of selecting MCP servers that appear semantically relevant but are less suitable from a structural or contextual perspective.

Another notable observation is that neither extreme setting yielded the best result. When the semantic weight was too low, the model relied excessively on structural signals, which weakened its ability to capture the core functional intent expressed in task descriptions. In contrast, when the structural weight was disabled entirely, the model lost the auxiliary guidance provided by compatibility relations, which also slightly degraded recommendation quality. In this sense, the setting of 0.9 and 0.1 provides the most effective trade-off, enabling T2MRec to exploit semantically dominant matching while still benefiting from structural support.

(2) \textit{Temperature in Contrastive Learning}.
Figure~\ref{fig:temperature} shows the effect of the temperature parameter in the contrastive objective used to train the dual-tower encoder. The results show that T2MRec is sensitive to this parameter, and that the best overall performance is obtained at a moderate temperature. This behavior is consistent with the role of temperature in controlling the sharpness of similarity distributions and, consequently, the discrimination strength of contrastive learning.

When the temperature value was too low, the contrastive objective became overly sharp, causing the model to overemphasize hard negatives and making optimization less stable. This issue may lead to an overly rigid representation space and weaker generalization. In contrast, when the temperature value was too high, the contrastive distribution became overly smooth, reducing the separation between positive and negative pairs. As a result, the learned embeddings became less discriminative, which weakened both retrieval and ranking performance.

The best overall performance was achieved at a temperature of 0.07, indicating that this value made the most effective balance between representation discrimination and optimization stability. In this setting, the contrastive objective was sharp enough to encourage clear semantic separation while remaining smooth enough to avoid overfitting to a narrow set of hard negatives. More broadly, these results confirm that the temperature parameter is a critical factor in shaping the quality of the shared semantic space learned by T2MRec. A properly calibrated temperature enables the model to capture task–MCP server compatibility more accurately, and 0.07 is the most suitable choice under our experimental setting.

(3) \textit{Learning Rate}.
Figure~\ref{fig:LR} illustrates the effect of the learning rate used to train the dual-tower encoder. The results show that T2MRec is clearly sensitive to this parameter, and that both excessively large and excessively small learning rates lead to inferior performance. In general, the learning rate directly affects the stability and efficiency of optimization and therefore substantially influences the quality of the learned task–MCP server representations.

When the learning rate was too large, parameter updates became excessively aggressive, making the optimization process unstable and causing the model to overshoot good local solutions. Under this condition, the encoder had difficulty learning a well-structured semantic space in which matched task–MCP pairs were placed close together while irrelevant pairs were effectively separated. As a result, both retrieval coverage and ranking quality were negatively affected. In contrast, when the learning rate was too low, the optimization process became overly conservative. Although training remained stable, the model failed to make sufficient progress within the fixed training budget and converged slowly to a suboptimal solution. This result also weakened the discriminative ability of the learned representations and led to inferior recommendation performance.

The best overall performance was achieved with a learning rate of 0.001. This setting provided the most effective balance between optimization stability and convergence efficiency, allowing the model to learn more discriminative semantic representations for the task–MCP matching. Overall, these results confirm that the learning rate is a key training hyperparameter for T2MRec.

\begin{tcolorbox}[title=\textbf{Answer to RQ3},colback=blue!3!white,colframe=blue!45!black]
T2MRec shows stable but non-negligible sensitivity to its key hyperparameters. The best results are achieved with a semantic–structural ratio of 0.9:0.1, showing that semantic relevance should dominate the matching process, while structural information remains a useful complement. The model also performs best with a temperature of 0.07 and a learning rate of 0.001, indicating the importance of balancing representation discrimination with optimization stability. %Overall, these findings support that the design of T2MRec is effective.
\end{tcolorbox}

\section{MCP server Recommendation Agent Prototype}
\label{6}

To demonstrate how the T2MRec model can support practical MCP tool selection, we instantiate the proposed recommendation approach as a conversational recommender AI agent, \textbf{T2MAgent}\footnote{https://github.com/ssea-lab/Task2MCP/T2MAgent}. The purpose of this case study of prototype development is not to introduce an additional algorithmic component, but to show how retrieval-based MCP server recommendation can be integrated into an interactive decision-support workflow for real development tasks.

\subsection{Design Objectives}

Sections~\ref{3} and~\ref{4} present the Task2MCP dataset and the T2MRec model, which together support the retrieval and ranking of candidate MCP servers for a given task. In practice, however, providing ranked results alone is often insufficient for adoption in real-world development scenarios. Developer requests are often brief and under-specified, with important constraints such as programming language, execution environment, privacy requirements, and deployment conditions left implicit. As a result, developers often need to inspect metadata further, compare alternatives, and determine how to incorporate a recommended MCP tool into their workflow.

To address this gap, we design a conversational MCP server recommendation agent built on top of T2MRec. Given a natural-language task description and optional constraints, the agent identifies suitable MCP servers, explains why they are appropriate, and guides practical adoption. Four objectives guide the design.
\begin{itemize}
\item \textit{Structured understanding of developer intent}.
The agent should convert natural-language requests into a structured task specification for the MCP server recommendation. In addition to identifying the core task intent, it should extract explicit constraints when they are provided, such as programming language, runtime environment, platform, or domain requirements. It should also recognize higher-level preferences, including latency, reliability, and privacy considerations. When essential constraints are missing, the agent should support concise follow-up interactions to reduce ambiguity before making a recommendation.

\item \textit{Evidence-grounded recommendation rationale}.
The agent should explain recommendations based on observable MCP tool information rather than unsupported free-form generation. Recommendation rationales should be grounded in MCP server metadata, including descriptions, tool lists, supported languages, system compatibility, licenses, and project activity signals. The explanation should explicitly connect task requirements with MCP server capabilities and clarify the principal trade-offs among recommended options.

\item \textit{Actionable adoption support}.
The agent should provide outputs that are directly useful for practical adoption. Beyond identifying suitable MCP servers, it should describe basic setup considerations, outline how the recommended MCP server can be incorporated into an agent workflow, and highlight common integration constraints or deployment issues. This design ensures that the output of our prototype system remains actionable rather than purely descriptive.

\item \textit{Interactive refinement and extensibility}.
The agent should support iterative refinement as user requirements evolve. Users may revise constraints, clarify priorities, or ask follow-up questions about deployment and configuration. The prototype system should therefore support multi-turn interaction while maintaining a modular design that can accommodate new MCP sources and evolving metadata schemas.
\end{itemize}

\subsection{Business Workflow}

The T2MAgent agent follows a retrieval-centered workflow in which candidate MCP servers are first identified by the T2MRec model and then explained by an LLM over a restricted evidence set. This design preserves controllability and reduces the risk of unsupported recommendations. The system workflow for users consists of four stages.

First, the user submits a task request via an interactive interface, optionally including practical constraints such as preferred programming language, platform environment, or deployment requirements. The input may be incomplete or partially specified.

Second, an LLM-based parsing module transforms the user request into a structured task specification. This module identifies the core development objective and normalizes the constraints that are relevant to MCP tool feasibility and integration. If critical information is missing, the prototype system would request clarification before retrieval.

Third, the structured task specification is passed to T2MRec for candidate retrieval and ranking. This recommender searches the MCP catalog over standardized metadata fields and returns a ranked shortlist of MCP server candidates. To support practical decision-making, the prototype system presents only the top-5 MCP servers as final recommendations. This design reflects a common usage pattern in which developers compare a small number of alternatives rather than inspect a long ranked list.

\begin{figure*}[t]
  \centering
  \includegraphics[width=\textwidth]{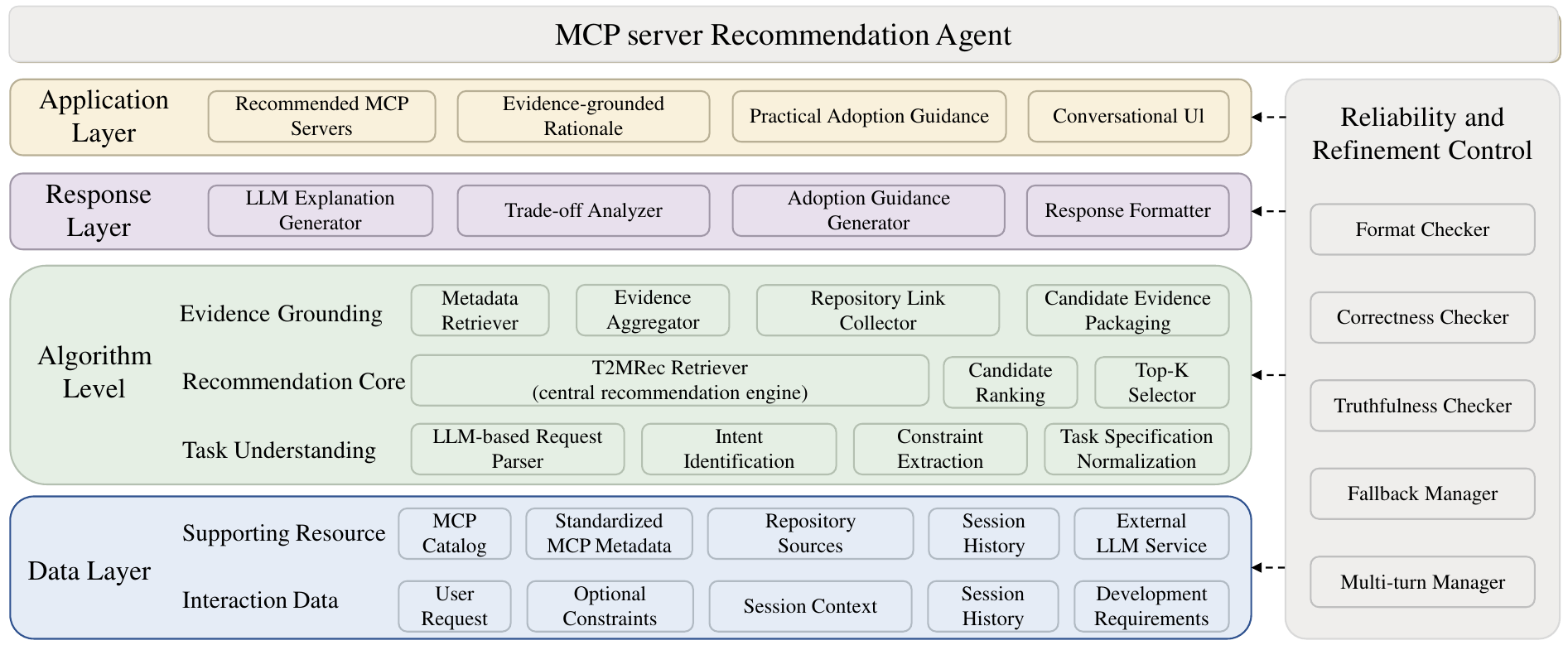}
  \caption{Overall system architecture of T2MAgent for retrieval-centered MCP server recommendation.}
  \label{fig:agent}
  \Description{System architecture of T2MAgent for retrieval-centered MCP server recommendation.}
\end{figure*}

Fourth, the prototype system augments the selected MCP tools with traceable evidence and invokes an LLM to produce a structured response. For each recommended MCP server, it retrieves metadata and provenance information, including descriptive fields and links to the corresponding repository or project webpage. According to this evidence, the LLM model generates a response that explains why each recommended MCP server is suitable, summarizes the key trade-offs, and provides basic guidance for adoption. Importantly, the LLM model operates only on the pre-filtered candidates and their associated evidence, rather than performing unconstrained retrieval.

The prototype system also supports multi-turn refinement. Users may revise requirements, ask about installation or configuration, or request comparisons under new constraints. To improve reliability, it includes fallback rules for response regeneration. If an LLM response fails format checks, omits required fields, or conflicts with retrieved evidence, the prototype system falls back to the retrieved shortlist. It then regenerates a new response under stricter constraints. In this way, the T2MAgent agent combines retrieval quality, evidence traceability, and interactive usability within a single workflow.

\subsection{System Architecture}

Figure~\ref{fig:agent} presents the overall architecture of T2MAgent. Following the retrieval-centered workflow described above, the prototype system is organized into four functional layers: data, algorithm, response, and application layers, together with a reliability and refinement control module. This design separates candidate retrieval from response generation, thereby improving controllability, traceability, and support for multi-turn interaction.

(1) \textit{The data layer} provides the information foundation of the prototype system. It consists of supporting resources and interaction data. Supporting resources include the MCP catalog, standardized MCP server metadata, repository sources, session history, and external LLM services. Interaction data includes the user request, optional constraints, session context, session history, and development requirements accumulated during interaction. This layer supports both recommendation and explanation by maintaining MCP-side knowledge and user-side context in a unified representation.

(2) \textit{The algorithm layer} implements the recommendation pipeline of T2MRec. It includes three main components: task understanding, recommendation core, and evidence grounding. The task understanding component converts natural-language user requests into structured task specifications through LLM-based request parsing, intent identification, constraint extraction, and task specification normalization. The recommendation core is centered on the T2MRec retriever, which retrieves and ranks candidate MCP servers according to the structured task specification. A top-$K$ selector then produces a shortlist for downstream processing. The evidence grounding component augments shortlisted candidates with traceable supporting information, including metadata, repository links, and related project-level evidence, and packages them into a bounded evidence set for response generation.

(3) \textit{The response layer} transforms retrieved MCP server candidates and grounded evidence into structured outputs. It includes an LLM explanation generator, a trade-off analyzer, an adoption guidance generator, and a response formatter. Importantly, the LLM model in this layer operates only on the shortlisted candidates and their associated evidence, rather than performing unconstrained retrieval on possible ones. This design ensures that recommendation rationales remain grounded in observable MCP information.

\begin{figure*}[t]
  \centering
  \includegraphics[width=0.49\textwidth]{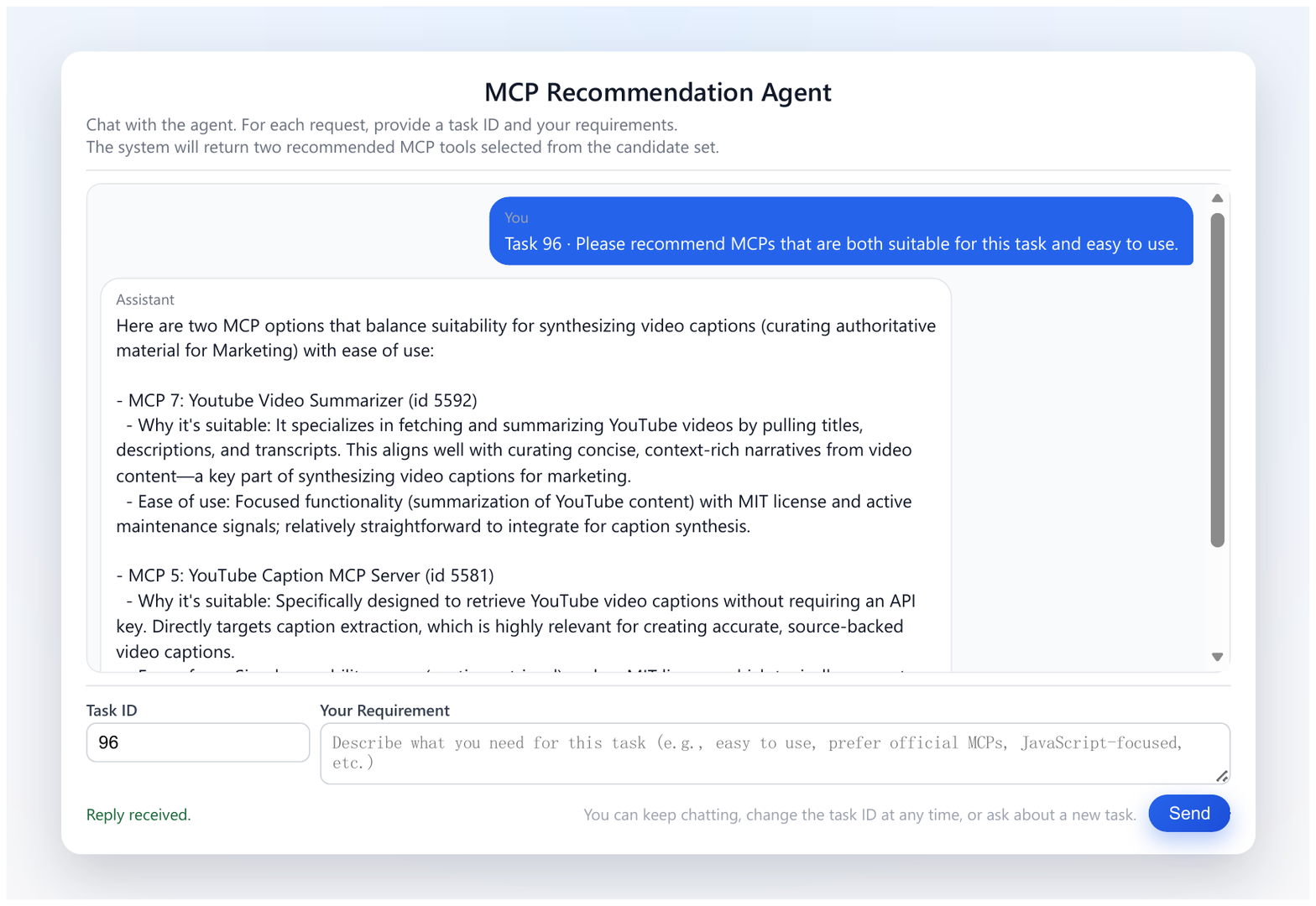}
  \includegraphics[width=0.49\textwidth]{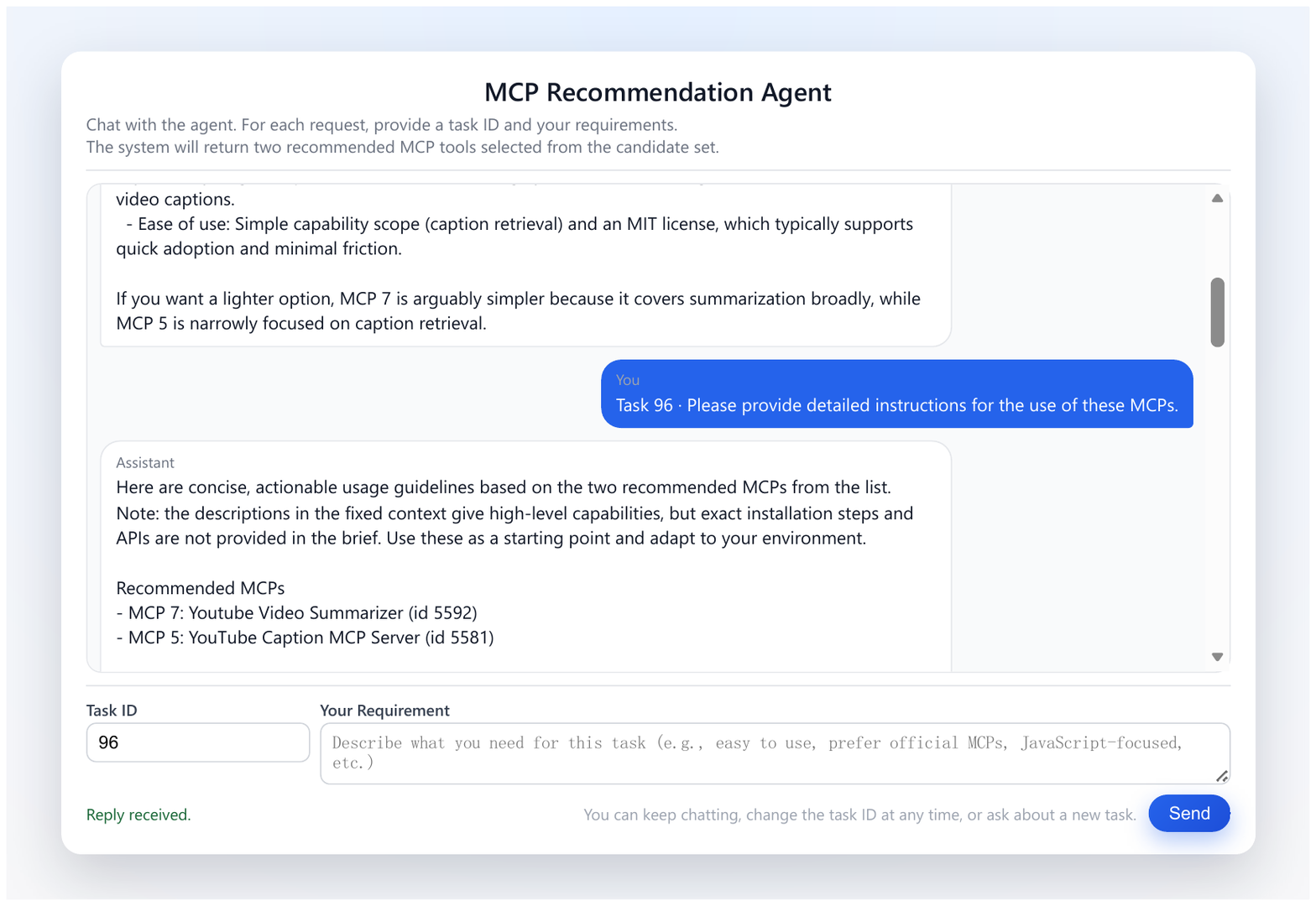}
  \caption{Interactive interface of the T2MAgent agent showing MCP server recommendations and evidence-grounded explanations.}
  \label{fig:youtube-ui}
  \Description{User interface examples of the MCP recommendation agent.}
\end{figure*}

(4) \textit{The application layer} presents the final outputs through a conversational interface (see Figure~\ref{fig:youtube-ui}), including recommended MCP servers, evidence-grounded rationales, and practical adoption guidance. The interface remains lightweight and delegates parsing, recommendation, and generation to the lower layers.

To improve robustness, the prototype system further includes a \textit{reliability \& refinement control module}. This module contains a format checker, a correctness checker, a truthfulness checker, a fallback manager, and a multi-turn manager. It verifies output quality, checks consistency between generated content and retrieved evidence, and supports refinement when users revise requirements or regenerate content.

Overall, the architecture of T2MAgent separates candidate retrieval and ranking from natural-language response generation. MCP server recommendation is handled by the retrieval pipeline, whereas LLM components are used for task understanding and evidence-grounded response construction. This separation helps maintain controllability and traceability while enabling the prototype system to support explainable recommendations and iterative interaction.

\subsection{Illustrative Case}

\begin{figure*}[t]
  \centering
  \includegraphics[width=\textwidth]{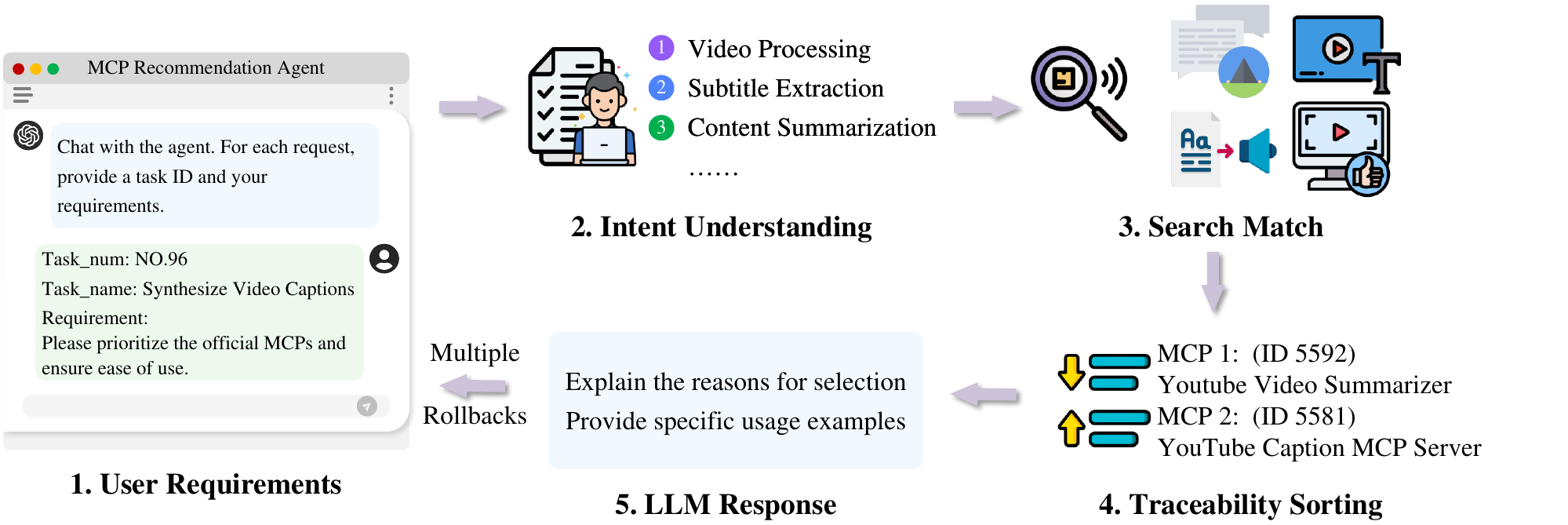}
  \caption{Illustrative task used in the case study: generating marketing-oriented captions for YouTube videos.}
  \label{fig:case-example}
  \Description{Illustrative task example for the MCP recommendation agent.}
\end{figure*}

We use a marketing-oriented development task to illustrate how the MCP server recommendation agent operates. As shown in Figure~\ref{fig:case-example}, a user aims to generate captions for YouTube videos. This task requires extracting video-level signals, such as titles, descriptions, and transcripts (or captions), and producing concise outputs suitable for downstream marketing use. In addition to functional matching, the user also values ease of adoption and practical usability.

Given this task specification, the T2MAgent agent returned two high-confidence MCP server candidates: \emph{Youtube Video Summarizer} (ID 5592) and \emph{YouTube Caption MCP Server} (ID 5581). These recommendations are complementary and correspond to different usage priorities. Figure~\ref{fig:youtube-ui} shows the two recommended MCP servers, along with evidence-based explanations and practical guidance, enabling users to compare alternatives and iteratively refine their requirements within the same interaction loop.
\begin{itemize}
\item \emph{Youtube Video Summarizer} is recommended as the primary tool option because it supports richer end-to-end caption generation. It can access video metadata and textual evidence, including titles, descriptions, and transcripts, thereby providing the contextual information needed to generate concise, informative, marketing-oriented captions. This advantage makes it more suitable for tasks that require not only caption extraction but also higher-level summarization.

\item \emph{YouTube Caption MCP Server} is recommended as an alternative tool because it focuses on lightweight caption retrieval. Its key advantage is that it can obtain caption text without requiring an API key, simplifying deployment when credentials are limited. This feature makes it appropriate when the main requirement is accurate caption extraction, and subsequent rewriting or summarization is handled elsewhere in the LLM-based agent pipeline.
\end{itemize}

The prototype system also makes the decision boundary explicit. If the user prioritizes richer contextual understanding and higher-level summarization, \emph{Youtube Video Summarizer} is the preferable choice. If the user instead prioritizes a simpler, more reliable caption-extraction pipeline, \emph{YouTube Caption MCP Server} is more suitable than the former. The two MCP servers can also be composed: the caption server supplies accurate caption text, and the summarizer contributes broader contextual information for downstream caption generation.

\section{Threats to Validity}
\label{7}

This section discusses potential threats to the validity of our research and experimental results.

\subsection{Internal Validity}

Threats to internal validity concern dataset construction, baseline replication, and model configuration. The first threat arose from the construction of Task2MCP. The task–MCP labels were obtained through taxonomy-guided matching, LLM-assisted checking, and manual verification rather than from naturally occurring usage logs. Although this process improved consistency, subjective bias might remain in judging whether an MCP server is relevant to a task. To mitigate this threat, we applied structured matching criteria, conducted multiple rounds of validation, and performed manual cross-checking during dataset construction. Another threat concerned the implementation of baselines. Some baseline methods were originally proposed for traditional recommendation or generative recommendation settings rather than task-oriented MCP server recommendation. Although we followed the settings reported in the original papers and used publicly available implementations whenever possible, differences in implementation details and adaptation strategies might still affect the comparison results. To reduce this threat, all methods were evaluated under the same data split, candidate space, and evaluation protocol. In addition, T2MRec includes several configurable components, including semantic–structural fusion, centroid expansion, and LLM re-ranking. To examine whether the reported gains depend heavily on these design choices, we conducted ablation studies and parameter sensitivity analysis. 

\subsection{External validity}

Threats to external validity concern the generalizability of our findings beyond the current benchmark. The Task2MCP dataset is constructed from public MCP directories and GitHub repositories. Although it covers thousands of MCP servers across multiple categories and programming languages, it may not fully represent private, proprietary, or rapidly evolving MCP ecosystems. In addition, our task corpus is taxonomy-grounded and manually validated, which improves coverage and consistency. However, real-world development requests may still involve organizational constraints, legacy dependencies, or multi-step workflows that are not fully captured in the benchmark. To partially mitigate this threat, we constructed the dataset from diverse public MCP sources and evaluated the T2MRec model on tasks covering multiple categories. Nevertheless, further validation on real usage logs, proprietary MCP server/client environments, and interactive deployment settings is still needed. 

\subsection{Construct Validity}

Threats to construct validity concern the appropriateness of evaluation metrics used to measure recommendation quality. In this study, we employed widely-used ranking metrics, such as Recall@$k$, Precision@$k$, F1@$k$, and NDCG@$k$, to assess the effectiveness of T2MRec in retrieving and ranking MCP servers. While these metrics are standard in the recommender systems literature, they measure alignment with annotated relevance, which may not directly correspond to real-world integration success or developer satisfaction. Additionally, the task corpus relies on a curated relevant set for each task, which may overlook useful MCP servers outside the annotated set, potentially underestimating the true coverage of recommendations. To address this, we focused on comparative evaluation under identical conditions, enabling us to draw meaningful conclusions about the relative performance of different methods while acknowledging the benchmark's inherent limitations.

\subsection{Conclusion Validity}

Threats to conclusion validity concern the extent to which the current experimental setting influences the observed results. In our study, the reported gains of T2MRec may vary with the benchmark composition, task-level data split, and hyperparameter settings. To address this threat, we evaluated all methods using the same experimental protocol and conducted ablation studies and parameter sensitivity analyses to verify the contributions of key components. In addition, because the LLM-based components rely on fixed model versions and decoding settings, the reported results should be interpreted within the specific configuration used in this study.

\section{Conclusion}
\label{8}

In this article, we investigate the problem of task-oriented MCP tool recommendation. As the MCP ecosystem continues to expand, selecting suitable MCP servers for concrete development tasks becomes increasingly challenging. Effective recommendations require not only accurate modeling of the semantic alignment between task requirements and MCP server capabilities, but also careful consideration of feasibility constraints in realistic development environments. To facilitate research on task-oriented MCP server recommendation, we construct Task2MCP, a dataset that links taxonomy-grounded development tasks to curated MCP servers. Based on this dataset, we propose T2MRec, a task-to-MCP server recommendation model. It jointly models semantic relevance and structural compatibility, and further improves ranking quality through constrained LLM-based re-ranking. Experimental results on the Task2MCP dataset show that the T2MRec model consistently outperforms competitive baseline methods. We further instantiate the T2MRec model as an AI recommender agent to demonstrate its practical use for interactive, evidence-grounded MCP server selection.

In the T2MRec model, task-oriented MCP server recommendation is formulated as a retrieval-and-ranking problem over structured MCP metadata. Although the present study demonstrates the effectiveness of this formulation in offline single-task Top-$K$ recommendation, several directions remain for future research. Therefore, we plan to extend the benchmark with more realistic task sources, richer MCP metadata, and continuously evolving MCP ecosystems. We also plan to investigate more complex recommendation settings, including multi-tool composition, iterative clarification, and tighter integration between recommendation and execution feedback. In addition, because multiple MCP tools may satisfy the same development requirement under different constraints, it is necessary to explore more flexible evaluation settings that better reflect practical utility. More broadly, future work will examine how task-oriented MCP server recommendations can provide stronger support for interactive LLM-based agent systems and a wider range of software development scenarios.

\begin{acks}
This work was supported by the National Natural Science Foundation of China (No. 62302536), the Postdoctoral Fellowship Program of China Postdoctoral Science Foundation (Nos. GZC20240571 and 2024M751052), the Doctoral Student Program of the Young S\&T Talents Cultivation Project, CAST, and the Open Research Fund of the State Key Laboratory for Novel Software Technology, Nanjing University (No. KFKT2025B08).
\end{acks}

\bibliographystyle{ACM-Reference-Format}
\bibliography{sample-base}

@String{Computing = "Computing" }

@article{LiuFLWZZX25,
  author       = {Shang Liu and
                  Wenji Fang and
                  Yao Lu and
                  Jing Wang and
                  Qijun Zhang and
                  Hongce Zhang and
                  Zhiyao Xie},
  title        = {RTLCoder: Fully Open-Source and Efficient LLM-Assisted {RTL} Code
                  Generation Technique},
  journal      = {{IEEE} Trans. Comput. Aided Des. Integr. Circuits Syst.},
  volume       = {44},
  number       = {4},
  pages        = {1448--1461},
  year         = {2025}
}

@inproceedings{0002WR0W025,
  author       = {Jinghan Zhang and
                  Xiting Wang and
                  Weijieying Ren and
                  Lu Jiang and
                  Dongjie Wang and
                  Kunpeng Liu},
  editor       = {Toby Walsh and
                  Julie Shah and
                  Zico Kolter},
  title        = {{RATT:} {A} Thought Structure for Coherent and Correct {LLM} Reasoning},
  booktitle    = {Proceedings of the 39th AAAI Conference on Artificial Intelligence, AAAI 2025},
  address = {Philadelphia, PA, {USA}},
  pages        = {26733--26741},
  publisher    = {{AAAI} Press},
  year         = {2025}
}

@inproceedings{ZhangY0HL25,
  author       = {Xinghua Zhang and
                  Haiyang Yu and
                  Cheng Fu and
                  Fei Huang and
                  Yongbin Li},
  editor       = {Wanxiang Che and
                  Joyce Nabende and
                  Ekaterina Shutova and
                  Mohammad Taher Pilehvar},
  title        = {{IOPO:} Empowering LLMs with Complex Instruction Following via Input-Output
                  Preference Optimization},
  booktitle    = {Proceedings of the 63rd Annual Meeting of the Association for Computational
                  Linguistics (Volume 1: Long Papers), {ACL} 2025},
  address = {Vienna, Austria},
  pages        = {22185--22200},
  publisher    = {Association for Computational Linguistics},
  year         = {2025}
}

@article{HeTL25,
  author       = {Junda He and
                  Christoph Treude and
                  David Lo},
  title        = {LLM-Based Multi-Agent Systems for Software Engineering: Literature
                  Review, Vision, and the Road Ahead},
  journal      = {{ACM} Trans. Softw. Eng. Methodol.},
  volume       = {34},
  number       = {5},
  pages        = {124:1--124:30},
  year         = {2025}
}

@inproceedings{MaH0WZL25,
  author       = {Zhiyuan Ma and
                  Zhenya Huang and
                  Jiayu Liu and
                  Minmao Wang and
                  Hongke Zhao and
                  Xin Li},
  editor       = {Toby Walsh and
                  Julie Shah and
                  Zico Kolter},
  title        = {Automated Creation of Reusable and Diverse Toolsets for Enhancing
                  {LLM} Reasoning},
  booktitle    = {Proceedings of the 39th AAAI Conference on Artificial Intelligence, AAAI 2025},
  address = {Philadelphia, PA, {USA}},
  pages        = {24821--24830},
  publisher    = {{AAAI} Press},
  year         = {2025}
}

@inproceedings{0014W025,
  author       = {Han Xu and
                  Xingyuan Wang and
                  Shihao Ji},
  editor       = {Meeyoung Cha and
                  Chanyoung Park and
                  Noseong Park and
                  Carl Yang and
                  Senjuti Basu Roy and
                  Jessie Li and
                  Jaap Kamps and
                  Kijung Shin and
                  Bryan Hooi and
                  Lifang He},
  title        = {Reliable and Efficient Container Orchestration of LLMs via {MCP}},
  booktitle    = {Proceedings of the 34th {ACM} International Conference on Information
                  and Knowledge Management, {CIKM} 2025},
  pages        = {6885--6886},
  address = {Seoul, Republic of Korea},
  publisher    = {{ACM}},
  year         = {2025}
}

@inproceedings{LinRLZ25,
  author       = {Zhiwei Lin and
                  Bonan Ruan and
                  Jiahao Liu and
                  Weibo Zhao},
  title        = {A Large-Scale Evolvable Dataset for Model Context Protocol Ecosystem
                  and Security Analysis},
  booktitle    = {Proceedings of the 40th {IEEE/ACM} International Conference on Automated Software Engineering,
                  {ASE} 2025},
  address = {Seoul, Republic of Korea},
  pages        = {3985--3988},
  publisher    = {{IEEE}},
  year         = {2025}
}

@misc{guo2025measurement,
      title={A Measurement Study of Model Context Protocol Ecosystem}, 
      author={Hechuan Guo and Yongle Hao and Yue Zhang and Minghui Xu and Peizhuo Lv and Jiezhi Chen and Xiuzhen Cheng},
      year={2025},
      eprint={2509.25292},
      archivePrefix={arXiv},
      primaryClass={cs.CY},
      url={https://arxiv.org/abs/2509.25292}, 
}

@article{abs-2503-23278,
    author = {Hou, Xinyi and Zhao, Yanjie and Wang, Shenao and Wang, Haoyu},
    title = {Model Context Protocol (MCP): Landscape, Security Threats, and Future Research Directions},
    year = {2026},
    publisher = {Association for Computing Machinery},
    address = {New York, NY, USA},
    issn = {1049-331X},
    note = {Just Accepted},
    journal = {ACM Trans. Softw. Eng. Methodol.},
    doi = {10.1145/3796519}
}

@inproceedings{WangZLFJZLX26,
  author       = {Zihan Wang and
                  Rui Zhang and
                  Yu Liu and
                  Wenshu Fan and
                  Wenbo Jiang and
                  Qingchuan Zhao and
                  Hongwei Li and
                  Guowen Xu},
  editor       = {Sven Koenig and
                  Chad Jenkins and
                  Matthew E. Taylor},
  title        = {{MPMA:} Preference Manipulation Attack Against Model Context Protocol},
  booktitle    = {Proceedings of the 40th {AAAI} Conference on Artificial Intelligence,
                  {AAAI} 2026},
  pages        = {35838--35846},
  publisher    = {{AAAI} Press},
  address = {Singapore},
  year         = {2026}
}

@misc{abs-2512-15144,
      title={MCPZoo: A Large-Scale Dataset of Runnable Model Context Protocol Servers for AI Agent}, 
      author={Mengying Wu and Pei Chen and Geng Hong and Baichao An and Jinsong Chen and Binwang Wan and Xudong Pan and Jiarun Dai and Min Yang},
      year={2025},
      eprint={2512.15144},
      archivePrefix={arXiv},
      primaryClass={cs.CR},
      url={https://arxiv.org/abs/2512.15144}
}

@misc{fei2025mcpzero,
      title={MCP-Zero: Active Tool Discovery for Autonomous LLM Agents}, 
      author={Xiang Fei and Xiawu Zheng and Hao Feng},
      year={2025},
      eprint={2506.01056},
      archivePrefix={arXiv},
      primaryClass={cs.AI},
      url={https://arxiv.org/abs/2506.01056}
}

@misc{laddha2025humanmcp,
      title={HumanMCP: A Human-Like Query Dataset for Evaluating MCP Tool Retrieval Performance}, 
      author={Shubh Laddha and Lucas Changbencharoen and Win Kuptivej and Surya Shringla and Archana Vaidheeswaran and Yash Bhaskar},
      year={2025},
      eprint={2602.23367},
      archivePrefix={arXiv},
      primaryClass={cs.AI},
      url={https://arxiv.org/abs/2602.23367}
}

@misc{abs-2505-16700,
      title={MCP-RADAR: A Multi-Dimensional Benchmark for Evaluating Tool Use Capabilities in Large Language Models}, 
      author={Xuanqi Gao and Siyi Xie and Juan Zhai and Shiqing Ma and Chao Shen},
      year={2025},
      eprint={2505.16700},
      archivePrefix={arXiv},
      primaryClass={cs.AI},
      url={https://arxiv.org/abs/2505.16700}
}

@misc{abs-2508-14704,
      title={MCP-Universe: Benchmarking Large Language Models with Real-World Model Context Protocol Servers}, 
      author={Ziyang Luo and Zhiqi Shen and Wenzhuo Yang and Zirui Zhao and Prathyusha Jwalapuram and Amrita Saha and Doyen Sahoo and Silvio Savarese and Caiming Xiong and Junnan Li},
      year={2025},
      eprint={2508.14704},
      archivePrefix={arXiv},
      primaryClass={cs.AI},
      url={https://arxiv.org/abs/2508.14704}
}

@misc{abs-2508-20453,
      title={MCP-Bench: Benchmarking Tool-Using LLM Agents with Complex Real-World Tasks via MCP Servers}, 
      author={Zhenting Wang and Qi Chang and Hemani Patel and Shashank Biju and Cheng-En Wu and Quan Liu and Aolin Ding and Alireza Rezazadeh and Ankit Shah and Yujia Bao and Eugene Siow},
      year={2025},
      eprint={2508.20453},
      archivePrefix={arXiv},
      primaryClass={cs.CL},
      url={https://arxiv.org/abs/2508.20453}
}

@misc{abs-2509-24002,
      title={MCPMark: A Benchmark for Stress-Testing Realistic and Comprehensive MCP Use}, 
      author={Zijian Wu and Xiangyan Liu and Xinyuan Zhang and Lingjun Chen and Fanqing Meng and Lingxiao Du and Yiran Zhao and Fanshi Zhang and Yaoqi Ye and Jiawei Wang and Zirui Wang and Jinjie Ni and Yufan Yang and Arvin Xu and Michael Qizhe Shieh},
      year={2025},
      eprint={2509.24002},
      archivePrefix={arXiv},
      primaryClass={cs.CL},
      url={https://arxiv.org/abs/2509.24002}
}

@inproceedings{GuoXZHWM26,
  author       = {Zikang Guo and
                  Benfeng Xu and
                  Chiwei Zhu and
                  Wentao Hong and
                  Xiaorui Wang and
                  Zhendong Mao},
  editor       = {Sven Koenig and
                  Chad Jenkins and
                  Matthew E. Taylor},
  title        = {MCP-AgentBench: Evaluating Real-World Language Agent Performance with
                  MCP-Mediated Tools},
  booktitle    = {Proceedings of the 40th {AAAI} Conference on Artificial Intelligence,                   {AAAI} 2026},
  pages        = {30888--30896},
  publisher    = {{AAAI} Press},
  address = {Singapore},
  year         = {2026}
}

@inproceedings{WangGWLSCSDL26,
  author       = {Zhiqiang Wang and
                  Yichao Gao and
                  Yanting Wang and
                  Suyuan Liu and
                  Haifeng Sun and
                  Haoran Cheng and
                  Guanquan Shi and
                  Haohua Du and
                  Xiangyang Li},
  editor       = {Sven Koenig and
                  Chad Jenkins and
                  Matthew E. Taylor},
  title        = {MCPTox: {A} Benchmark for Tool Poisoning on Real-World {MCP} Servers},
  booktitle    = {Proceedings of the 40th {AAAI} Conference on Artificial Intelligence,
                  {AAAI} 2026},
  pages        = {35811--35819},
  publisher    = {{AAAI} Press},
  address = {Singapore},
  year         = {2026}
}

@inproceedings{YangLMLLGYZC025,
  author       = {Ningyuan Yang and
                  Guanliang Lyu and
                  Mingchen Ma and
                  Yiyi Lu and
                  Yiming Li and
                  Zhihui Gao and
                  Hancheng Ye and
                  Jianyi Zhang and
                  Tingjun Chen and
                  Yiran Chen},
  title        = {IoT-MCP: Bridging LLMs and IoT Systems Through Model Context Protocol},
  booktitle    = {Proceedings of the {ACM} Workshop on Wireless Network Testbeds, Experimental
                  evaluation {\&} Characterization, WiNTECH 2025},
  pages        = {73--80},
  publisher    = {{ACM}},
  address = {Hong Kong, China},
  year         = {2025}
}

@misc{abs-2508-13220,
      title={MCPSecBench: A Systematic Security Benchmark and Playground for Testing Model Context Protocols}, 
      author={Yixuan Yang and Cuifeng Gao and Daoyuan Wu and Yufan Chen and Yingjiu Li and Shuai Wang},
      year={2026},
      eprint={2508.13220},
      archivePrefix={arXiv},
      primaryClass={cs.CR},
      url={https://arxiv.org/abs/2508.13220}
}

@misc{abs-2504-03767,
      title={MCP Safety Audit: LLMs with the Model Context Protocol Allow Major Security Exploits}, 
      author={Brandon Radosevich and John Halloran},
      year={2025},
      eprint={2504.03767},
      archivePrefix={arXiv},
      primaryClass={cs.CR},
      url={https://arxiv.org/abs/2504.03767}
}

@article{LiuZPYWS23,
  author       = {Mingwei Liu and
                  Chengyuan Zhao and
                  Xin Peng and
                  Simin Yu and
                  Haofen Wang and
                  Chaofeng Sha},
  title        = {Task-Oriented {ML/DL} Library Recommendation Based on a Knowledge
                  Graph},
  journal      = {{IEEE} Trans. Software Eng.},
  volume       = {49},
  number       = {8},
  pages        = {4081--4096},
  year         = {2023}
}

@inproceedings{MaWKBP024,
  author       = {Mingyu Derek Ma and
                  Xiaoxuan Wang and
                  Po{-}Nien Kung and
                  P. Jeffrey Brantingham and
                  Nanyun Peng and
                  Wei Wang},
  editor       = {Michael J. Wooldridge and
                  Jennifer G. Dy and
                  Sriraam Natarajan},
  title        = {{STAR:} Boosting Low-Resource Information Extraction by Structure-to-Text
                  Data Generation with Large Language Models},
  booktitle    = {Proceedings of the 38th {AAAI} Conference on Artificial Intelligence, {AAAI}
                  2024},
  pages        = {18751--18759},
  publisher    = {{AAAI} Press},
  address = {Vancouver, Canada},
  year         = {2024}
}

@article{YuYXCLH24,
  author       = {Junliang Yu and
                  Hongzhi Yin and
                  Xin Xia and
                  Tong Chen and
                  Jundong Li and
                  Zi Huang},
  title        = {Self-Supervised Learning for Recommender Systems: {A} Survey},
  journal      = {{IEEE} Trans. Knowl. Data Eng.},
  volume       = {36},
  number       = {1},
  pages        = {335--355},
  year         = {2024}
}

@article{HeZLM26SPE,
  author       = {Shiyu He and
                  Yuqi Zhao and
                  Qibo Li and
                  Yutao Ma},
  title        = {RGPRec: {A} RAG-Enhanced {GNN} for Personalized Task Recommendations
                  in Open-Source Communities},
  journal      = {Softw. Pract. Exp.},
  volume       = {56},
  number       = {1},
  pages        = {3--25},
  year         = {2026}
}

@inproceedings{LiaoSW22,
  author       = {Mengqi Liao and
                  S. Shyam Sundar and
                  Joseph B. Walther},
  editor       = {Simone D. J. Barbosa and
                  Cliff Lampe and
                  Caroline Appert and
                  David A. Shamma and
                  Steven Mark Drucker and
                  Julie R. Williamson and
                  Koji Yatani},
  title        = {User Trust in Recommendation Systems: {A} comparison of Content-Based,
                  Collaborative and Demographic Filtering},
  booktitle    = {Proceedings of the {CHI} Conference on Human Factors in Computing Systems, {CHI} 2022},
  pages        = {486:1--486:14},
  publisher    = {{ACM}},
  address = {New Orleans, LA, USA},
  year         = {2022}
}

@article{ZengGC23,
  author       = {Liang Zeng and
                  Jiewen Guan and
                  Bilian Chen},
  title        = {{MSBPR:} {A} multi-pairwise preference and similarity based Bayesian
                  personalized ranking method for recommendation},
  journal      = {Knowl. Based Syst.},
  volume       = {260},
  pages        = {110165},
  year         = {2023}
}

@inproceedings{WangR0YLCXYZRC024,
  author       = {Yidan Wang and
                  Zhaochun Ren and
                  Weiwei Sun and
                  Jiyuan Yang and
                  Zhixiang Liang and
                  Xin Chen and
                  Ruobing Xie and
                  Su Yan and
                  Xu Zhang and
                  Pengjie Ren and
                  Zhumin Chen and
                  Xin Xin},
  editor       = {Edoardo Serra and
                  Francesca Spezzano},
  title        = {Content-Based Collaborative Generation for Recommender Systems},
  booktitle    = {Proceedings of the 33rd {ACM} International Conference on Information
                  and Knowledge Management, {CIKM} 2024},
  pages        = {2420--2430},
  publisher    = {{ACM}},
  address = {Boise, ID, USA},
  year         = {2024}
}

@inproceedings{wei2022clear,
  author       = {Moshi Wei and
                  Nima Shiri Harzevili and
                  Yuchao Huang and
                  Junjie Wang and
                  Song Wang},
  title        = {{CLEAR:} Contrastive Learning for {API} Recommendation},
  booktitle    = {Proceedings of the 44th {IEEE/ACM} International Conference on Software Engineering,
                  {ICSE} 2022},
  pages        = {376--387},
  publisher    = {{ACM}},
  address = {Pittsburgh, PA, USA},
  year         = {2022}
}

@article{LiLTHGLNWHZ24,
  author       = {Zhihao Li and
                  Chuanyi Li and
                  Ze Tang and
                  Wanhong Huang and
                  Jidong Ge and
                  Bin Luo and
                  Vincent Ng and
                  Ting Wang and
                  Yucheng Hu and
                  Xiaopeng Zhang},
  title        = {PTM-APIRec: Leveraging Pre-trained Models of Source Code in {API}
                  Recommendation},
  journal      = {{ACM} Trans. Softw. Eng. Methodol.},
  volume       = {33},
  number       = {3},
  pages        = {72:1--72:30},
  year         = {2024}
}

@article{SilvaMH24,
  author       = {Miguel G. Silva and
                  Sara C. Madeira and
                  Rui Henriques},
  title        = {A Comprehensive Survey on Biclustering-based Collaborative Filtering},
  journal      = {{ACM} Comput. Surv.},
  volume       = {56},
  number       = {12},
  pages        = {313:1--313:32},
  year         = {2024}
}

@article{AhmadianBRFMJ25,
  author       = {Sajad Ahmadian and
                  Kamal Berahmand and
                  Mehrdad Rostami and
                  Saman Forouzandeh and
                  Parham Moradi and
                  Mahdi Jalili},
  title        = {Recommender Systems Based on Nonnegative Matrix Factorization: {A}
                  Survey},
  journal      = {{IEEE} Trans. Artif. Intell.},
  volume       = {6},
  number       = {10},
  pages        = {2554--2574},
  year         = {2025}
}

@inproceedings{PuC00LC24,
  author       = {Yuanhao Pu and
                  Xiaolong Chen and
                  Xu Huang and
                  Jin Chen and
                  Defu Lian and
                  Enhong Chen},
  editor       = {Ruslan Salakhutdinov and
                  Zico Kolter and
                  Katherine A. Heller and
                  Adrian Weller and
                  Nuria Oliver and
                  Jonathan Scarlett and
                  Felix Berkenkamp},
  title        = {Learning-Efficient Yet Generalizable Collaborative Filtering for Item
                  Recommendation},
  booktitle    = {Proceedings of the 41st International Conference on Machine Learning, {ICML} 2024},
  series       = {Proceedings of Machine Learning Research},
  volume       = {235},
  pages        = {41183--41203},
  address = {Vienna, Austria},
  publisher    = {{PMLR} / OpenReview.net},
  year         = {2024}
}

@article{ZhengNZL24,
  author       = {Xiaoyao Zheng and
                  Zhen Ni and
                  Xiangnan Zhong and
                  Yonglong Luo},
  title        = {Kernelized Deep Learning for Matrix Factorization Recommendation System
                  Using Explicit and Implicit Information},
  journal      = {{IEEE} Trans. Neural Networks Learn. Syst.},
  volume       = {35},
  number       = {1},
  pages        = {1205--1216},
  year         = {2024}
}

@article{YuenKL21,
  author       = {Man{-}Ching Yuen and
                  Irwin King and
                  Kwong{-}Sak Leung},
  title        = {Temporal context-aware task recommendation in crowdsourcing systems},
  journal      = {Knowl. Based Syst.},
  volume       = {219},
  pages        = {106770},
  year         = {2021}
}

@article{YuYWYH24,
  author       = {Ting Yu and
                  Dongjin Yu and
                  Dongjing Wang and
                  Quanxin Yang and
                  Xueyou Hu},
  title        = {Iterative framework based on multi-task learning for service recommendation},
  journal      = {J. Syst. Softw.},
  volume       = {207},
  pages        = {111873},
  year         = {2024}
}

@article{WuHWZW23,
  author       = {Le Wu and
                  Xiangnan He and
                  Xiang Wang and
                  Kun Zhang and
                  Meng Wang},
  title        = {A Survey on Accuracy-Oriented Neural Recommendation: From Collaborative
                  Filtering to Information-Rich Recommendation},
  journal      = {{IEEE} Trans. Knowl. Data Eng.},
  volume       = {35},
  number       = {5},
  pages        = {4425--4445},
  year         = {2023}
}

@article{AnandM25,
  author       = {Vineeta Anand and
                  Ashish Kumar Maurya},
  title        = {A Survey on Recommender Systems Using Graph Neural Network},
  journal      = {{ACM} Trans. Inf. Syst.},
  volume       = {43},
  number       = {1},
  pages        = {9:1--9:49},
  year         = {2025}
}

@article{TianHZCLY25,
  author       = {Yuanhui Tian and
                  Shenquan Huang and
                  Yong Zhang and
                  Zirui Chen and
                  Panfeng Li and
                  Luchuan Yu},
  title        = {Crowdsourcing task recommendation model based on collaborative knowledge
                  graph},
  journal      = {Comput. Ind. Eng.},
  volume       = {210},
  pages        = {111516},
  year         = {2025}
}

@article{YangY25,
  author       = {Kang Yang and
                  Ruiyun Yu},
  title        = {Cybertron: task-adaptive intent attention graph neural networks for
                  few-shot recommendation},
  journal      = {Knowl. Inf. Syst.},
  volume       = {67},
  number       = {12},
  pages        = {12029--12053},
  year         = {2025}
}

@inproceedings{CaoYWLPYY23,
  author       = {Yuwei Cao and
                  Liangwei Yang and
                  Chen Wang and
                  Zhiwei Liu and
                  Hao Peng and
                  Chenyu You and
                  Philip S. Yu},
  editor       = {Jie Zhang and
                  Li Chen and
                  Shlomo Berkovsky and
                  Min Zhang and
                  Tommaso Di Noia and
                  Justin Basilico and
                  Luiz Pizzato and
                  Yang Song},
  title        = {Multi-task Item-attribute Graph Pre-training for Strict Cold-start
                  Item Recommendation},
  booktitle    = {Proceedings of the 17th {ACM} Conference on Recommender Systems, RecSys
                  2023},
  pages        = {322--333},
  address = {Singapore},
  publisher    = {{ACM}},
  year         = {2023}
}

@article{KimK23,
  author       = {Mintae Kim and
                  Wooju Kim},
  title        = {Task-Oriented Collaborative Graph Embedding Using Explicit High-Order
                  Proximity for Recommendation},
  journal      = {Big Data Res.},
  volume       = {33},
  pages        = {100382},
  year         = {2023}
}

@inproceedings{Wang0WFC19,
  author       = {Xiang Wang and
                  Xiangnan He and
                  Meng Wang and
                  Fuli Feng and
                  Tat{-}Seng Chua},
  editor       = {Benjamin Piwowarski and
                  Max Chevalier and
                  {\'{E}}ric Gaussier and
                  Yoelle Maarek and
                  Jian{-}Yun Nie and
                  Falk Scholer},
  title        = {Neural Graph Collaborative Filtering},
  booktitle    = {Proceedings of the 42nd International {ACM} {SIGIR} Conference on
                  Research and Development in Information Retrieval, {SIGIR} 2019},
  pages        = {165--174},
  publisher    = {{ACM}},
  address = {Paris, France},
  year         = {2019}
}

@inproceedings{he2020lightgcn,
  author       = {Xiangnan He and
                  Kuan Deng and
                  Xiang Wang and
                  Yan Li and
                  Yong{-}Dong Zhang and
                  Meng Wang},
  editor       = {Jimmy X. Huang and
                  Yi Chang and
                  Xueqi Cheng and
                  Jaap Kamps and
                  Vanessa Murdock and
                  Ji{-}Rong Wen and
                  Yiqun Liu},
  title        = {LightGCN: Simplifying and Powering Graph Convolution Network for Recommendation},
  booktitle    = {Proceedings of the 43rd International {ACM} {SIGIR} Conference on
                  Research and Development in Information Retrieval, {SIGIR} 2020},
  pages        = {639--648},
  address = {Virtual Event, China},
  publisher    = {{ACM}},
  year         = {2020}
}

@inproceedings{SunLCYC20,
  author       = {Zhensu Sun and
                  Yan Liu and
                  Ziming Cheng and
                  Chen Yang and
                  Pengyu Che},
  editor       = {Kostas Kontogiannis and
                  Foutse Khomh and
                  Alexander Chatzigeorgiou and
                  Marios{-}Eleftherios Fokaefs and
                  Minghui Zhou},
  title        = {Req2Lib: {A} Semantic Neural Model for Software Library Recommendation},
  booktitle    = {Proceedings of the 27th {IEEE} International Conference on Software Analysis, Evolution
                  and Reengineering, {SANER} 2020},
  pages        = {542--546},
  publisher    = {{IEEE}},
  address = {London, ON, Canada},
  year         = {2020}
}

@inproceedings{LiHC0LGY21,
  author       = {Bo Li and
                  Qiang He and
                  Feifei Chen and
                  Xin Xia and
                  Li Li and
                  John C. Grundy and
                  Yun Yang},
  editor       = {Diomidis Spinellis and
                  Georgios Gousios and
                  Marsha Chechik and
                  Massimiliano Di Penta},
  title        = {Embedding app-library graph for neural third party library recommendation},
  booktitle    = {Proceedings of the 29th {ACM} Joint European Software Engineering Conference
                  and Symposium on the Foundations of Software Engineering, {ESEC/FSE} 2021},
  pages        = {466--477},
  publisher    = {{ACM}},
  address = {Athens, Greece},
  year         = {2021}
}

@article{LiGWQLXZ24,
  author       = {Duantengchuan Li and
                  Yuxuan Gao and
                  Zhihao Wang and
                  Hua Qiu and
                  Pan Liu and
                  Zhuoran Xiong and
                  Zilong Zhang},
  title        = {Homogeneous graph neural networks for third-party library recommendation},
  journal      = {Inf. Process. Manag.},
  volume       = {61},
  number       = {6},
  pages        = {103831},
  year         = {2024}
}

@article{JuYWXMLGQYWLYZ26,
  author       = {Wei Ju and
                  Siyu Yi and
                  Yifan Wang and
                  Zhiping Xiao and
                  Zhengyang Mao and
                  Hourun Li and
                  Yiyang Gu and
                  Yifang Qin and
                  Nan Yin and
                  Senzhang Wang and
                  Xinwang Liu and
                  Philip S. Yu and
                  Ming Zhang},
  title        = {A Survey of Graph Neural Networks in Real World: Imbalance, Noise,
                  Privacy and {OOD} Challenges},
  journal      = {{IEEE} Trans. Pattern Anal. Mach. Intell.},
  volume       = {48},
  number       = {3},
  pages        = {3036--3055},
  year         = {2026}
}

@inproceedings{ZhaoWWTWR24,
  author       = {Yuyue Zhao and
                  Jiancan Wu and
                  Xiang Wang and
                  Wei Tang and
                  Dingxian Wang and
                  Maarten de Rijke},
  editor       = {Grace Hui Yang and
                  Hongning Wang and
                  Sam Han and
                  Claudia Hauff and
                  Guido Zuccon and
                  Yi Zhang},
  title        = {Let Me Do It For You: Towards {LLM} Empowered Recommendation via Tool
                  Learning},
  booktitle    = {Proceedings of the 47th International {ACM} {SIGIR} Conference on
                  Research and Development in Information Retrieval, {SIGIR} 2024},
  pages        = {1796--1806},
  publisher    = {{ACM}},
  address = {Washington DC, USA},
  year         = {2024}
}

@article{JinLHJJH24,
  author       = {Bowen Jin and
                  Gang Liu and
                  Chi Han and
                  Meng Jiang and
                  Heng Ji and
                  Jiawei Han},
  title        = {Large Language Models on Graphs: {A} Comprehensive Survey},
  journal      = {{IEEE} Trans. Knowl. Data Eng.},
  volume       = {36},
  number       = {12},
  pages        = {8622--8642},
  year         = {2024}
}

@inproceedings{LuoGHKLM25,
  author       = {Ne Luo and
                  Aryo Pradipta Gema and
                  Xuanli He and
                  Emile van Krieken and
                  Pietro Lesci and
                  Pasquale Minervini},
  editor       = {Luis Chiruzzo and
                  Alan Ritter and
                  Lu Wang},
  title        = {Self-Training Large Language Models for Tool-Use Without Demonstrations},
  booktitle    = {Proceedings of the Association for Computational Linguistics: {NAACL}
                  2025},
  series       = {Findings of {ACL}},
  pages        = {1253--1271},
  address = {Albuquerque, New Mexico, USA},
  publisher    = {Association for Computational Linguistics},
  year         = {2025}
}

@inproceedings{ParamanayakamKA25,
  author       = {Varatheepan Paramanayakam and
                  Andreas Karatzas and
                  Iraklis Anagnostopoulos and
                  Dimitrios Stamoulis},
  title        = {Less is More: Optimizing Function Calling for {LLM} Execution on Edge
                  Devices},
  booktitle    = {Proceedings of the Design, Automation {\&} Test in Europe Conference 2025, {DATE} 2025},
  pages        = {1--7},
  publisher    = {{IEEE}},
  address = {Lyon, France},
  year         = {2025}
}

@inproceedings{RajputMSKVHHT0S23,
  author       = {Shashank Rajput and
                  Nikhil Mehta and
                  Anima Singh and
                  Raghunandan Hulikal Keshavan and
                  Trung Vu and
                  Lukasz Heldt and
                  Lichan Hong and
                  Yi Tay and
                  Vinh Q. Tran and
                  Jonah Samost and
                  Maciej Kula and
                  Ed H. Chi and
                  Mahesh Sathiamoorthy},
  editor       = {Alice Oh and
                  Tristan Naumann and
                  Amir Globerson and
                  Kate Saenko and
                  Moritz Hardt and
                  Sergey Levine},
  title        = {Recommender Systems with Generative Retrieval},
  pages = {10299--10315},
  booktitle    = {Proceedings of the 37th Annual Conference
                  on Neural Information Processing Systems, NeurIPS 2023},
  publisher = {Curran Associates, Inc.},
  address = {New Orleans, LA, USA},
  year         = {2023}
}

@inproceedings{XuHZ24,
  author       = {Shuyuan Xu and
                  Wenyue Hua and
                  Yongfeng Zhang},
  editor       = {Grace Hui Yang and
                  Hongning Wang and
                  Sam Han and
                  Claudia Hauff and
                  Guido Zuccon and
                  Yi Zhang},
  title        = {OpenP5: An Open-Source Platform for Developing, Training, and Evaluating
                  LLM-based Recommender Systems},
  booktitle    = {Proceedings of the 47th International {ACM} {SIGIR} Conference on
                  Research and Development in Information Retrieval, {SIGIR} 2024},
  pages        = {386--394},
  publisher    = {{ACM}},
  address = {Washington DC, USA},
  year         = {2024}
}

@inproceedings{ZhengHLCZCW24,
  author       = {Bowen Zheng and
                  Yupeng Hou and
                  Hongyu Lu and
                  Yu Chen and
                  Wayne Xin Zhao and
                  Ming Chen and
                  Ji{-}Rong Wen},
  title        = {Adapting Large Language Models by Integrating Collaborative Semantics
                  for Recommendation},
  booktitle    = {Proceedings of the 40th {IEEE} International Conference on Data Engineering, {ICDE} 2024},
  pages        = {1435--1448},
  publisher    = {{IEEE}},
  address = {Utrecht, The Netherlands},
  year         = {2024}
}

@inproceedings{Dong0L25,
  author       = {Hoang V. Dong and
                  Yuan Fang and
                  Hady W. Lauw},
  editor       = {Wolfgang Nejdl and
                  S{\"{o}}ren Auer and
                  Meeyoung Cha and
                  Marie{-}Francine Moens and
                  Marc Najork},
  title        = {A Contrastive Framework with User, Item and Review Alignment for Recommendation},
  booktitle    = {Proceedings of the 18th {ACM} International Conference on Web
                  Search and Data Mining, {WSDM} 2025},
  pages        = {117--126},
  address = {Hannover, Germany},
  publisher    = {{ACM}},
  year         = {2025}
}

@article{YangWR25,
  author       = {Jingbo Yang and
                  Wenjun Wu and
                  Jian Ren},
  title        = {{DSKIPP:} {A} Prompt Method to Enhance the Reliability in LLMs for
                  Java {API} Recommendation Task},
  journal      = {Softw. Test. Verification Reliab.},
  volume       = {35},
  number       = {2},
  year         = {2025},
  pages   =  {e1913}
}

@inproceedings{ZhangCWLWKJWY26,
  author       = {Fan Zhang and
                  Jinpeng Chen and
                  Tao Wang and
                  Huan Li and
                  Senzhang Wang and
                  Feifei Kou and
                  Ye Ji and
                  Kaimin Wei and
                  Zhenye Yang},
  editor       = {Sven Koenig and
                  Chad Jenkins and
                  Matthew E. Taylor},
  title        = {DiMA: Distinguishing Resident and Tourist Preferences via Multi-Modal
                  {LLM} Alignment for Out-of-Town Cross-Domain Recommendation},
  booktitle    = {Proceedings of the 40th {AAAI} Conference on Artificial Intelligence,
                  {AAAI} 2026},
  pages        = {16262--16270},
  address = {Singapore},
  publisher    = {{AAAI} Press},
  year         = {2026}
}

@inproceedings{PengLHYYS25,
  author       = {Qiyao Peng and
                  Hongtao Liu and
                  Hua Huang and
                  Jian Yang and
                  Qing Yang and
                  Minglai Shao},
  editor       = {Christos Christodoulopoulos and
                  Tanmoy Chakraborty and
                  Carolyn Rose and
                  Violet Peng},
  title        = {A Survey on LLM-powered Agents for Recommender Systems},
  booktitle    = {Proceedings of the Association for Computational Linguistics: {EMNLP}
                  2025},
  series       = {Findings of {ACL}},
  pages        = {11574--11583},
  address = {Suzhou, China},
  publisher    = {Association for Computational Linguistics},
  year         = {2025}
}

@article{WuZQWGSQZZLXC24,
  author       = {Likang Wu and
                  Zhi Zheng and
                  Zhaopeng Qiu and
                  Hao Wang and
                  Hongchao Gu and
                  Tingjia Shen and
                  Chuan Qin and
                  Chen Zhu and
                  Hengshu Zhu and
                  Qi Liu and
                  Hui Xiong and
                  Enhong Chen},
  title        = {A survey on large language models for recommendation},
  journal      = {World Wide Web {(WWW)}},
  volume       = {27},
  number       = {5},
  pages        = {60},
  year         = {2024}
}

@misc{abs-2601-04690,
      title={Do LLMs Benefit from User and Item Embeddings in Recommendation Tasks?}, 
      author={Mir Rayat Imtiaz Hossain and Leo Feng and Leonid Sigal and Mohamed Osama Ahmed},
      year={2026},
      eprint={2601.04690},
      archivePrefix={arXiv},
      primaryClass={cs.LG},
      url={https://arxiv.org/abs/2601.04690}
}

@misc{abs-2505-06416,
      title={ScaleMCP: Dynamic and Auto-Synchronizing Model Context Protocol Tools for LLM Agents}, 
      author={Elias Lumer and Anmol Gulati and Vamse Kumar Subbiah and Pradeep Honaganahalli Basavaraju and James A. Burke},
      year={2025},
      eprint={2505.06416},
      archivePrefix={arXiv},
      primaryClass={cs.CL},
      url={https://arxiv.org/abs/2505.06416}
}

@misc{xing2025hgmf,
      title={HGMF: A Hierarchical Gaussian Mixture Framework for Scalable Tool Invocation within the Model Context Protocol}, 
      author={Wenpeng Xing and Zhipeng Chen and Changting Lin and Meng Han},
      year={2026},
      eprint={2508.07602},
      archivePrefix={arXiv},
      primaryClass={cs.AI},
      url={https://arxiv.org/abs/2508.07602}
}

@misc{li2025netmcp,
      title={NetMCP: Network-Aware Model Context Protocol Platform for LLM Capability Extension}, 
      author={Enhan Li and Hongyang Du and Kaibin Huang},
      year={2025},
      eprint={2510.13467},
      archivePrefix={arXiv},
      primaryClass={cs.NI},
      url={https://arxiv.org/abs/2510.13467}
}

@misc{antonioni2025jsplit,
      title={JSPLIT: A Taxonomy-based Solution for Prompt Bloating in Model Context Protocol}, 
      author={Emanuele Antonioni and Stefan Markovic and Anirudha Shankar and Jaime Bernardo and Lovro Markovic and Silvia Pareti and Benedetto Proietti},
      year={2025},
      eprint={2510.14537},
      archivePrefix={arXiv},
      primaryClass={cs.AI},
      url={https://arxiv.org/abs/2510.14537}
}

@misc{ahmadi2025mcp,
      title={MCP Bridge: A Lightweight, LLM-Agnostic RESTful Proxy for Model Context Protocol Servers}, 
      author={Arash Ahmadi and Sarah Sharif and Yaser M. Banad},
      year={2026},
      eprint={2504.08999},
      archivePrefix={arXiv},
      primaryClass={cs.CR},
      url={https://arxiv.org/abs/2504.08999}
}

@misc{abs-2508-07575,
      title={MCPToolBench++: A Large Scale AI Agent Model Context Protocol MCP Tool Use Benchmark}, 
      author={Shiqing Fan and Xichen Ding and Liang Zhang and Linjian Mo},
      year={2025},
      eprint={2508.07575},
      archivePrefix={arXiv},
      primaryClass={cs.AI},
      url={https://arxiv.org/abs/2508.07575}
}

@misc{luo2025mcp,
      title={MCP-Universe: Benchmarking Large Language Models with Real-World Model Context Protocol Servers}, 
      author={Ziyang Luo and Zhiqi Shen and Wenzhuo Yang and Zirui Zhao and Prathyusha Jwalapuram and Amrita Saha and Doyen Sahoo and Silvio Savarese and Caiming Xiong and Junnan Li},
      year={2025},
      eprint={2508.14704},
      archivePrefix={arXiv},
      primaryClass={cs.AI},
      url={https://arxiv.org/abs/2508.14704}
}

@misc{song2025help,
      title={Help or Hurdle? Rethinking Model Context Protocol-Augmented Large Language Models}, 
      author={Wei Song and Haonan Zhong and Ziqi Ding and Jingling Xue and Yuekang Li},
      year={2025},
      eprint={2508.12566},
      archivePrefix={arXiv},
      primaryClass={cs.AI},
      url={https://arxiv.org/abs/2508.12566}, 
}

@inproceedings{HeLZNHC17,
  author       = {Xiangnan He and
                  Lizi Liao and
                  Hanwang Zhang and
                  Liqiang Nie and
                  Xia Hu and
                  Tat{-}Seng Chua},
  editor       = {Rick Barrett and
                  Rick Cummings and
                  Eugene Agichtein and
                  Evgeniy Gabrilovich},
  title        = {Neural Collaborative Filtering},
  booktitle    = {Proceedings of the 26th International Conference on World Wide Web,
                  {WWW} 2017},
  pages        = {173--182},
  publisher    = {{ACM}},
  address = {Perth, Australia},
  year         = {2017}
}

@article{LinDXLCZLWLZGYTZ25,
  author       = {Jianghao Lin and
                  Xinyi Dai and
                  Yunjia Xi and
                  Weiwen Liu and
                  Bo Chen and
                  Hao Zhang and
                  Yong Liu and
                  Chuhan Wu and
                  Xiangyang Li and
                  Chenxu Zhu and
                  Huifeng Guo and
                  Yong Yu and
                  Ruiming Tang and
                  Weinan Zhang},
  title        = {How Can Recommender Systems Benefit from Large Language Models: {A}
                  Survey},
  journal      = {{ACM} Trans. Inf. Syst.},
  volume       = {43},
  number       = {2},
  pages        = {28:1--28:47},
  year         = {2025}
}

@ARTICLE{11400603,
  author={Gao, Shanquan and Wang, Yihui and Ou, Zhenwei},
  journal={IEEE Trans. Software Eng.}, 
  title={CVH-REC: A novel method for web API recommendation based on cross-view HGNNs}, 
  year={2026},
  volume={},
  number={},
  note={Early Access},
  doi={10.1109/TSE.2026.3666184},
  pages={1-16}
}

@article{BaiHYYHZS25,
  author       = {Ting Bai and
                  Le Huang and
                  Yue Yu and
                  Cheng Yang and
                  Cheng Hou and
                  Zhe Zhao and
                  Chuan Shi},
  title        = {Efficient Multi-task Prompt Tuning for Recommendation},
  journal      = {{ACM} Trans. Inf. Syst.},
  volume       = {43},
  number       = {4},
  pages        = {104:1--104:21},
  year         = {2025}
}

@article{ChenXLHL23,
  author       = {Jiajia Chen and
                  Xin Xin and
                  Xianfeng Liang and
                  Xiangnan He and
                  Jun Liu},
  title        = {GDSRec: Graph-Based Decentralized Collaborative Filtering for Social
                  Recommendation},
  journal      = {{IEEE} Trans. Knowl. Data Eng.},
  volume       = {35},
  number       = {5},
  pages        = {4813--4824},
  year         = {2023}
}

@inproceedings{adamw,
  author       = {Ilya Loshchilov and
                  Frank Hutter},
  title        = {Decoupled Weight Decay Regularization},
  booktitle    = {Proceedings of the 7th International Conference on Learning Representations, {ICLR} 2019},
  publisher    = {OpenReview.net},
  address = {New Orleans, LA, USA},
  year         = {2019},
  url          = {https://openreview.net/forum?id=Bkg6RiCqY7},
  pages = {1--18}
}

% \appendix



\end{document}